\documentclass[aps, pra, a4paper, showpacs, twocolumn, english, 10pt]{revtex4-1}
\usepackage{bbm, amsthm, bm,textcomp, nicefrac,geometry,ragged2e}
\usepackage[dvipsnames]{xcolor}
\usepackage{float}
\usepackage[bbgreekl]{mathbbol}
\usepackage{graphicx,epstopdf,color,verbatim,enumitem,ulem}
\usepackage{pbox,array}
\usepackage{mathrsfs}
\usepackage{hyperref}
\usepackage[euler]{textgreek}
\hypersetup{
    colorlinks = true,
    linkcolor = Black,
    citecolor = Black,
    filecolor = Black,      
    urlcolor = Black,
}
\usepackage{amssymb}
\usepackage{mathtools}
\usepackage{stackrel}
\usepackage[thinlines]{easytable}
\usepackage{amsmath}
\makeatletter
\usepackage[caption=false]{subfig}
\usepackage[T1]{fontenc}
\usepackage[caption=false]{subfig}
\usepackage{babel}
\addto\captionsenglish{}

\newcommand\id{\mathbbm{1}}
\newcommand{\ketbra}[2]{\left| #1\right\rangle\!\left\langle#2\right|}

\newcommand{\ket}[1]{\left|#1\right\rangle}

\newcommand{\bra}[1]{\left\langle #1\right|}

\newcommand{\proj}[1]{\ket{#1}\!\bra{#1}}
\newcommand{\projf}[1]{\big|#1\big\rangle\big\langle#1\big|}

\definecolor{brickred}{rgb}{0.8, 0.0, 0.0}

\begin{document}

\title{Entanglement Purification by Counting and Locating Errors with Entangling Measurements}
\author{F. Riera-S\`abat$^1$, P. Sekatski$^2$, A. Pirker$^{1}$ and W.~D\"ur$^1$}
\affiliation{$^1$Institut f\"ur Theoretische Physik, Universit\"at Innsbruck, Technikerstra{\ss}e 21a, 6020 Innsbruck, Austria\\
$^2$Departement Physik, Universit\"at Basel, Klingelbergstra{\ss}e 82, 4056 Basel, Switzerland}
\date{\today}

\begin{abstract}
We consider entanglement purification protocols for multiple copies of qubit states. We use high-dimensional auxiliary entangled systems to learn about the number and positions of errors in the noisy ensemble in an explicit and controlled way, thereby reducing the amount of noise in the ensemble and purifying the remaining states. This allows us to design entanglement purification protocols for any number of copies that work particularly well for a small number of expected errors, i.e., high fidelity of initial states. The main tool is a counter gate with which the required nonlocal information can be transferred into the high-dimensional entangled qudit auxiliary states. We compare our schemes to standard recurrence protocols that operate on pairs of copies, and hashing and breeding protocols that operate on a (asymptotically) large number of copies. Our protocols interpolate between these two regimes, leading to higher achievable fidelity and yield. We illustrate our approach for bipartite qubit states and generalize it to purify multiparty GHZ states.
\end{abstract}

\maketitle
\section{Introduction}
\label{sec Intro}

Entanglement is a central resource for many applications in quantum information processing. Entangled states play a key role in quantum networks \cite{kimble2008quantum,wehner2018quantum}, for security applications such as secret key agreement or secret sharing \cite{ekert1991quantum,pirandola2017fundamental}, but also in distributed quantum computation \cite{dur2003entanglement,li2012high} or distributed metrology \cite{sekatski2020optimal,wolk2020noisy}. However, entanglement is fragile under noise and decoherence, and needs to be properly protected to be maintained. In particular, the distribution of entanglement over noisy channels poses a challenge, and schemes have been introduced to achieve this under realistic conditions. The usage of quantum error correction \cite{nielsen2002quantum,steane1998introduction,gottesman1997stabilizer,campbell2008measurement} is one possibility, but since error correction is universally applicable to protect arbitrary unknown quantum information, requirements are stringent and error thresholds are rather low. In contrast, entanglement purification \cite{bennett1996purification, bennett1996mixed,deutsch1996quantum,dur2007entanglement,rozpkedek2018optimizing,krastanov2019optimized,vollbrecht2005interpolation,ruan2018adaptive,dehaene2003local,bombin2005entanglement,buscemi2010distilling,bratzik2013quantum,fang2019non,brandao2011one,bombin2006topological,de2020protocols,fang2020no,fang2020no2,SDP-bound,hu2021long,Zhou:20} is a direct way to establish high-fidelity entangled states from multiple copies of noisy entangled states. In contrast to error correction, target states are known and it was shown that such schemes have a higher error tolerance \cite{dur2007entanglement, fujii2009entanglement, zwerger2013universal, nickerson2014freely, zwerger2016measurement, yamamoto2001concentration}. In particular for long-distance communication based on quantum repeaters \cite{briegel1998quantum,dur1999quantum,Ladd_2006,RevModPhys.83.33,sheng2013hybrid,azuma2015all,zwerger2016measurement,zwerger2018long,pirandola2017fundamental,pirandola2019end}, entanglement purification is an essential element and represents the bottleneck for rates and error tolerance due to rather large overheads. It is hence of outermost importance to design entanglement purification schemes with high error thresholds and high yield.

Here we introduce a class of entanglement purification protocols for qubit entangled states. In contrast to all previous schemes, the protocols are based on the usage of high-dimensional auxiliary states that allow one to purify qubit entanglement. These extra dimensions and levels turn out to offer new possibilities and lead to more efficient schemes for entanglement purification. The basic idea is similar as in hashing and breeding protocols \cite{bennett1996mixed,bennett1996purification}, where extra entanglement is used to read-out nonlocal information about the noisy ensemble. This information allows one to detect (and eventually correct) the presence and position of errors within the ensemble, and thereby purifying it. While in hashing and breeding this information is gathered by measuring parities of randomly chosen subsets, thereby excluding more and more incompatible state configurations, here we gather the required information in a controlled and specific way. To this aim, we introduce a so-called counter gate as a central tool. This gate allows one to transfer information about specific kinds of errors to an entangled, high-dimensional auxiliary system in such a way that not only parities can be extracted, but also the number of errors--or even the position of errors directly--can be obtained. This provides a much more efficient and direct way to design entanglement purification protocols for systems with a small or moderate number of copies. In fact, one can establish deterministic entanglement purification protocols that can detect up to a certain number of errors in the ensemble. If the number of errors in the ensemble is bounded, our scheme produces perfect states with unit fidelity in a deterministic way provided local operations are noiseless. Clearly, noise in local operations influences the achievable fidelity, as we demonstrate. The protocols can run in two modes, deterministic and probabilistic. In deterministic mode, the protocol always succeeds in determining the error configuration. In probabilistic mode, the protocol may abort if one encounters a situation with too many errors. In this case, determining the exact configuration is too costly and would require more entanglement than can be gained. The achievable fidelity in probabilistic mode is typically higher, but also an improved yield is possible in some cases.

We introduce our scheme for a particularly relevant subclass of noise channels, namely amplitude damping channels. In this case, sending particles of a maximally entangled state with a single excitation, e.g., the Bell state $|\Psi_{01}\rangle = \big( |01\rangle + |10\rangle \big) / \sqrt{2}$, through a channel, or trying to store the entangled state in an (imperfect) quantum memory, the excitation may eventually decay and leaves one with a product state $|0\rangle \otimes |0\rangle$. For many physical set-ups, this might actually be the dominant noise source. For such states, we introduce a counter gate that allows one to count the number of error states within an ensemble using some maximally entangled auxiliary state of high dimension. We show that this tool is sufficient to learn all required information about a noisy ensemble of multiple copies, thereby purifying it. To understand the underlying idea, one may describe the noisy ensemble of $n$ identical copies of a mixed state $\hat{\rho} (F) = F | \Psi_{11}\rangle \langle \Psi_{11}| + ( 1 - F ) |00\rangle \langle 00 |$ as a mixture of $k$ noisy states $|00\rangle$, and $n - k$ maximally entangled states $|\Psi_{11}\rangle$, where each particular configuration has a probability of $F^{n - k}( 1 - F )^k$ to occur. For a given $k$, there are $\binom{n}{k}$ configurations. The goal is to figure out in which configuration the system is. If this can be achieved, the mixedness of the noisy ensemble is removed and it is described by a pure state, corresponding to one of the possible configurations. The method is inspired by so-called breeding protocols \cite{bennett1996purification,bennett1996mixed,dur2007entanglement}, where already prepurified entangled states are used to learn information about the ensemble thereby purifying it. Entangled states that were used up by the protocol need to be returned in the end, and only if more states are produced than used during the process, the protocol has a nonzero yield and is meaningful. In contrast to our scheme, breeding is an asymptotic scheme that collects parity information about the ensemble using entangled qubit pairs, and requires a (asymptotically) large number of copies to work. We, however, provide an explicit scheme that works for any finite number of copies and collects the required information directly in a high-dimensional auxiliary system.

A second important family of noisy states are ones that are generated by dephasing. In this case, one ends up with a mixture of two Bell states, which we also show how to purify directly. In fact the generalized protocol we introduce is capable to deal with two types of errors, $|01\rangle$ and $|10\rangle$ where now $|\Psi_{00}\rangle = \big( |00\rangle +|11 \rangle \big) / \sqrt{2}$ is the desired state, or equivalently with mixtures of three different Bell states. The counter gate now acts in such a way that for the desired maximally entangled state the counter state is unchanged, while it ``counts down'' for $01$ errors and up for $10$ errors. Still, it is possible with a slightly more involved scheme to read out all required information and purify the ensemble. Finally, we show that also for completely general mixed states, our scheme is applicable. We show this by treating so-called Werner states, i.e., mixtures of the desired maximally entangled state and the identity. Since {\em any} bipartite state of two qubits can be brought to Werner form without changing the fidelity, our protocol allows one to purify any state with sufficiently high initial fidelity.

A crucial feature of an EPP is if it can still provide purified states in presence of noise. We show that the EIPs that we introduce here can be implemented with noisy auxiliary states that only contain $X$-errors. Such noisy states can be obtained from several copies of a rank-2 Bell diagonal state, what provides a way to obtain the auxiliary states from the initial ensemble, as always one can use a different EPP to pre-purify part of the ensemble and obtain Bell-diagonal rank-2 states. This allows us to drop the requirement of having access to a set of pure maximally entangled states of qudits, what makes the protocol more readily applicable. Also, our EIP is robust under imperfect operations, as the number of gates applied is small enough.

We also generalize our approach to multipartite entangled states, namely Greenberger-Horne-Zeilinger (GHZ) states. We also show that a more realistic version of the scheme, inspired by hashing \cite{bennett1996mixed,dur2007entanglement} is possible. In this case, rather than using noiseless maximally entangled auxiliary states, noisy states that are formed from states of the initial ensemble are used. This also allows us to show that the schemes are robust against noise in local operations, and can compete with previously known purification protocols such as breeding, hashing, or recurrence protocols. Recurrence protocols \cite{dur2007entanglement} typically operate on only two copies of a state at a time \footnote{There exist variants that act on more copies, constructed from quantum error correction codes.}, and increase the fidelity in each step probabilistically. Hashing and breeding protocols operate on a (asymptotically) large ensemble. They are deterministic but require a particular measurement-based implementation to be applicable in realistic situations where local noise and decoherence in operations are also taken into account \cite{zwerger2014robustness}. With our schemes, we interpolate between the two regimes. We not only provide explicit protocols for any number of copies, but also variants that work deterministic providing a high yield at the prize of slightly reduced output fidelity, or probabilistically where output fidelity and possibly yield can be increased.

The paper is organized as follows. In Sec.~\ref{sec Backg}, we provide some background information and the required definitions and methods. In Sec.~\ref{sec new gadgets}, we introduce a bilateral controlled gate that operates between qubits and qudits systems. We also describe an alternative depolarization procedure for qubits. In Sec.~\ref{sec error identification protocol}, we introduce our entanglement purification protocol. We describe the detailed procedure and we analyze the yield and the fidelity of the purified ensemble for three different classes of states. In Sec.~\ref{sec Noise model}, we analyze the protocol under the presence of noise in the auxiliary states and in the operations. In Sec.~\ref{sec multipartite}, we extend the protocol to purify GHZ states using auxiliary GHZ states of qudits. A discussion and a summary are given in Sec.~\ref{sec Conclusions}. In Ref. \cite{riera2020assisted}, we provide a condensed description of the protocols and the main results of this paper.

\section{Background}
\label{sec Backg}

\subsection{Entanglement purification protocols}

Entanglement purification protocols are procedures that decrease the noise of ensembles of copies of a noisy entangled state by means of local operation and classical communication (LOCC). Since entanglement can not increase under such transformations, the main idea is to concentrate the entanglement of an ensemble of $n$ noisy entangled states into a few pairs of qubits, $n\rightarrow m$ where $m<n$, and to discard the others.

\subsubsection{Bell states}
\textit{Bell states} are pure maximally entangled states of two qubits which form an orthonormal basis of the Hilbert space $\mathcal{H} = \mathbb{C}^2 \otimes \mathbb{C}^2$. The four Bell states are defined as
\begin{equation}
    |\Psi_{ij}\rangle_{AB} \equiv \id \otimes \hat{\sigma}^j_x \hat{\sigma}^i_z \, \left( \frac{ |00\rangle_{AB} + |11\rangle_{AB} }{\sqrt{2}} \right),
\end{equation}
where $\hat{\sigma}_k$ are the Pauli matrices and, $i$, $j\in\mathbb{Z}_2$ are called the phase bit and the amplitude bit respectively. Any two qubit state can thus be expressed as
\begin{equation}
    \hat{\rho} = \sum_{i, j, m, n} \alpha_{ijmn} |\Psi_{ij}\rangle \langle\Psi_{mn}|.
\end{equation}
If one is ready to sacrifice the entanglement of the state, the value of the amplitude or the phase bit can be obtained by LOCC. One determines the value of the bit by measuring both qubits in the $Z$ or $X$ basis respectively and then comparing the two outcomes of the measurements. When one bit of information is obtained the state collapses and the information about the other bit is destroyed, i.e., only one of the bits is accessible by local measurements. Accessing both bits requires joint measurements on both qubits, which is not possible without additional entanglement in a nonlocal setting.

\subsubsection{Fidelity}

The entanglement of a two-qubit state $\hat{\rho}_{AB}$ can be quantified by its \textit{fidelity} $F$ with a maximally entangled state. The fidelity is given by
\begin{equation}
    F = \langle \Psi_{00}| \; \hat{\rho}_{AB} \; |\Psi_{00}\rangle,
\end{equation}
where w.l.o.g we assume the maximally entangled state is $|\Psi_{00}\rangle$, keeping in mind that the fidelity shall be maximized by means of local unitary (LU) transformations of the state. The fidelity range is $0 \leq F \leq 1$, where $F = 1$ iff the mixed state is $\hat{\rho}_{AB} = |\Psi_{00}\rangle \langle\Psi_{00}|$, and $F = 0$ iff $\hat{\rho}_{AB}$ is orthogonal to $|\Psi_{00}\rangle \langle\Psi_{00}|$.

After applying a purification protocol, in general, the post-purified states are correlated, i.e., the density operator of the ensemble is transformed as $\hat{\rho}^{\otimes n}_{AB}\rightarrow\hat{\Gamma}$. In these situations, the fidelity of the $k^{\text{th}}$ state is defined as
\begin{equation}
    F^{(k)}=\langle\Psi_{00}|\;\hat{\Gamma}_{k}\;|\Psi_{00}\rangle,
\end{equation}
where $\hat{\Gamma}_{k} \equiv \text{tr}_{\neg k}\hat{\Gamma}$ is the reduced density matrix of the $k^{\text{th}}$ state. In this cases we refer to the average value of the fidelity of each state as the \textit{local fidelity} of the ensemble, i.e., 
\begin{equation}
    \label{eq local fidelty}
    F=\frac{1}{m}\sum_{k=1}^mF^{(k)}.
\end{equation}
In addition, since all the ensemble is described with a single density operator $\hat{\Gamma}$, it is also necessary to quantify the noise of $\hat{\Gamma}$ with respect to the pure ensemble of $m$ states $|\Psi_{00}\rangle^{\otimes m}$. This is done via the \textit{global fidelity}, $F_g$, defined as
\begin{equation}
    F_g = \langle\Psi_{00}|^{\otimes m} \; \hat{\Gamma} \; |\Psi_{00}\rangle^{\otimes m}.
\end{equation}
For $m$ identical copies, i.e., $\hat \Gamma = \hat \rho_{AB}^{\otimes m}$, the two fidelities are simply related by $F_g = F^m$. When the final state of the $m$ qubit pairs can be correlated the two quantities can deviate a lot from this simple relation, depending on the nature of correlations. We show in Appendix~\ref{app fidelities}, that the possible values of the fidelities satisfy the tight bounds
\begin{equation}
\label{eq: F tradeof}\begin{split}
        F_g \leq \, & F \leq 1 - \frac{1 - F_g}{m},
        \\ 1 - m ( 1 - F ) \leq \, & F_g \leq F 
\end{split}
\end{equation}
Here the left inequality is saturated by the states of the ensemble where the errors maximally bunch, and the right inequality is saturated by the states where the errors maximally anti-bunch. Hence, for a general state $\hat \Gamma$ both $F$ and $F_g$ are important quantifiers of the entanglement in the ensemble, providing complementary insights on the distribution of errors.

Notably, for product states with a large local fidelity $F = 1 -\epsilon \approx 1$, the global fidelity
\begin{equation}
    F_g = F^m = 1 - m \epsilon + O \left( \epsilon^2 \right)
\end{equation}
is saturating its worst-case value $F_g$ of Eq.~\eqref{eq: F tradeof} to the leading order of $\epsilon$.

\subsubsection{Bell diagonal states and the depolarization map}
\label{sec belld depolarization}

An important family of two-qubit states are the so-called Bell-diagonal states, taking the form
\begin{equation}
    \hat{\rho}_{BD} = \sum_{i, j} p_{ij} \proj{\Psi_{ij}}.
\end{equation}
These are all the states diagonal in the Bell basis (up to local basis change LU), or all the states that have random marginals $\hat{\rho}_A = \hat{\rho}_B = \frac{1}{2} \id$. Without loss of generality we can assume that $p_{00} \geq p_{01},p_{10},p_{11}$. The fidelity of a Bell diagonal state is thus simply given by the maximal weight $F = p_{00}$.

Any two-qubit state $\hat{\rho}_{AB}$ can be brought to a Bell diagonal form by means of a \textit{depolarization map} $\mathcal{D}$, given by the Kraus operators $\mathcal{D}_1:\{\frac{1}{2}\hat{\sigma}_{i} \otimes \hat{\sigma}_{i} \}^{3}_{ i = 0 }$ \cite{bennett1996mixed,bennett1996purification,deutsch1996quantum}. It transforms a general two qubit state as
\begin{equation}
    \begin{split}
        \mathcal{D}_1 : \hat{\rho} &= \sum_{i, j, m, n} \alpha_{ ijmn } \ketbra{\Psi_{ij}}{\Psi_{mn}}\\ &\mapsto \sum_{i, j}\underbrace{\alpha_{ ijij }}_{ = \, p_{ij}} \ketbra{ \Psi_{ij} }{\Psi_{ij}}.
    \end{split}
\end{equation}
In practice, depolarization can be realized by applying one of the four local unitary transformations $\hat{\sigma}_i \otimes \hat{\sigma}_i$ at random, and thus only requires local operations and shared randomness (LOSR). 

Crucially, the depolarized process does not change the fidelity of a state, nor does it affect the local and global fidelity of an ensemble (if applied on any of the pairs). For this reason in the construction of a purification protocol, it is convenient to assume that Bell-diagonal states are supplied to start with. A protocol that can purify Bell-diagonal states, also works for any kind of mixed state, as one always can run a preprocessing depolarization step to reach the starting situation without lowering the fidelity.

\subsubsection{Yield}

The second parameter we use to evaluate the performance of an entanglement purification protocol is the yield $Y$. The yield of protocol $\mathcal{P}$ which maps $n$ copies of a fixed state $\hat{\rho}$ to $m$ ideal target states (Bell pairs $\ket{\Psi_{00}}$) is defined as $Y_{\mathcal{P},\hat{\rho}} = \frac{m}{n}$. In other words, it corresponds to the ratio of the number of maximally entangled states obtained over the total number of initial copies.

In some protocols, auxiliary maximally entangled states are used and destroyed during the execution. In these cases, the number of consumed auxiliary states has to be given back from the purified ensemble. The number of used auxiliary ebits is called the \textit{resources} $R$. Therefore, the effective number of obtained target states is given by $m=\#\text{purified states}-R$.

In realistic situations, it is not possible to obtain pure Bell states due to the limitations of each protocol. Therefore in these cases, we define a target local fidelity or global fidelity, $F^{(t)}=1-\epsilon$, and we consider the states whose fidelity is larger than the target fidelity as successfully purified. For protocols that rely on auxiliary maximally entangled states, this implies that one can not give maximally entangled states back. In these cases, for computing the yield, we assume that the entanglement of a mixed state $\hat{\rho}$ is equivalent to $1-S(\hat{\rho}_{BD})$ ebits, where $S(\hat{\rho}) = - \text{tr} (\hat{\rho} \log_2 \hat{\rho} )$ is the von Neumann entropy, and $S(\hat{\rho}_{BD})$ the entropy of one purified state in its Bell-diagonal from. This factor corresponds to the yield reached with the hashing protocol in the asymptotic limit, which one can use to obtain pure Bell states \cite{bennett1996purification}. Note that any protocol that provides Bell states with unit fidelity could be used in general. Then we consider that $n$ copies of a mixed state are equivalent to $n [ 1 - S(\hat{\rho}_{BD}) \, ]$ maximally entangled states since this is the number of perfect Bell-states which hashing would purify. Therefore, since the resources are maximally entangled states the yield is given by
\begin{equation}
    Y_\epsilon = \frac{\#\text{purified states}_\epsilon - [1 - S(\hat{\rho}'_{BD}) \, ]^{-1} R}{n},
\end{equation}
as we use $( 1 - S (\hat{\rho}'_{BD}) \, )^{-1}R$ states of the purified ensemble to give back the $R$ auxiliary e bits lost in the process, where $\hat{\rho}'_{BD}$ is the Bell-diagonal form of the purified states.

\subsubsection{Bottom to top approach to purification -- recurrence protocols}

An important class of protocols are the recurrence protocols \cite{bennett1996purification,bennett1996mixed,deutsch1996quantum,dur2003entanglement}. These are probabilistic protocols which typically take the states to purify in pairs $\hat{\rho}_{A_1 B_1}\otimes \hat{\rho}_{A_2 B_2}$ and apply local gates $U_{A_1 A_2}$ and $V_{B_1 B_2}$ on each side. After that, one of the two systems, say $A_2$ and $B_2$, is measured out on both sides revealing information about the other copy. Depending on the outcome of the measurement, the fidelity of the state $\hat{\rho}'_{A_1B_1}$ of the remaining system is increased, otherwise, it is simply discarded. This step is iterated using for each round two copies of the state $\hat{\rho}'_{A_1 B_1}$ for which the purification was successful at the previous step. These protocols allow one to obtain states with fidelity arbitrarily close to 1. However, an infinite number of steps are necessary to obtain a pure Bell state. Moreover, in each step, half of the states of the ensemble are measured out, and hence, at least half of the ensemble is destroyed. This leads to a vanishing yield in the asymptotic limit if states with unit fidelity are required. Recurrence protocols are a good choice for purifying ensembles of small sizes and when only states with a certain target fidelity are demanded.

Recurrence protocols are highly robust under noisy operations \cite{dur2007entanglement,zwerger2016measurement}. A noisy gate-based implementation of such protocols produces states with a lower fidelity as compared to the ideal case. In the asymptotic limit, the output fidelity is also reduced which means only states of a certain fidelity can be obtained.

\subsubsection{Top to bottom approach to purification -- ensemble interpretation}

An alternative approach to recurrence protocols is to deal with the state of the $n$-copy ensemble as a whole. The basic idea is as follows: An ensemble of $n$ copies of a mixed state $\hat{\rho}$ is a mixture of products of pure states where each possible configuration occurs with a certain probability. Precisely, the overall distribution describing the ensemble is a product of pure state distributions for each pair, i.e., the state $\hat{\rho}^{\otimes n}$ is an ensemble $\{ p_{\mu_1} \cdots p_{\mu_n}, \, |\phi_{\mu_1} \rangle \cdots | \phi_{\mu_n} \rangle\}$ where $p_{\mu}$ and $|\phi_{\mu} \rangle$ are the eigenvalues and eigenstates of $\hat{\rho}$.

By performing measurements one can determine the configuration and obtain $n$ pure states. However, when the states of the distribution $|\phi_{\mu}\rangle$ are entangled the measurement required to reveal information about the configuration is \textit{entangling}, and hence some entanglement has to be consumed in order to perform it. In principle, this can be done by consuming auxiliary Bell states. For the case where the ensemble is made of Bell-diagonal states, we show in Appendix~\ref{app sec dichotomic measurement} that any dichotomic measurement, that can distinguish two arbitrary configurations, requires at least one Bell state (ebit) to be performed. In fact, we also show that some dichotomic measurements cost more entanglement. Thus, the general rule of thumb is that at most one bit of information about the ensemble can be obtained from each auxiliary Bell state. This clearly suggests that it is unfeasible to learn all possible configurations--the required resources would exceed $n$. However, to get a positive yield one can try to only discriminate between \textit{the most probable configurations} versus all the others.

\subsubsection{The bilateral CNOT}

Let us now review an important tool that allows one to transfer one bit of information about the ensemble onto an auxiliary Bell state, and is at the core of breeding and hashing entanglement protocols discussed in the next paragraph \cite{bennett1996mixed,dur2007entanglement}. To this end consider a two-qubit state $\hat{\rho}_{A_1 B_1}$ and a known auxiliary Bell state $|\Psi_{vw}\rangle_{A_2 B_2}$. The tool consists of applying a bilateral CNOT gate, that is a CNOT on both sides,
\begin{equation}
    b \text{CNOT}_{1 \rightarrow 2} = \text{CNOT}^{A_1 A_2}_{1 \rightarrow 2} \otimes \text{CNOT}^{B_1 B_2}_{1 \rightarrow 2}
\end{equation}
where $\text{CNOT}_{1\rightarrow2} = \proj{0}_1\otimes \id_2 + \proj{1}_1 \otimes \hat{X}_2$. To understand the role of the bilateral CNOT consider its action on Bell states $\hat{\rho}_{A_1 B_1}=\proj{\Psi_{ij}}$
\begin{equation}
    \label{eq backaction}
    \begin{split}
        b \text{CNOT}_{1 \rightarrow 2}|\Psi_{ij}\rangle_{A_1 B_1} |\Psi_{vw} \rangle_{A_2 B_2} & \\ = \, |\Psi_{i\oplus v, j} \rangle_{A_1 B_1}|\Psi_{v, w\oplus j} \rangle_{A_2 B_2} &,
    \end{split}
\end{equation}
where $i \oplus j \equiv (i + j) \text{mod 2}$. $b \text{CNOT}$ adds the value of the phase bit of the control Bell state onto the phase bit of the auxiliary Bell state, and the value of the amplitude bit of the auxiliary state onto the amplitude bit of the control state. At this point local measurements on the auxiliary systems allow one to learn the value of $j$ consuming one ebit, provided that $w$ is known.

However, it also allows performing more general dichotomic measurements on the ensemble. If $b$CNOT is applied repeatedly on different pairs of qubits $\{ k_1, k_2, \dots, k_\ell \}$ in the ensemble as a control system and the same auxiliary pair as target system (initially in the state $|\Psi_{00}\rangle$), the final value of the phase bit of the latter encodes the parity (the binary sum) of the phase bits of the states from the ensemble on which the gate was applied $w=j_{k_1}\oplus j_{k_2}\oplus\dots\oplus j_{k_\ell}$. Furthermore, it is easy to see that combining the bilateral CNOT with QFT in Eq.~\eqref{eq QFT} on the control qubits allows one to map the value of the amplitude bit onto the control state $w\mapsto w\oplus i$. It follows, that a proper combination of $b$CNOT and QFT gates offers the possibility to measure the value of any function
\begin{equation}
    w = j_{k_1} \oplus j_{k_2} \oplus \dots \oplus j_{k_\ell} \oplus i_{r_1} \oplus i_{r_2} \oplus \dots \oplus i_{r_{\ell'}},
\end{equation}
by consuming one auxiliary Bell pair. 

\subsubsection{Breeding and hashing protocols}

In breeding protocols \cite{bennett1996purification,bennett1996mixed,dur2007entanglement}, the states of the ensemble are first depolarized to their Bell-diagonal form. The ensemble is then a mixture of products of $n$ Bell states, i.e., the product distribution given by $\{p_{i_1 j_1}\cdots p_{i_nj_n}, |\Psi_{i_1 j_1} \rangle\cdots|\Psi_{i_n j_n}\rangle\}$. As already mentioned, trying to identify all possible configurations, given by the $2n$-bit string $(i_1, j_1, \dots, i_n, j_n)$, would consume $2n$ Bell pairs and give a negative yield. Instead, the breeding protocol relies on the asymptotic equipartition property and only discriminates the configuration that belongs to the typical set \cite{cover1999}. The typical set is composed of $2^{n S(\hat{\rho}_{BD})}$ configurations, where $\hat{\rho}_{BD}$ are the states of the initial copies in its Bell-diagonal form and $S$ is the von Neumann entropy. Hence the procedure only requires to consume $n S( \hat{\rho}_{BD})$ Bell pairs, implying a yield of $Y = 1 - S(\hat{\rho}_{BD})$. Furthermore, in the asymptotic limit, $n \to \infty$, the probability of occurrence of a configuration outside the typical set approaches zero. Therefore the fidelity of the purified ensemble approaches one $F'_g \rightarrow 1$.

The hashing protocol \cite{bennett1996mixed,dur2007entanglement} is based on the same procedure as breeding, however, it does not use pure auxiliary states. The information about the string is encoded in the state of the ensemble itself instead of auxiliary Bell states. The usage of Bell states with a nonzero phase bit as a target of the bilateral controlled-NOT gate introduces a back action to the control states, see Eq.~\eqref{eq backaction}. However, in the asymptotic limit, the initial bit string of the whole ensemble can be determined and the back action corrected providing the same results as the breeding protocol.

In the case of breeding and hashing protocols, a noisy gate-based implementation is unfeasible. This is because the number of operations applied to the target states approaches infinity in the asymptotic limit. Therefore the noise accumulates making the information about the ensemble unreadable. Nevertheless, this problem can be overcome by considering a measurement-based implementation of the protocols \cite{raussendorf2001one,zhou2003quantum,briegel2009measurement,zwerger2013universal,zwerger2014robustness}.

\subsection{Maximally entangled states for bipartite systems of qudits}\label{sec backg qudits}

In different fields, most notable in quantum communication and computation, the advantage to use qudits rather than qubits has been pointed out. In addition, many experimental demonstrations regarding the control of $d$-level systems, including the generation of $d$-dimensional entanglement, have been reported \cite{cozzolino2019high,erhard2020advances,wang2020qudits,ecker2019overcoming}.

For a bipartite system of qudits, i.e., the Hilbert space $\mathcal{H}_{AB} = \mathbb{C}_A^d\otimes\mathbb{C}_B^d$, we can obtain an orthonormal basis of maximally entangled states. For the qudit teleportation scheme \cite{bennett1993teleporting}, the following basis was introduced,
\begin{equation}
    \label{eq generalized bell}
    \big|\Psi^{(d)}_{mn}\big\rangle_{AB} = \frac{1}{\sqrt{d}} \sum_{k = 0}^{d-1} e^{i \frac{2 \pi}{d} km} \big| k \big\rangle_A \big| k\ominus n\big\rangle_B,
\end{equation}
where $m$, $n\in\mathbb{Z}_d$ are called the phase and amplitude index respectively, $k\ominus n\equiv (k-n)\text{mod}\,d$ and $d$ is the dimension of the systems (qudits). We also refer to the states of Eq.~\eqref{eq generalized bell} as $d$ level states. To avoid overloading the notation, the $d$ will be avoided when we refer to Bell states $(d = 2)$. Analogously to the Bell states, the value of one of the two indices, $m$ or $n$, can be obtained via local measurements and classical communication.

The von Neumann entanglement entropy of one copy of a maximally entangled state is $E \big( |\Psi^{(d)}_{mn} \rangle_{AB} \big) \equiv S(\text{tr}_A |\Psi^{(d)}_{mn} \rangle \langle \Psi^{(d)}_{mn}|) = \log_2 d$ ebits, which is the conversion rate in the asymptotic limit \cite{horodecki2009quantum}. For that reason we consider that one maximally entangled state of $d$ levels is equivalent to $\log_2 d$ Bell states. Observe that with $n$ Bell states one can locally obtain one maximally entangled state of $d=2^{n}$ levels, and vice versa.

The elements of the basis of maximally entangled states of Eq.~\eqref{eq generalized bell} are related to each other by the local application of the generalized Pauli operators $\hat{X}$, $\hat{Z}$. They are defined via their action on the computational basis states,
\begin{equation}
    \hat{X} \, | \!\; j \!\; \rangle = | j \ominus 1\rangle \quad \text{and} \quad \hat{Z} \, | \!\; j \!\; \rangle  =e^{i\frac{2\pi}{d} j} | \!\; j \!\; \rangle.
\end{equation}
The local application of $\hat{X}$ and $\hat{Z}$ to the maximally entangled states Eq.~\eqref{eq generalized bell} is given by
\begin{equation}
    \label{eq local XZ}
    \begin{gathered}
        \hat{X}^v \hat{Z}^w \otimes \id \, \big |\Psi^{(d)}_{mn} \big \rangle = e^{i\frac{2\pi}{d} v ( m + w )}\big| \Psi^{(d)}_{m\oplus w, n \ominus v} \big \rangle, \\ \id \otimes \hat{X}^v \hat{Z}^w  \, \big|\Psi^{(d)}_{mn} \big\rangle = e^{i\frac{2\pi}{d}w n} \big| \Psi^{(d)}_{m\oplus w, n\oplus v} \big\rangle.
    \end{gathered}
\end{equation}
An important single-qudit operation that we make use of is the quantum Fourier transformation (QFT), defined as
\begin{equation}
    \label{eq QFT}
    \text{QFT} = \frac{1}{\sqrt{d}} \sum_{ m, n = 0 }^{d - 1}e^{i\frac{2\pi}{d} mn } |m\rangle \langle n|,
\end{equation}
where $d$ is the dimension of the system. The QFT allows one to locally interchange the phase and amplitude indices in the following way:
\begin{equation}
    \label{eq QFTinterchange ij}
    \text{QFT}\otimes\text{QFT}^\dagger\big|\Psi^{(d)}_{mn}\big\rangle=e^{i\frac{2\pi}{d}nm}\big|\Psi^{(d)}_{n,d\ominus m}\big\rangle.
\end{equation}

\section{New techniques}
\label{sec new gadgets}

In the following, we introduce a new bilateral controlled operation between a bipartite system of qubits and a bipartite system of qudits, and an alternative depolarization technique for bipartite systems of qubits. These two tools are central for the new entanglement purification protocol for qubits that we introduce in this paper.

\subsection{The counter gate}
\label{sec counter gate}

In our entanglement purification protocols, the information about the states of the ensemble is transferred to an auxiliary maximally entangled system of qudits by means of the implementation of the \textit{counter gate}, $b\text{CX}^{(d)}$. It is a bilateral controlled-$X$ gate that is applied between a bipartite system of qubits and a bipartite system of qudits. Formally, it is defined as
\begin{equation}
    \label{eq c}
    b \text{CX}^{(d)}_{1 \to 2} = \text{CX}_{1 \rightarrow 2}^{A_1 A_2} \otimes \text{CX}_{1 \rightarrow 2}^{B_1 B_2},
\end{equation}
where $\text{CX}_{1\rightarrow 2}$ is the \textit{controlled-X} gate,
\begin{equation}
    \begin{aligned}
        \text{CX}_{1 \rightarrow 2} = \Big( |0\rangle\langle0|\otimes\id + | 1 \rangle \langle 1 | \otimes\hat{X} \Big),
    \end{aligned}
\end{equation}
with a two-dimensional control system and as a $d$-dimensional target system. The name of the counter gate is given due to its functionality. If the qudits target system is in one of the maximally entangled basis states with zero phase index, the effect of the application of the counter gate $b$CX Eq.~\eqref{eq c} is a change of the amplitude index of this qudit system which depends on the control qubit state. Specifically, the action of the counter gate between the computational basis and a maximally entangled state of qudits of the form $|\Psi^{(d)}_{0\ell}\rangle$ Eq.~\eqref{eq generalized bell}, is given by
\begin{equation}
    \label{eq action bCX}
    \begin{aligned}
       b\text{CX}_{1\to 2} \big|mn\big\rangle_1 \big|\Psi^{(d)}_{0 \ell}\big\rangle_2 \! = \! \big| mn\big\rangle_1 \Big( \hat{X}^m \otimes \hat{X}^n \big|\Psi^{(d)}_{0\ell} & \big\rangle_2 \! \Big) \! \\ = \big|mn\big\rangle_1\big|\Psi^{(d)}_{0, \ell\ominus m \oplus n} & \big \rangle_2.
    \end{aligned}
\end{equation}
Here and in the following, we do not write the dimension of the auxiliary system $(d)$ on which $b$CX acts, when it is clear from the equation. In the second equality, we make use of Eq.~\eqref{eq local XZ} with $w=0$. Note that if the qubits state is the $|01\rangle$, the amplitude index of the entangled state of qudits is increased by one. In opposition, if the state is the $|10\rangle$ then the index is decreased by one. Crucially, there exists a subspace spanned by $|00\rangle$ and $|11\rangle$ which leaves the state $|\Psi^{(d)}_{0\ell}\rangle$ invariant after the application of $b$CX, i.e., $b \text{CX} \, |\Psi_{00}\rangle |\Psi^{(d)}_{0 \ell} \rangle = |\Psi_{00}\rangle|\Psi^{(d)}_{0 \ell} \rangle$. Hence, for the maximally entangled state $|\Psi_{00}\rangle$, the amplitude index remains unchanged. On the other hand, if the target state has a nonzero phase index, the application of $b$CX adds a phase depending on this nonzero phase index, i.e., $b \text{CX} \, |mn\rangle |\Psi^{(d)}_{k \ell} \rangle = e^{i\frac{2 \pi}{d} m k} |mn\rangle |\Psi^{(d)}_{k, \ell \ominus m \oplus n} \rangle$. Due to this phase, if the control state is a general superposition of $|00\rangle$ and $|11\rangle$, then it does not remain invariant. A relative phase is introduced.

In an ensemble of $n$ bipartite systems of qubits together with an auxiliary bipartite system of qudits, one can implement the so-called \textit{error number gate} (ENG). It consists in applying the counter gate between each state of the ensemble and the auxiliary state. In a similar way, one also can implement the \textit{error position gate} (EPG), which consists in applying the counter gate between the auxiliary state and each state of the ensemble a number of times corresponding to its position in the ensemble. The explicit form of these two gates is given by
\begin{equation}
    \begin{aligned}
        \label{eq ENG EPG}
        & \text{ENG} : \; \prod_{i = 1}^{n} b\text{CX}_{i \to \text{aux}} \\ 
        & \text{EPG} : \; \prod_{i = 1}^n (b \text{CX}_{i \to \text{aux} })^i.
    \end{aligned}
\end{equation}

\subsection{Depolarization procedure}
\label{sec new depolarization procedure}

The new protocol requires a different depolarization protocol apart from the Bell diagonal depolarization map, see Sec.~\ref{sec bell diagonal deploraization}. The new depolarization procedure is a LOSR (local operation and shared randomness) transformation which leaves the subspace of two Bell states invariant and transforms the other two Bell states into separable states, Eq.~\eqref{eq separant Bell states}. This operation is given by the channel $\mathcal{D}_2$, the Kraus operators of which are
\begin{equation}
    \mathcal{D}_2 : \left\{ \frac{1}{\sqrt{2}} \, \id_2 \otimes \id_2, \, \frac{1}{\sqrt{2}} \, e^{i\frac{\pi}{2} |1\rangle\langle1| }\otimes e^{i\frac{\pi}{2} |0\rangle\langle0| } \right\}.
\end{equation}
The action of $\mathcal{E}$ on the Bell states is given by
\begin{equation}
    \label{eq separant Bell states}
    \begin{aligned}
        \mathcal{D}_2 : & \; |\Psi_{m0}\rangle \langle\Psi_{m0}| \mapsto |\Psi_{m0}\rangle \langle\Psi_{m0}| , 
        \\ \mathcal{D}_2: & \; |\Psi_{m1}\rangle \langle\Psi_{m1}| \mapsto \frac{1}{2} \Big( |01\rangle \langle01| + |10\rangle \langle10| \Big).
    \end{aligned}
\end{equation}

\section{Error identification protocol}
\label{sec error identification protocol}

In this section, we introduce and analyze a new class of entanglement purification protocols that we denote as \textit{error identification protocols} (EIP). They consist of several steps. First, a depolarization of an ensemble of $n$ copies of noisy Bell states is carried out, preparing the ensemble in a mixture of pure states of the desired form. Then, the protocol identifies and discards the states of the distribution that are not the Bell state, which we call \textit{error states}.

This protocol is inspired by the hashing and breeding protocols, but here the information about the ensemble is transferred to maximally entangled systems of qudits. These larger-dimensional systems contain more entanglement than a single Bell state, and hence, more information about the ensemble can be obtained by measurements, in opposition to breeding and hashing where only parity information is obtained. Besides that, the protocol is designed to work in regimes of small number of copies, unlike the standard breeding and hashing.

\subsection{Ensemble interpretation and structure of the protocol}
\label{sec protocol structure}

The initial scenario of the protocols consist of two parties sharing an ensemble of $n$ copies of a noisy Bell state
\begin{equation}
    \label{eq initial state}
    \hat{\rho}_{AB}=\sum_{\mu=0}^3 p_\mu|\phi_\mu\rangle\langle\phi_\mu|,
\end{equation}
where $|\phi_{\mu}\rangle \in \{\,|\Psi_{00}\rangle, |01\rangle,|10\rangle,|\Psi_{10}\rangle\,\}$, $0\leq p_\mu \leq 1$, $\sum_{\mu}p_\mu = 1$ and $|\phi_0\rangle \equiv |\Psi_{00}\rangle$. Therefore, the fidelity of the state Eq.~\eqref{eq initial state} is $F = p_0$. The other states $|\phi_{\mu \neq0}\rangle$ are so-called \textit{error states}. This is a general scenario, as we can always depolarize any state to this form: first applying channel $\mathcal{D}_1$ and then channel $\mathcal{D}_2$, Secs.~\ref{sec belld depolarization} and \ref{sec new depolarization procedure}.

The whole ensemble is then described by
\begin{equation}
    \label{eq collection of states}
    \hat{\rho}_{AB}^{\otimes n} = \sum_{\mu_1,\dots\!\,, \,\mu_n} p_{\mu_1} \! \cdots \, p_{\mu_n} \big|\Phi_{\boldsymbol{\mu}}\big\rangle \big \langle \Phi_{\boldsymbol{\mu}}\big|,
\end{equation}
where $\boldsymbol{\mu} = (\mu_1, \dots, \mu_n)$ and $|\Phi_{\boldsymbol{\mu}} \rangle \equiv \bigotimes_{i = 1}^n |\phi_{\mu_i} \rangle$. We call the subindex $i$ of $|\phi_{\mu_i}\rangle$ the $position$ of the state. The ensemble is prepared in some unknown pure state labeled by $\boldsymbol{\mu}$, where each configuration occurs with a certain probability, i.e.,  $\hat{\rho}_{AB}^{ \otimes n} \simeq \{ p_{\mu_1} \! \cdots p_{\mu_n}, |\Phi_{\boldsymbol{\mu}} \rangle\}$. The numbers of errors of each kind in the ensemble follow a multinomial distribution. However, the total number of errors is binomial distributed, i.e., the probability of the ensemble containing $k$ errors of any kind is given by
\begin{equation}
    \label{eq pk}
    p(k) = \binom{n}{k} F^{n - k} \big( 1 - F \big)^{k}.
\end{equation}
Then, if we do not differentiate between the different kind of errors, we can decompose the ensemble with respect to the total number of errors. We can group the possible configurations by the total number of error states that they contain and write the density operator as 
\begin{equation}
    \label{eq lambdas}
    \hat{\rho}^{\otimes n}_{ AB } = \sum_{ k = 0 }^n p ( k ) \, \hat{\Lambda}_k,
\end{equation}
where $\hat{\Lambda}_k$ is the density operator describing the ensemble for a fixed total number of errors $k$, i.e,
\begin{equation}
    \label{eq: Lmbda k}
    \hat{\Lambda}_k = \frac{ k! ( n - k )! }{ n! } \sum_{\substack{\boldsymbol{\mu} 
    \\ k \text{ errors}}} \big|\Phi_{\boldsymbol{\mu}} \big\rangle\big\langle \Phi_{\boldsymbol{\mu}} \big|.
\end{equation}
The idea of the error identification protocols is to learn information about the errors in the ensemble to identify and discard \footnote{As it will be clear later, in our protocol only separable errors are detected at each step. So that it is never the case that we can correct an error with LOCC, as it happens for the hashing and breeding protocols.} the qubit pairs containing an error, only keeping the $|\Psi_{00}\rangle$ states. In the simplest case the procedure is divided into three steps.
\begin{enumerate}
    \item Depolarization of the initial ensemble into the required form $\rho_{AB}^{\otimes n}$ of Eq.~\eqref{eq initial state}.
    \item Detection of the number of errors $k$, reducing the ensemble into the state $\hat \Lambda_k$ of Eq.~\eqref{eq: Lmbda k}.
    \item Identification of the pure-state configuration $\ket{\Phi_{\bm \mu}}$ within $\hat \Lambda_k$.
\end{enumerate}
For steps 2 and 3 locally inaccessible information about the ensemble is required. Therefore we have to expend some entanglement in the process. For this purpose, the parties initially share a pool of $|\Psi_{00}^{(d)}\rangle$ auxiliary states of different dimensions. By applying the counter gate $b$CX, Eq.~\eqref{eq c}, in an appropriate way, one can transfer some information about the number and the position of the errors states to the auxiliary state. Then, the auxiliary state is measured, destroying its entanglement but revealing information about the errors. One important feature of the protocol is that each step is determined by the outcomes of the previous measurements since the number of errors of each kind influences the procedure.

To understand our motivation to introduce the intermediate step 2, one notes that the resources needed to locate all the errors increase with the total number of errors $k$ in the ensemble. On the one hand, when $k$ is low we are only a few bits away from knowing the exact configuration. On the other, if $k$ is large the missing information can be so vast that the resources required for purification exceed the potential number of Bell pairs. In such a case one can either discard the ensemble altogether, leading to a drop of the expected yield, or leave it untouched, leading to a drop in fidelity. Note that if the protocol is requested to output a constant number of qubit pairs deterministically, one can always locally prepare any missing number of pairs in separable states with $F = 1/2$ (e.g., $\ket{00}$).

In addition, the information cost of knowing the total number of errors is low as compared to the entropy of the whole ensemble so that in principle theory step 2 has a low entanglement cost. In reality, depending on the form of the original state $\hat \rho_{AB}$ it is not always possible (or desirable) to perform such a nonlocal measurement of the total number of errors efficiently. Nevertheless, in these cases reading out \emph{some} information about the total number of different errors can be very beneficial.

To see this we analyze the probability of each configuration, $p_{\mu_1} \! \cdots p_{\mu_n}$, since, for certain ensembles, the probabilities of some particular configurations are small enough and can be neglected, reducing the required information to purify the ensemble. Specifically, the amount of errors in the ensemble is given by a binomial distribution, i.e., the probability of $\hat{\Lambda}_k$ is given by $p(k)$ Eq.~\eqref{eq pk} and hence the expected number of errors is given by $\langle k\rangle = n( 1 - F )$. Note that for $k>\langle k\rangle$ errors, the probability of finding $\hat{\Lambda}_{k}$ decreases with the number of errors it contains, i.e., $p( k_2 ) < p( k_1 )$ for $\langle k\rangle < k_1 < k_2$. Then, depending on the expected number of errors $\langle k\rangle$, we can assume that the possible number of errors is upper bound by a certain value $\lambda$, such that $\sum_{k = \lambda + 1}^n p( k ) < \epsilon$. This means we can restrict the set of possible configurations to only those containing $k \leq \lambda$ errors. With this assumption, the resources required to purify the ensemble are reduced and the yield is increased, sometimes becoming positive instead of negative.

However, under the assumption $k \leq \lambda$ no states with fidelity one are obtained, due to the nonzero probability of a configuration with more than $\lambda$ errors. For a given ensemble there are different values of $\lambda$ which lead to a successful purification. The higher $\lambda$, the better is our assumption (higher fidelity of the purified states). However, the amount of information required increases with $\lambda$. Therefore, for lower values of $\lambda$, fewer auxiliary states are needed and the higher is the yield.

In addition, for a fixed number of errors $k$ randomly distributed on all the pairs we observe that determining the pure state configuration is not necessarily the best way to purify the ensemble. As we show in Sec.~\ref{sec toy model 2 errors}, it can be beneficial to obtain partial information about the error configuration, as it may allow locating the errors in a subensemble which is then discarded. In such a scenario the exact configuration of the discarded subensemble is never determined, which explains why it can improve the average yield.

\subsection{Purification of noisy states resulting from amplitude damping channel}
\label{sec toy model}

In this section, we introduce an illustrative example of how to purify a specific family of states with an EIP. We call this protocol EIP$_{\text{damp}}$, as we consider the noise modelled by an \textit{amplitude damping channel} $\mathcal{N}_p$ \cite{nielsen2002quantum}, which Kraus operators are given by
\begin{equation}
    \begin{aligned}
        K_0 = \! |0\rangle\langle0| \! + \! \sqrt{p} \, |1\rangle\langle1| \quad \text{and} \quad K_1 = \sqrt{1 - p} \, |0\rangle\langle 1|.
    \end{aligned}
\end{equation}
This noise describes energy dissipation in many quantum systems, such as spontaneous emission.

The result of sending the state $|\Psi_{11}\rangle$ through an amplitude damping channel, i.e., $\big(\mathcal{N}_F \otimes \mathcal{N}_F \big)|\Psi_{11}\rangle\langle\Psi_{11}|$, is LU equivalent to the state of Eq.~\eqref{eq initial state} with $p_2 = p_3 = 0$, i.e.
\begin{equation}
    \label{eq toy state}
    \hat{\rho}_{AB} = F \, |\Psi_{00}\rangle\langle\Psi_{00}| + \big( 1 - F \big) |01\rangle \langle 01|.
\end{equation}
We assume that two parties share an ensemble of $n$ copies of this state Eq.~\eqref{eq toy state}. This ensemble only contains one kind of error state, since $|\phi_{\mu_i}\rangle$, Eq.~\eqref{eq initial state}, can only be a Bell state with probability $F$ or the error state $|01\rangle$ with probability $1 - F$. 

This is the simplest case, where we can count how many errors are there in the ensemble with minimal resources. Considering an unknown distribution, the number of errors ranges from 0 to $n$, giving $n + 1$ possibilities that we have to discriminate. Assuming that the required resource grows at least as the logarithm of the number of possibilities to discriminate (the worst case entropy), see Appendix~\ref{app sec dichotomic measurement} for details, implies that at least $\log_2( n + 1 )$ e-bit are required to learn the number of errors. Since the ensemble only contains errors of 01 kind, we can saturate this limit by transferring the number of errors on a $d = n + 1$ level state $|\Psi^{( n + 1 )}_{00}\rangle$. So, for this kind of states, we can learn the exact number of errors efficiently. For this reason, in this case, we can (and we will) avoid the assumption on the maximal number of errors $\lambda$, which allows us to purify the ensemble to $F = 1$. Nevertheless, introducing $\lambda$ could be interesting if one is ready to increase the yield at the price of lowering the fidelity, especially if the initial number of copies is very low.

To count the errors of a certain configuration with only one kind of errors, i.e., $\boldsymbol{\mu} = (\mu_1, \dots, \mu_n)$ where $\mu_i \in \{0, 1 \}$ of Eq.~\eqref{eq initial state}, we apply the ENG with the ensemble and the auxiliary state $|\Psi^{ ( n + 1 )}_{00}\rangle$, i.e.
\begin{equation}
    \label{eq C sum erros}
    \prod_{ i = 1 }^n b\text{CX}_{i \to \text{aux}} \big| \Phi_{\boldsymbol{\mu}} \big\rangle\big|\Psi^{( n + 1 )}_{00} \big\rangle_{\text{aux}} \! \! = \big|\Phi_{\boldsymbol{\mu}} \big\rangle \big|\Psi^{( n + 1 )}_{0k} \big \rangle_{\text{aux}},
\end{equation}
where, from Eq.~\eqref{eq action bCX}, $k = \#\text{errors}$ in $\boldsymbol{\mu}$. Observe that, crucially, the information of the amplitude index corresponds to the exact number of errors, since in this case, the only possible error 01 always ``counts up''. The action of this operation to the ensemble $ \hat{\rho}^{\otimes n}$ is given by
\begin{equation}
    \label{eq C sum erros 2}
    \begin{aligned}
        \prod_{i=1}^{n}b\text{CX}_{i\to \text{aux}}\,\hat{\rho}^{\otimes n}\otimes\big|\Psi^{(n+1)}_{00}\big\rangle\big\langle\Psi^{(n+1)}_{00}\big|\,b\text{CX}^{\dagger}_{i\to \text{aux}}&\\=\sum_{k=0}^{n}\binom{n}{k}F^{n-k}\big(1-F\big)^k\hat{\Lambda}_k\otimes\big|\Psi^{(n+1)}_{0k}\big\rangle\big\langle\Psi^{(n+1)}_{0k}\big|&.
    \end{aligned}
\end{equation}
Note that the ENG leaves the states of the ensemble invariant. However, the measurement of the auxiliary state reveals the total number of errors in the ensemble and changes our description thereof to $\hat{\rho}^{\otimes n} \to \hat{\Lambda}_k$.

The next step consists in isolating each error state in a different subensemble. This procedure depends on the number of errors $k$. The minimum amount of information necessary to identify $k$ errors is $\log_2 \binom{n}{k}$ bits, since we have to distinguish between all possible configurations (distributed uniformly). By our assumption on the entanglement cost of nonlocal measurement this task requires at least the same amount of e-bits.

For this particular case, as pure Bell states are obtained, i.e., $F' = F'_{g} = 1$, the yield of the protocol is given by
\begin{equation}
    \label{eq Y toy model det}
    Y = \frac{n - \sum^n_{ k = 1 } p (k) \, k - R_t}{n}
\end{equation}
where $p(k)$ is the probability of detecting $k$ errors given in Eq.~\eqref{eq pk} and $\sum^{n}_{k = 1} p(k) k = \langle k \rangle = n (1 - F)$. $R_t$ is the expected number of resources needed to detect and locate the errors. It is given by
\begin{equation}
    \label{eq Rt}
    R_t = \log_2(n + 1) + \sum_{ k = 0 }^n p( k ) R( k ),
\end{equation}
where $R(k)$ is the number of resources needed to locate all $k$ errors. In Eq.~\eqref{eq Rt} we consider that one maximally entangled state of $d$ levels is equivalent to $\log_2 d$ Bell states. To compute the yield we need to provide a procedure for identifying the exact configuration for each $k$, and compute the resource $R(k)$ it requires.

\subsubsection*{0.\texorpdfstring{$\;\;$}{TEXT} No errors; \texorpdfstring{$k = 0$}{TEXT}}

The simplest instance is where no errors are detected. In this case, the protocol is already completed
\begin{equation}
    R(0) = 0,
\end{equation} 
and $n$ pure states are obtained. 

\subsubsection{One error; \texorpdfstring{$k = 1$}{TEXT}}
\label{sec toy model 1 error}

When one error is detected $k = 1$ we have to distinguish between $n$ possible configurations since the error can be located at any position. Once again, we can optimize the resources, since we can transfer the position of the error into an auxiliary state of $d=n$ levels. This is achieved by means of EPG, as
\begin{equation}
    \label{eq C gate position}
    \prod_{i=1}^n\,(b\text{CX}_{i\to \text{aux}})^i\big|\Phi_{\boldsymbol{\mu}}\big\rangle\big|\Psi^{(n)}_{00}\big\rangle_{\text{aux}} = \big|\Phi_{\boldsymbol{\mu}}\big\rangle\big|\Psi^{(n)}_{0j}\big\rangle_{\text{aux}},
\end{equation}
where, $j$ corresponds to the position of the error state [see Eq.~\eqref{eq action bCX}], i.e., $|\phi_{\mu_{j}}\rangle = |01\rangle$, by measuring the auxiliary state we can obtain the amplitude index, $j$, therefore identifying the position of the error and distilling $n-1$ states in the end. In this case, the number of resources to locate the error is given by
\begin{equation}
    \label{eq Ressource1}
    R(1) = \log_2 (n).
\end{equation}

\subsubsection{Two errors; \texorpdfstring{$k = 2$}{TEXT}}
\label{sec toy model 2 errors}

In case that two errors are detected, one has to distinguish between $\frac{1}{2} n ( n - 1 )$ configurations. The minimum information needed to identify them is hence $\log_2 \big( \frac{1}{2} n ( n - 1 ) \big)$ bits. The strategy we propose does not reach this minimum, but it is close (ratio between the resources used by EIP$_{\text{damp}}$ and the minimum amount is smaller than 1.03 for $n < 300$, see Appendix~\ref{sec app ident 2 identical errors} for details). In the first step we determine the sum of the positions of the two errors transferring this information into an auxiliary state by applying the EPG, Eq.~\eqref{eq C gate position}. For that purpose, we require an auxiliary state of $d = 2n - 3$ levels to encode all the possible values for the sum of both error positions. Then we can use this information to isolate each error state in a subensemble (see Appendix~\ref{sec app ident 2 identical errors}), thereby reducing the situations to the one error case, described in Sec.~\ref{sec toy model 1 error}. When the position of one error is identified, the position of the other error is directly determined, since we know the sum of the positions.

The size of the subensembles where we isolate each error depends on the position of the two errors and ranges from $1$ to $n/2$. Consequently, the number of resources varies depending on the position of the errors. The expected value of $R(2)$ is given by
\begin{equation}
    R(2) = \log_2( 2n - 3 ) + \langle \, \log_2 d_3 \, \rangle
\end{equation}
where $d_3$ is the size of the two subensembles where each error is isolated. The expected value of the amount of entanglement necessary to identify the states once they are isolated is $\langle \log_2 d_3 \rangle$ (see Appendix~\ref{sec app ident 2 identical errors} for details).
\\ \\
\textbf{Alternative procedure}
\\ \\
In some cases, determining the corresponding configuration of the ensemble is not the most efficient way to purify it. We show this by introducing an alternative procedure to purify an ensemble with two 01 errors.

We start splitting the ensemble into two subensemble of $n/2$ states (we assume $n$ even for the matter of the example, if $n$ is odd one splits the ensemble into unequal parts $(n \pm 1) / 2$). Then, we count the number of errors in each subensemble as we described in Sec.~\ref{sec toy model}, using one three-level auxiliary state, i.e., $|\Psi^{(3)}_{00}\rangle_{\text{aux}}$. After this step, there are two different possible scenarios: $i)$ we find one error in each subensemble, or $ii)$ two errors in the same subensemble. The probability of each scenario is given by $p_i = n/( 2n - 2 )$ and $p_{ii} = ( n - 2 )/( 2n - 2 )$ respectively. In scenario $i)$, the procedure consists in locating each error as described in Sec.~\ref{sec toy model 1 error}, and therefore using two auxiliary states of $d=n/2$ levels. In this case, we obtain $n-2$ Bell states. In scenario $ii)$, if $n/2\geq 8$, we iterate the procedure with the subensemble with the errors as input. Otherwise, we discard the whole subensemble and we keep the other $n/2$ Bell states. Note that this alternative procedure is probabilistic--in some branches, it returns less than $n-2$ Bell states from the ensemble. For an ensemble of $n=8$, the yield of this procedure would be
\begin{equation}
    Y_a = \frac{p_i \left(8 - 2 - 2 \log_2 4 \right) + p_{ii} \, 4 - \log_2 3}{8} = 0.159.
\end{equation}
On the other hand--assuming that to obtain one bit of information of the configuration one needs to consume one ebit--any procedure that completely determines the configuration at least requires $S(\hat{\Gamma})$ resources. Therefore, the yield of such procedures for this particular case of an ensemble containing two errors, $\hat{\Lambda}_2$, is upper bound by
\begin{equation}
    Y_k^* = \frac{ n - k - S ( \hat{\Lambda}_k )}{n},
\end{equation}
where $S(\hat{\Lambda}_2) = \log_2 \binom{n}{2} = \log_2 ( \frac{1}{2} n ( n - 1 ) )$. From this bound we prove our initial statement, as for ensembles of $n=8$ states with two errors, the yield $Y_a$ is higher than the upper bound $Y_2^* = 0.149$.

\subsubsection{General number of errors}
\label{sec:toy:general:protocol}

In this section, we present a technique to deal with an arbitrary number of errors in the ensemble. For that purpose, suppose the ensemble of $n$ states contains in total $k$ errors, which is known after the first step of the error identification protocol. If $k\leq2$, then the procedures of Secs.~\ref{sec toy model 1 error} and \ref{sec toy model 2 errors} can be directly applied. In contrast, if $k > 2$, we apply the following steps:
\begin{enumerate}
	\item\label{enum:toy:general:1} Divide the ensemble into $a$ blocks of size $n/a$.
	\item\label{enum:toy:general:2} Identify the number of errors in each block.
	\item\label{enum:toy:general:3} If the number of errors in a block is less than or equal to two, locate the errors using the methods of Sec.~\ref{sec toy model 1 error} (one error) and Sec.~\ref{sec toy model 2 errors} (two errors).
	\item \label{enum:toy:general:4} If the number of errors in a block is larger than two, then the procedure is applied again for this block, i.e., go back to step \ref{enum:toy:general:1} with $n'=n/a$.
\end{enumerate}
Formally speaking, the number of errors distributed over $a$ blocks of size $n/a$ corresponds to random variables $k_i$ for $1 \leq i \leq a$ where 
\begin{equation}
    k_i \sim \mathrm{Hyp} \left (n - \frac{(i - 1) n}{a}, k - \sum \limits^{i - 1}_{j = 1} k_j, \frac{n}{a} \right)
\end{equation} 
and $\sum^{a}_{i = 1} k_i = k$, where $\mathrm{Hyp}$ is the hypergeometric distribution \footnote{$k_i\sim \text{Hyp} (x, y, z)$ means that $k_i$ is given with probability $P(k_i | x, y, z ) = \binom{y}{k_i} \binom{x - y}{z - k_i} / \binom{x}{z}$}.

In step \ref{enum:toy:general:2} of the procedure, we have to identify the number of errors in each block, leading to a configuration of errors $(k_1, \ldots, k_a)$. This can be done using the technique of Sec.~\ref{sec toy model} using the counter gate $b$CX and additional auxiliary states $|\Psi^{(d_{i})}_{00}\rangle_{\text{aux}}$ of $d_{i}$-levels for each block $i$. We denote the number of resources necessary to identify the error configuration $(k_1, \ldots, k_a)$ where $\sum^{a}_{i = 1} k_i = k$ for a given $k$ by $R_{F}(k)$. The computation of the levels $d_{i}$ and the number of resources $R_{F}(k)$ can be found in Appendix~\ref{app:sec:amp:general}.

Steps \ref{enum:toy:general:3} and \ref{enum:toy:general:4} of the procedure are used to locate the errors in the individual blocks of the ensemble. We differentiate two situations:
 \begin{itemize}
 	\item $k_i \leq 2$: In this case, the resources for locating two or less errors have to be used, see Sec.~\ref{sec toy model 1 error} (one error) and Sec.~\ref{sec toy model 2 errors} (two errors).
	\item $k_i > 2$: In this case, the $i-$th block is again divided into $a$ blocks, and the procedure is applied recursively to block $i$, until only one or two errors are left.
\end{itemize}
We denote the resources necessary to execute the steps \ref{enum:toy:general:3} and \ref{enum:toy:general:4} of the procedure by $R_{L}(k)$. We provide the computation of $R_{L}(k)$ in Appendix~\ref{app:sec:amp:general}.

In summary, the total number of resources for the protocol if $k$ errors are detected are
\begin{equation}
    \label{eq:toy:general:resources:k}
    \begin{aligned}
        R(1) & = \log_2(n), \\ R(2) & = \log_2 (2n - 3) + \langle \, \log_2 d_3 \, \rangle, \\ R(k) & = R_{F}(k) + R_{L}(k),
    \end{aligned}
\end{equation}
and the expected value of resources $R_t$ is given by Eq.~\eqref{eq Rt}.
\begin{figure}
    \centering
    \textbf{EIP$_{\text{damp}}$}\par\medskip\vspace{-3pt}
    \includegraphics[width=\columnwidth]{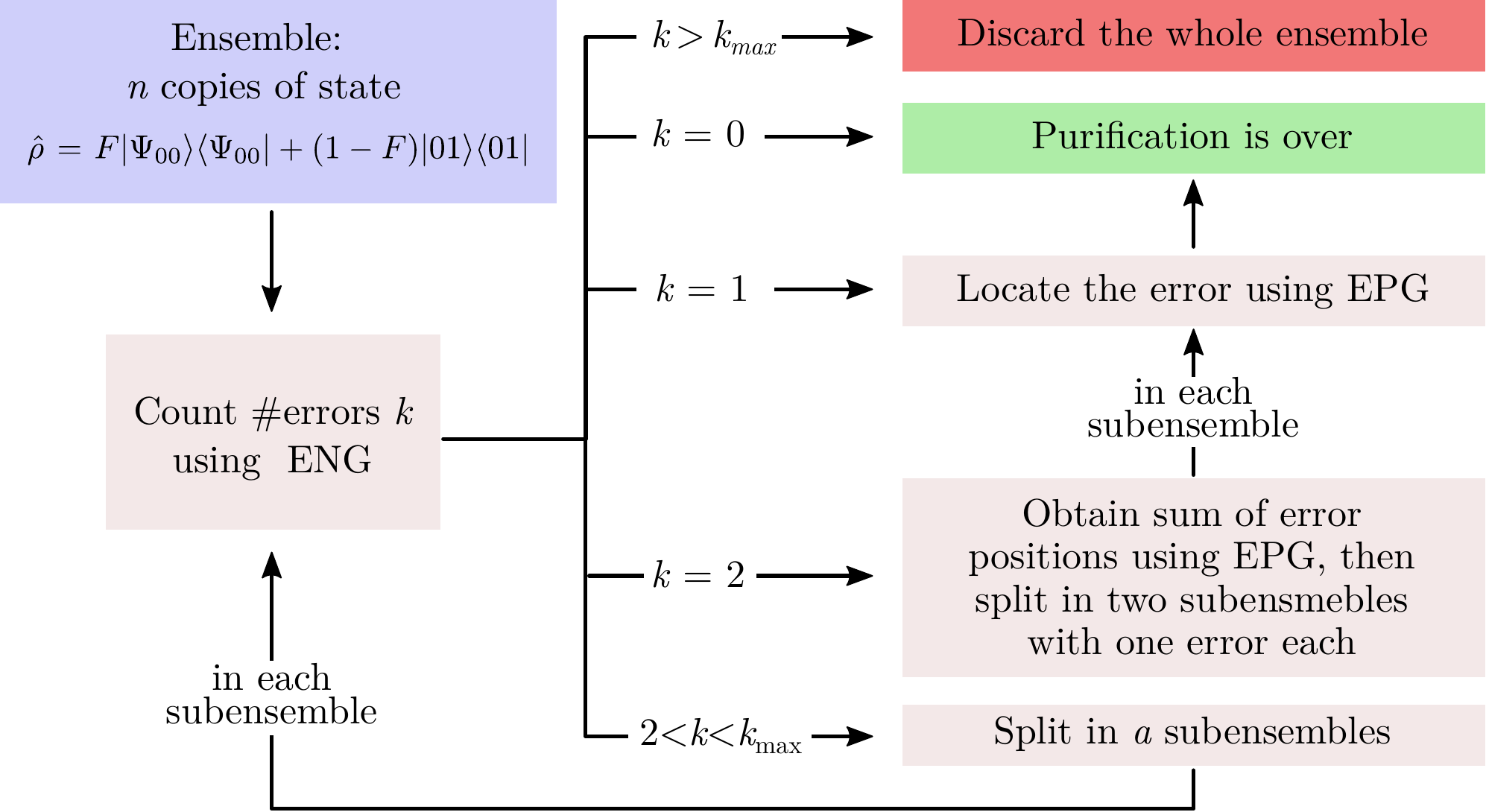}
    \caption{Diagram illustrating the steps of the EIP$_{\text{damp}}$. The blue box corresponds to the stating point, green box to the situation where one end up with a successful purification and red boxes to situations where too many errors are found and the procedure aborts.}
    \label{fig:tree_EIPdamp}
\end{figure}
\subsubsection{Too many error: possibility to abort}
\label{sec toy model probabilistic}

In case the ensemble contains too many errors, the required resources may exceed the number of obtained Bell states leading to a negative yield. In this case, the most efficient option is to abort the protocol, discard the whole ensemble, and output the remaining auxiliary entangled states only. Including the aborting possibility, we can increase the yield.

To compute the yield suppose that the protocol aborts if more than a certain number of errors $k_{\max}$ are detected. This means that for $k>k_{\max}$ the $n$ states of the ensemble are discarded. In this case, the yield is given by
\begin{equation}
    \label{eq: abort yield}
    Y=\frac{-\log_2(n+1)+\sum_{k=0}^{k_{\max}}p(k)\left[n - k - R(k)\right]}{n},
\end{equation}
where the resource costs $R(k)$ of purifying the ensemble with a total number of errors $k$ are still given by the Eq.~\eqref{eq:toy:general:resources:k}.

The optimal value of $k_{\max}$ (maximizing the yield) is the maximum number of errors that we can locate with a positive yield, i.e.
\begin{equation}
    k^{\text{opt}}_{\max}(n) = \max k \quad\text{s.t.}\quad n - k - R(k) > 0.
\end{equation}
The protocol with abortion still achieves unit fidelity for all states it outputs.

In Fig.~\ref{fig:tree_EIPdamp}, we show a schematic representations of the steps of the protocol.

\subsubsection{Yield of the purified ensemble}

In the following, we present the results of the protocol with abortion for different initial fidelity $F$ and $k_{\max}$.
\begin{figure}
    \centering
    \subfloat[\centering]{\includegraphics[width=1\columnwidth]{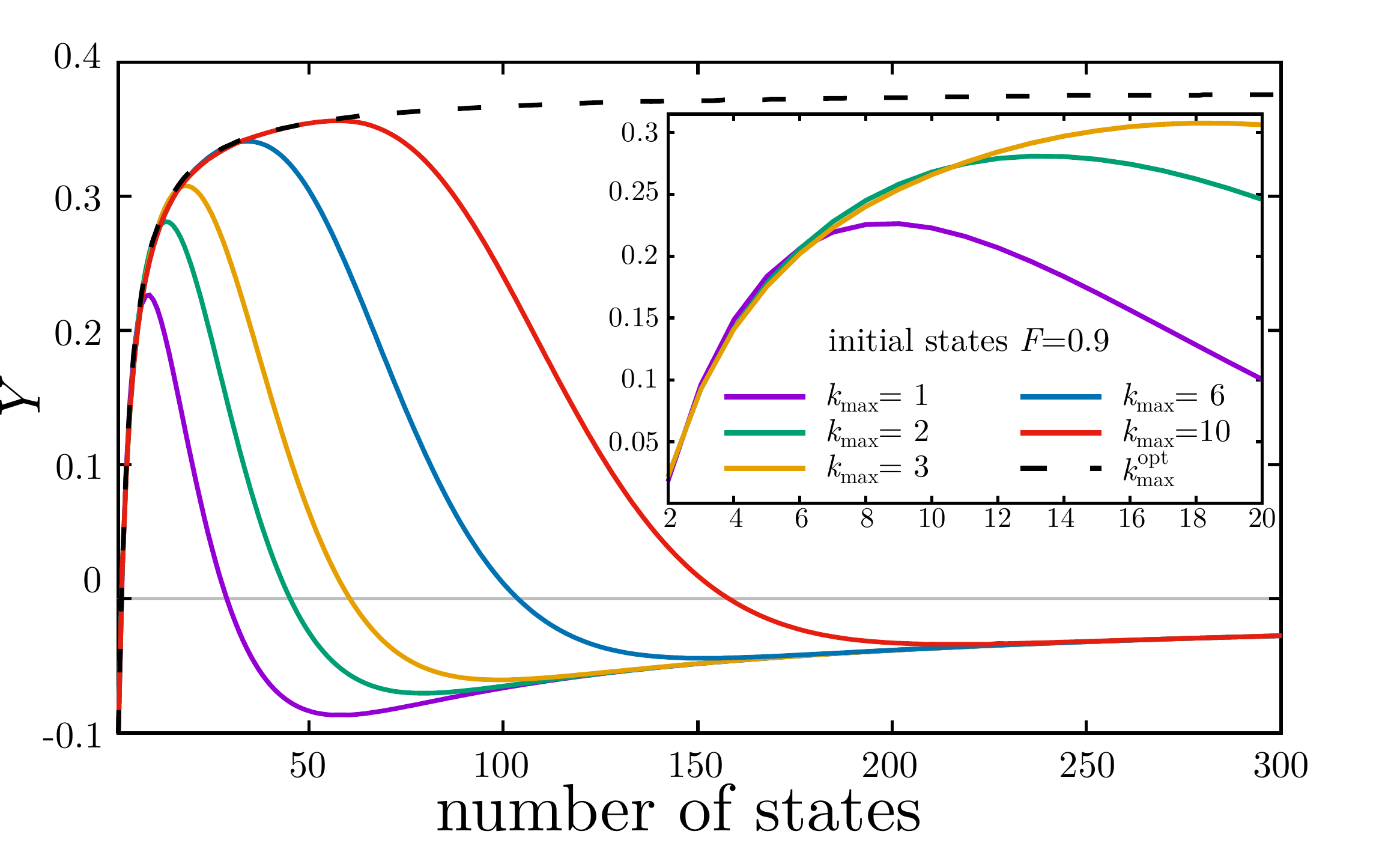} \label{fig:toy:general:y:f0,9}}
    \hfill \vspace{-0.cm}
    \subfloat[\centering]{\includegraphics[width=1\columnwidth]{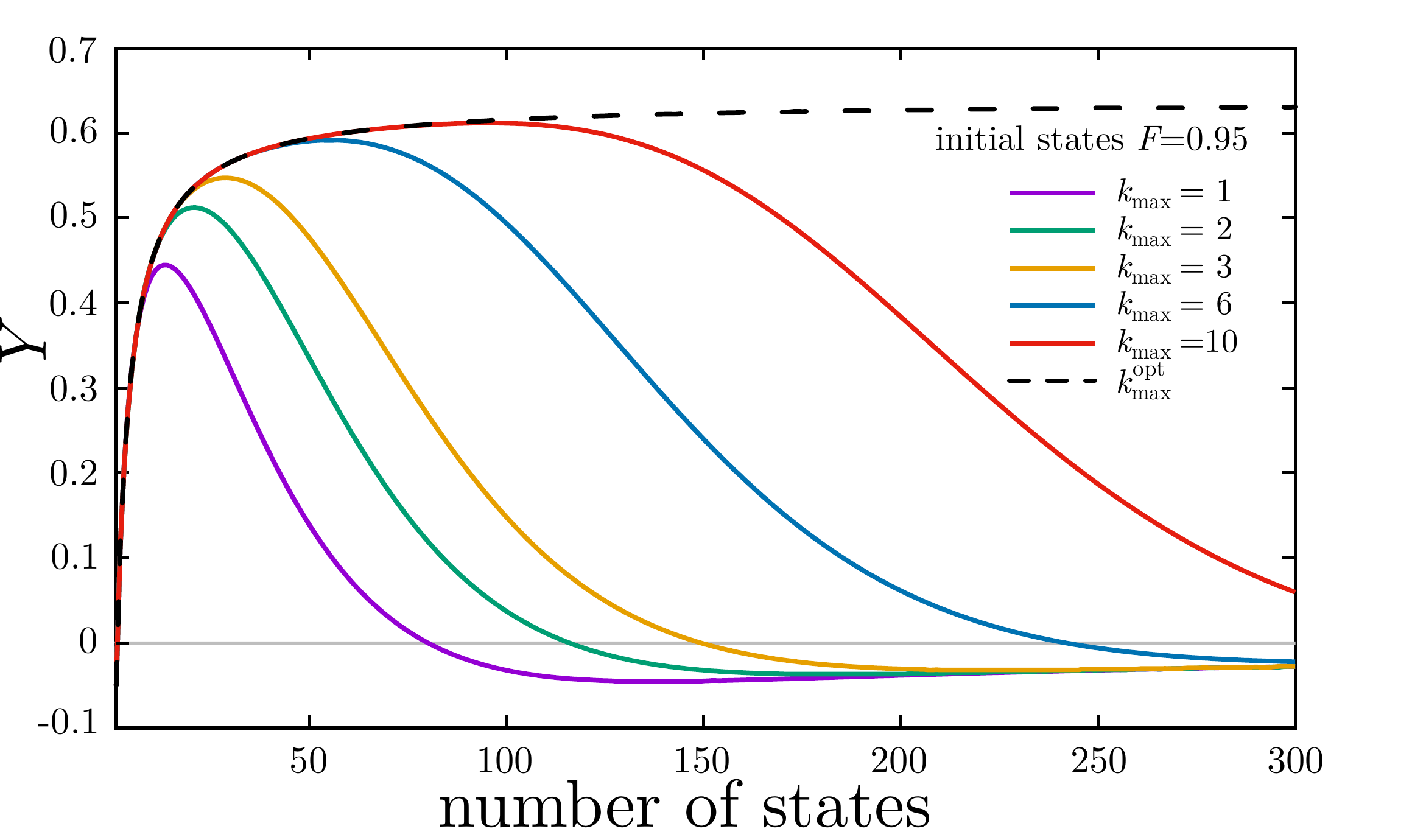} \label{fig:toy:general:y:f095}}
    \caption{The figures show the yield of the probabilistic protocol for $a=2$ as a function of the initial ensemble size. Different colors correspond to different values of $k_{\max}$. Dashed curve correspond to the $k_{\max}^{\text{opt}}$. In (a), the initial fidelity is $F = 0.90$. In the zoom for $n \in [2, 20]$, we only plot the curves corresponding to $k_{\max}=1,\,2$ and $3$. In (b), the initial fidelity is $F=0.95$.}
\label{fig:toy:general:y}
\end{figure}
The plots in Fig.~\ref{fig:toy:general:y} show that for each $k_{\max}$ and initial fidelity $F$ there exists a maximum yield $Y_{\mathrm{max}}$ which depends on the number of initial states $n$, i.e., $Y_{\mathrm{max}} = Y (n_{\mathrm{max}} )$. The maximum $Y_{\mathrm{max}}$ as well as $n_{\mathrm{max}}$ both increase with $k_{\max}$. Note that if for each size of the ensemble $n$ we choose $k_{\max} = k^{\text{opt}}_{\max}$, the curve of the yield reaches all $Y_{max}$, as it is the optimal value. In the asymptotic limit all curves with different fixed values of $k_{\max}$ approach 0. On the other hand, the yield for $k_{\max}^{\text{opt}}$ approaches a fixed value. Furthermore we also find that the yield increases with the initial fidelity $F$ of the ensemble.

\subsection{Purification of rank-3 states}\label{sec rank3 states}

In analogy to Sec.~\ref{sec toy model}, here we analyze the purification of another kind of noisy Bell state, given by
\begin{equation}
    \label{eq rank3 state}
        \hat{\rho}_{AB} = F |\Psi_{00}\rangle \langle\Psi_{00}| + \frac{ 1 - F }{2} \Big( |01\rangle \langle01| + |10\rangle \langle10| \Big).
\end{equation}
This is a more general scenario as considered in the previous section as it corresponds to the state of Eq.~\eqref{eq initial state} with $p_3=0$. The state of Eq.~\eqref{eq rank3 state} can be locally obtained from any state which is not full rank in its Bell diagonal form. This includes rank-2 and rank-3 Bell-diagonal states, resulting e.g., from sending a Bell state through a dephasing channel. An ensemble of $n$ copies of this state contains two different kinds of errors, which complicates their identification. Since the counter gate does not allow us to equally treat the two different kinds of errors, i.e., the 01 error ``counts up'' while the 10 ``counts down'', see Eq.~\eqref{eq action bCX}, we can not efficiently count the total number of errors $\# 01 + \# 10$. In this case, we have to reduce the set of possible configurations to purify the ensemble with a positive yield. We restrict the ensemble to the most probable configurations, those with at most a certain number of errors $\lambda$, i.e., we ignore the possibility of the ensemble containing more than $\lambda$ errors (justified in the introduction of Sec.~\ref{sec error identification protocol}). Unlike the procedure discussed in Sec.~\ref{sec toy model probabilistic}, in this case, we do not abort the protocol if the ensemble contains too many errors, as it can not be efficiently verified. Instead, the procedure is executed assuming that the ensemble contains at most $\lambda$ errors. As at any point of the protocol we do not verify if the initial assumption is correct, the fidelity of the purified ensemble is reduced due to the small but nonzero probability for the ensemble to contain more than $\lambda$ errors. We call these protocols \textit{EIP}($\lambda$) (where when we name them we specify the assumed value of $\lambda$).

\subsubsection{First step: determine the error difference}
\label{sec 0 step}

\begin{figure}
    \centering
    \includegraphics[width=1\columnwidth]{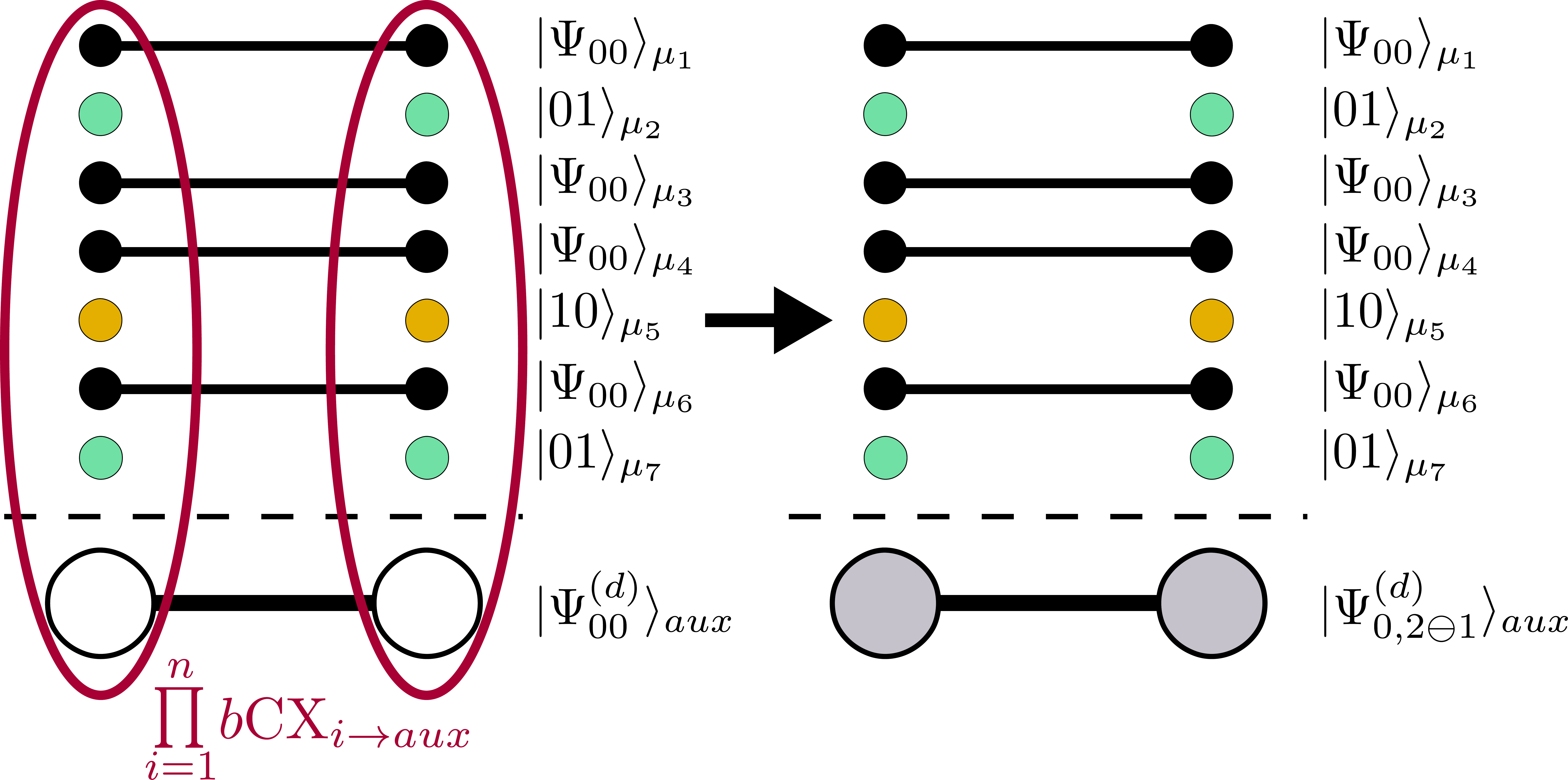}
    \caption{\label{fig 1}Representation of the application of the ENG, Eq.~\eqref{eq C sum erros}, between a specific configuration $\boldsymbol{\mu} = (\mu_1, \dots, \mu_7)$ of bipartite qubit states and one auxiliary state of $d$ levels. The ensemble contains four $|\Psi_{00}\rangle$ states (black), two $|01\rangle$ errors (green) and one $|10\rangle$ error (yellow). Therefore, the amplitude index of the auxiliary state is increased by $2\ominus 1 = 1$.}
\end{figure}

Consider an ensemble of $n$ copies of the state of Eq.~\eqref{eq rank3 state} for which we assume a certain maximal number of errors $\lambda$. The first step of all EIP($\lambda$) consists in applying the ENG, Eq.~\eqref{eq ENG EPG}. Due to the different actions of the counter gate for each kind of error, by measuring the target state, we obtain the \textit{difference} of error states in the ensemble, i.e., $\#01-\#10$, instead of the total number of errors. The possible values for the difference range from $-\lambda$ to $\lambda$, and hence we need an auxiliary state of $d=2\lambda+1$ levels in order to encode the value of this difference. The new amplitude index of the auxiliary state is given by $j=(\#01-\#10)\text{mod}(2\lambda+1)$, see Fig.~\ref{fig 1} for an example. By measuring the auxiliary state, we determine the value of $j$, gaining some information about the kind and the number of errors. For instance, if $j=0$ we can conclude that there are either no errors, or the same amount of 01 and 10 errors. Once $j$ is obtained, the following configurations are possible
\renewcommand{\arraystretch}{1.5}
\begin{table}[H]
    \centering
    \begin{tabular}{c|c}
        \textbf { Difference of errors $\;$}&\textbf{ Errors in the ensemble }\\\hline $j = 0$ & $\big\{01^{\otimes \alpha}\text{ and }10^{\otimes \alpha}\big\}_\alpha$ \\ $1\leq j \leq \lambda$ & $\big\{01^{\otimes j + \beta} \, \text{ and } \, 10^{\otimes \beta} \big\}_\beta$ \\ $\lambda < j\leq 2\lambda$ & $\big\{01^{\otimes \gamma}\text{ and } 10^{\otimes \gamma + 2 \lambda + 1 - j }\big\}_\gamma$,
    \end{tabular}
\end{table}
\noindent
where $\alpha \in [ 0, \dots, \frac{\lambda}{2}]$, $\beta\in[ 0, \dots, \frac{1}{2}(\lambda - j) ]$ and $\gamma \in [0, \dots, \frac{1}{2}(j - 1 - \lambda) ]$. In other words, if $0 \leq j \leq \lambda$, the ensemble can be in any configuration such that the difference between the number of 01 and 10 errors is $j$. For $- \lambda \leq j < 0$ the role of 01 and 10 errors is exchanged. In order to determine the exact number of errors (up to $\lambda$) and identify them, a specific procedure depending on $j$ and $\lambda$, is required. We call $R_{\lambda}(j)$ the expected number of resources needed to identify the errors, once a specific difference $j$ is obtained.

\subsubsection{Up to one error; \texorpdfstring{$\lambda= 1$}{TEXT}}

First, we introduce the EIP(1) procedure, for which we assume that the ensemble contains at most one error, i.e., $\lambda = 1$. After the first measurement (described above), one determines the number and the kind of error contained in the ensemble: $j = 0$ corresponds to no errors, $j=1$ to one 01 error and $j = ( -1 ) \text{mod} \, 3 = 2$ to one 10 error. If one error is detected, it can be identified with the procedure described in Sec.~\ref{sec toy model 1 error}. The only difference appears in the case that one 10 error is detected, i.e., $j=2$. Here we have to take into account that the error state 10 ``counts down''.

The number of resources $R_{\lambda}(j)$ needed to locate the error in each case is given by
\begin{equation}
    \begin{aligned}
        R_{1}(0) &= 0\,, \\
        R_{1}(1) &= \log_2 n\,, \\
        R_{1}(2) &= \log_2 n\,.
    \end{aligned}
\end{equation}
A detailed analysis of the performance of the protocol can be found in Appendix~\ref{app sec one error}. We find that the fidelity of the ensemble is always increased after the protocol, but in the limit of large $n\to \infty$ the fidelity does not change, i.e., $F'\rightarrow F$ and $F'_{g}\rightarrow F^n$, as in this limit the probability of having at most one error vanishes. This effect can be obviated by ``blocking'', i.e., analyzing the $n$ initial pairs as several smaller ensembles that are purified independently.

\subsubsection{Up to two errors; \texorpdfstring{$\lambda= 2$}{TEXT}}
\label{sec up to 2 errors}

Next, we present the EIP(2), which purifies the ensemble under the assumption of at most two errors, i.e., $\lambda=2$. The main difference with the EIP(1) procedure is that one does not know the total numbers of different errors from the error difference $j$ measured at the first step. The possible scenarios after obtaining the difference $j$ are
\renewcommand{\arraystretch}{1.2}
\begin{table}[H]
    \centering
    \begin{tabular}{c|c}
        \textbf{ Difference of errors \textit{j} $\,$}&\textbf{ Errors in the ensemble }\\ \hline 2 & two 01 \\ 1 & 01 \\ 0& no errors$\;$ OR $\;$01 and 10 \\ $(-1)$mod $5 = 4$ & $\,$10 \\ $(-2)$mod $5 = 3$ & \,two 10.
    \end{tabular}
\end{table}
\noindent
Note that if the value of the amplitude index is $j\neq0$, we determine the number of errors of each kind. If one or two identical errors are detected, the procedures become analogous to the ones described in Sec.~\ref{sec toy model 1 error} and Sec.~\ref{sec toy model 2 errors} respectively. However, if the value $j=0$ is found, the ensemble can contain either no errors or two different errors, i.e., one 01 and one 10 error. The procedure in this scenario consists in applying the EPG, Eq.~\eqref{eq C gate position}, with an auxiliary state of $d=2n-1$ levels followed by another measurement of this state. We find that only in the case of no errors, the new amplitude index is $j'=0$. On the other hand, if there are two different errors, then $j'=(r-\ell)\text{mod}(2n-1)\neq 0$, where $|\phi_{\mu_r}\rangle=|01\rangle$ and $|\phi_{\mu_\ell}\rangle=|10\rangle$. From the value of $j'$ we can find two subensembles with one error in each (see Appendix~\ref{sec app 2 different}). Then, we can identify one of the two errors with the procedure described in Sec.~\ref{sec toy model 1 error}, and the other error state is directly identified from the value of $j'$.

\begin{figure}
    \centering
    \textbf{EIP(2)}\par\medskip\vspace{-10pt}
    \subfloat[\centering]{\includegraphics[width=1\columnwidth]{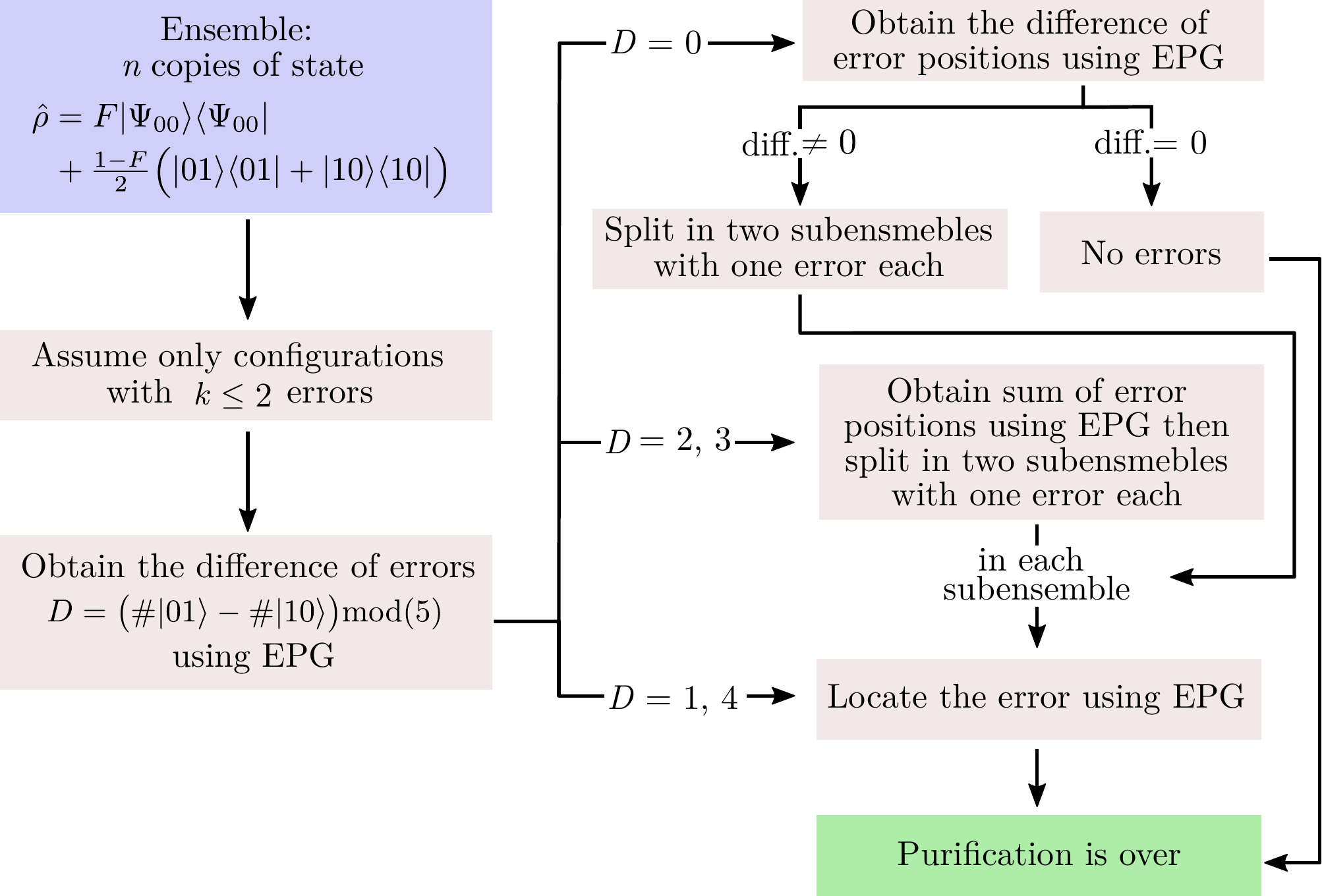} \label{fig:tree_EIP2}}
    \caption{Diagram illustrating the steps of the EIP(2). The blue box corresponds to the stating point and the green box to final situation where one end up with a purified ensemble.}
\end{figure}

In Fig.~\ref{fig:tree_EIP2}, we illustrate the steps of EIP(2) in a diagram. Equations for the expected value of the number of resources are shown in Appendix~\ref{sec app FL} and the yield is shown in Appendix~\ref{sec Y and F}.

\subsubsection{Up to a general number \texorpdfstring{$\lambda$}{TEXT} of errors}
\label{sec general number of errors}

Following the same reasoning as in the previous sections, one can design an efficient procedure in order to identify the position of all errors up to a certain number $\lambda$ of them, defining a general protocol EIP($\lambda$). However, due to the opposite action of the errors states the situation quickly becomes intractable when increasing $\lambda$. In general, too much auxiliary entanglement is required, and when $\lambda\to n$ the protocol becomes inefficient. Nevertheless, in this section, we introduce a non-optimal general strategy that allows us to identify up to any number of errors $\lambda$. The number of auxiliary states required depends on the number of states $n$ and the maximum number of errors $\lambda$ considered.

The goal is to isolate each error in a subensemble and to subsequently identify them as it is explained in Sec.~\ref{sec toy model 1 error}. The protocol begins by applying the ENG, Eq.~\eqref{eq C sum erros}, with an auxiliary state of $d=2\lambda+1$ levels. From this operation, followed by the measurement of the auxiliary state, we can infer the difference of errors, i.e., $(\#01-\#10)\text{mod}(2\lambda+1)$, which provides a lower bound for the number of errors. The strategy now consists in splitting the ensemble and repeating the previous step for each subensemble, therefore determining the difference of errors in each subensemble. Note that, as the total difference of errors is known, now an auxiliary state of $d=\lambda+1$ is needed to perform this measurement. Iterating this step we can divide the ensemble into subensembles until the difference of error in each subensemble is 0, 1, or $-1$. Finally, if $\lambda$ or $\lambda-1$ errors are detected, these errors are already isolated. Otherwise, we check in each subensemble if it contains zero or one error. This is done by performing a random splitting of the ensemble and repeating the same step for each new subensemble. If there is more than one error the probability for detecting them is larger than $\frac{1}{2}$. Therefore this step can be iteratively repeated until the probability to find more than one error is arbitrarily small.

\subsubsection{Yield and fidelity of the purified ensemble}
\label{sec Y and F}

\begin{figure}
    \centering
    \subfloat[\centering]{\includegraphics[width=1\columnwidth]{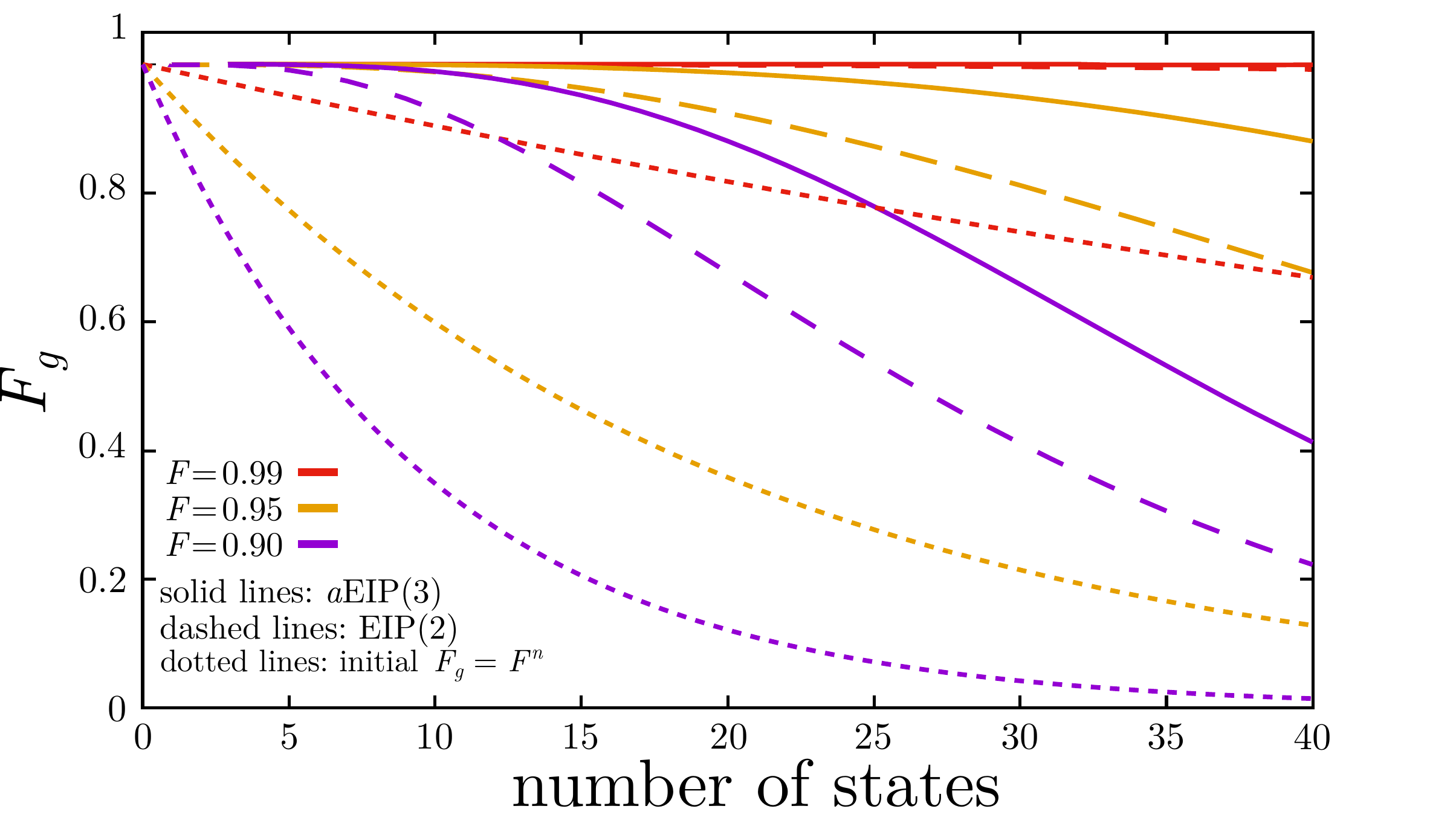} \label{fig det2 Fg}} \hfill
    \subfloat[\centering]{\includegraphics[width=1\columnwidth]{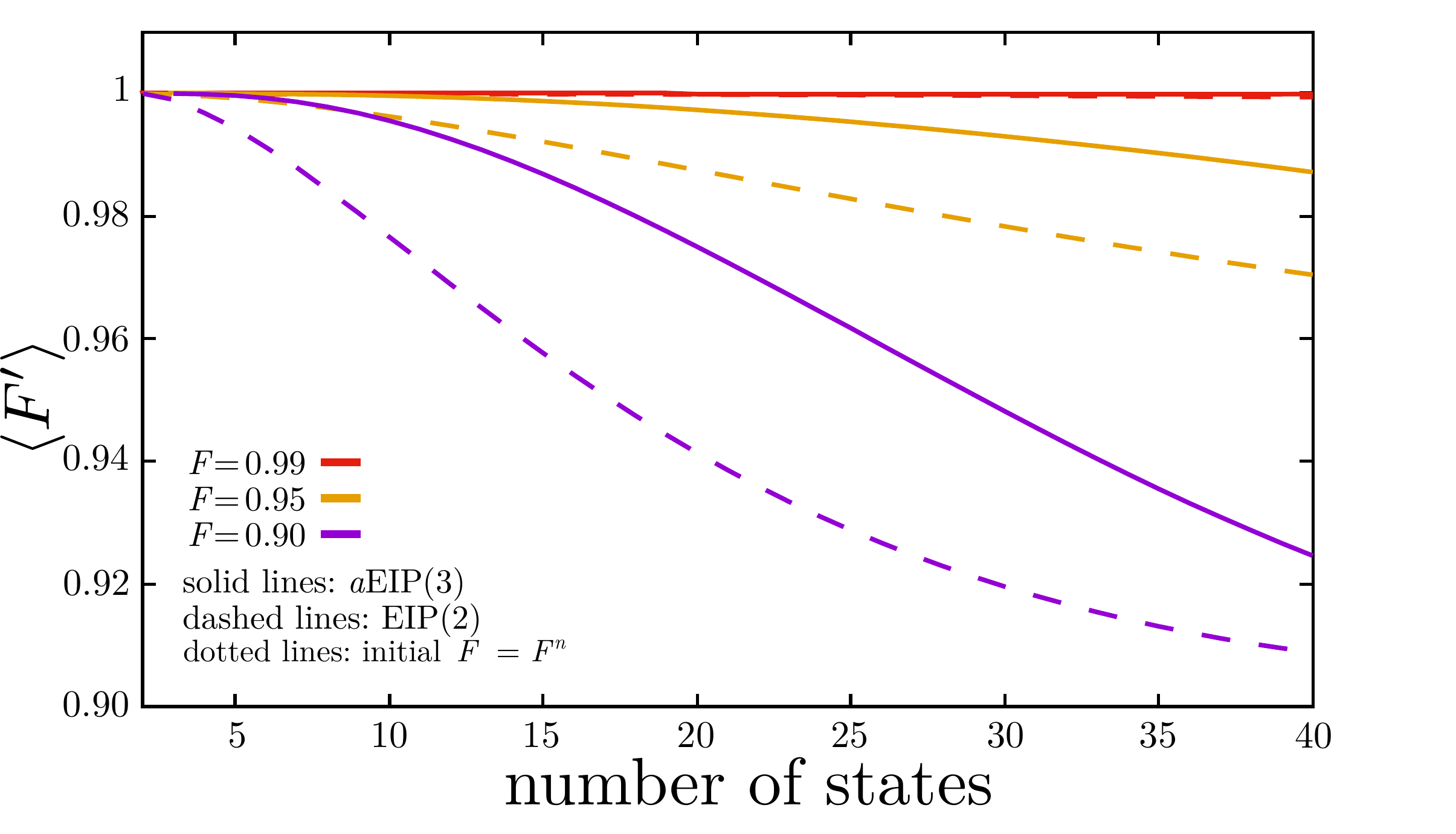} \label{fig det vs prob Fid}}
    \hfill
    \subfloat[\centering]{\includegraphics[width=1\columnwidth]{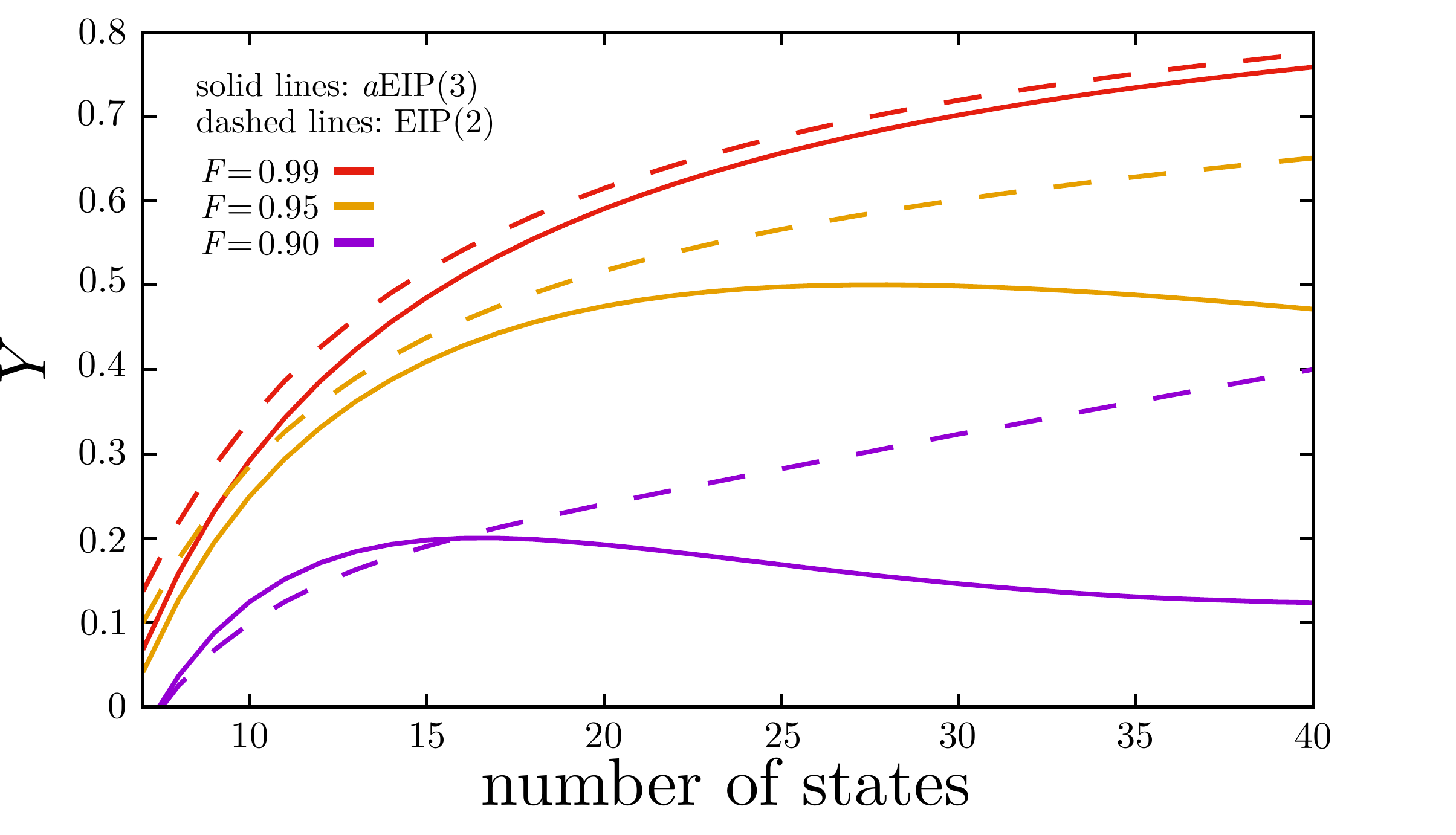} \label{fig det vs prob Y}}
    \centering
    \caption{Output fidelity and yield for the EIP(2) and the \textit{a}EIP(3) for an ensemble of Bell-diagonal rank-3 states, Eq.~\eqref{eq rank3 state}. Solid lines correspond to the \textit{a}EIP(3), dashed lines to the EIP(2). Each color represents a different value of the initial fidelity $F$. In (a), we plot the global fidelity as a function of the number of states. Dotted lines correspond to the initial ensemble. In (b), we plot the local fidelity of the purified ensemble as a function of the number of states. In (c), we plot the yield of the protocols as a function of the number of states.}
\end{figure}

The initial local fidelity and the global fidelity of the ensemble are given by $F$ and $F_g = F^n$ respectively. The global fidelity of the purified ensemble $F'_g$ corresponds to the probability of identifying all the errors of the ensemble. Since the identification of the errors is only correct if the ensemble contains at most $\lambda$ errors, $F'_g$ is equal to the probability of the ensemble to contain $k\leq\lambda$ errors. Since if this is not the case, at least one of the errors will be neither identified nor discarded such that the global final state is orthogonal to $\ket{\Psi_{00}}^{\otimes m}$. Therefore the expression for the global fidelity only depends on $\lambda$ and is given by
\begin{equation}
    F'_g = \sum^{\lambda}_{ k = 0 } \binom{n}{k} F^{n - k} \big(1 - F\big)^{k}.
\end{equation}
In Fig.~\ref{fig det2 Fg} we plot $F_g$ of the initial and the purified ensemble as a function of the number of states. The purification is done via the EIP(2) (Sec.~\ref{sec up to 2 errors}). Note that the global fidelity always increases after the purification, i.e., $F_g<F'_g$, as $F_g$ is given by the probability of the ensemble containing no-errors, and $F_g'$ by the probability of the initial ensemble containing $k\leq\lambda$ errors. For the same reason, when both values approach zero when $n\to\infty$.

Unlike for the global fidelity, the contribution of the non-considered configurations (those with $k>\lambda$ errors) to the local fidelity is not zero. If case the ensemble contains more than $\lambda$ errors one would incorrectly locate $k$ errors, and the probability of each state of being an error does not drop to zero, and it has to be explicitly computed, see Appendix~\ref{sec app FL}. In Fig.~\ref{fig det vs prob Fid} we plot the local fidelity $F'$ of the ensemble purified with EIP(2) as a function to the initial number of states $n$. From this figure, one can see again that the local fidelity always increases, i.e., $F<F'$, but in the asymptotic limit $n\to \infty$ it approaches the initial local fidelity, $F'\to F$, for the same reason that $F'_g\to0$. For $n > (\lambda+1)/(1-F)$, the assumption about the maximum number of errors is on average wrong, and for $n\gg(\lambda+1)/(1-F)$ it is almost certainly wrong. In this regime, the information obtained from the ensemble is not correctly interpreted and random states are identified as errors. This behavior can be clearly observed considering an ensemble of $F=0.8$ and $n=20$, where the expected number of errors is already 4.

Another important feature of the behavior of $F'$, is that it significantly depends on the number of identified errors. $F'$ is higher in the case that no errors are detected, and it decreases for the two different errors case, for the one error, and finally for the two identical errors case. These differences can be explained due to the compatible configurations that are not considered. The probability of falling outside of the considered configurations depends on the number of errors detected, since after each measurement, the probability distribution of all possible configurations changes because several of them are no longer compatible. For instance, if two identical errors are detected, the second most probable configuration contains three identical errors, whereas if no errors are detected, the second most probable configuration contains four. Therefore, the conditional probability that the assumption $k\leq \lambda$ is false after two errors are detected is different from when no errors are detected. See Appendix~\ref{sec app 2 errors} for more details.

The yield is given by the initial number of states $n$, minus the expected number of resources and errors identified over all possible scenarios of different values of $j$ up to $\lambda$ errors divided by $n$, i.e.
\begin{equation}
    \label{eq yepslion}
    Y_\epsilon = \frac{n - [ 1 - S(\hat{\rho}'_{BD}) \, ]^{-1} R_{t} - \sum_{j=-\lambda}^{\lambda}p_\lambda(j)\,k(j)}{n}\,,
\end{equation}
where $p_\lambda(j)$ is the probability of obtaining a difference $j$ and it depends on $\lambda$. $k(j)$ is the expected number of errors identified if a difference $j$ is obtained, see Appendix~\ref{app eq yields 2 errors} for a concrete example for $\lambda=2$. The expected number of resources depends on $\lambda$, and it is given by
\begin{equation}
    R_{t}=\log_2(2\lambda+1)+\sum_{j=-\lambda}^\lambda p_\lambda(j)R_{\lambda}(j),
\end{equation}
where $p(j)$ is the probability of obtaining a difference of $j$ and it is given in Appendix~\ref{sec app FL}. In this case, as no pure states are obtained, we include the discount factor $(1-S(\hat{\rho}'_{BD}))^{-1}$ to quantify the number of purified qubit pairs that have to be given back in order to compensate for the consumed resource states. Besides, we recall that here we assume that the target fidelity is low enough, such that the purification is always successful and the yield is computed on all output states. Therefore the yield plotted in Fig.~\ref{fig det vs prob Y} describes the clear increase of the amount of entanglement in the ensemble.

In Appendix~\ref{app sec Y2} we provide the expressions of $R_{\lambda}(j)$ for $\lambda=2$ errors. Fig.~\ref{fig det vs prob Y} shows the yield of purifying an ensemble of rank-3 Bell states using EIP(2) protocol. One can see that for states with $F\geq 0.9$ a purification with a positive yield can be successfully performed. In the asymptotic limit $n\to\infty$ the yield approaches $Y_\epsilon\to1$. However, we recall that in the asymptotic limit the fidelity of the states is not increased. Equation \eqref{eq yepslion} is only useful for regimes (given by $F$ and $n$) where the assumption about the maximum number of errors is justified.

\subsubsection{Protocol with abortion; \texorpdfstring{$\lambda= 3$}{TEXT}}

Here we describe a protocol to purify Bell-diagonal rank-3 states that detects up to three errors, $\lambda=3$. Introducing two modifications to the EIP(2), we obtain a new protocol that allows one to detect the three errors situations as well. When three errors are detected, one can identify them with the procedure described in Sec.~\ref{sec general number of errors}. However, if we detect more errors, the procedure requires much more resources. For that reason, one option is to abort the protocol if three errors are detected. Following this strategy, we provide an alternative protocol that assumes that the ensemble contains at most three errors, and performs well if this assumption is justified. These three errors configurations are the most probable of all the configurations neglected by EIP(2), and discarding them increases the output fidelity. On the other hand, in most cases, it reduces the yield. We call this protocol \textit{a}EIP(3).

\begin{figure}
    \centering
    \textbf{$a$EIP(3)}\par\medskip\vspace{-0pt}
    \includegraphics[width=1\columnwidth]{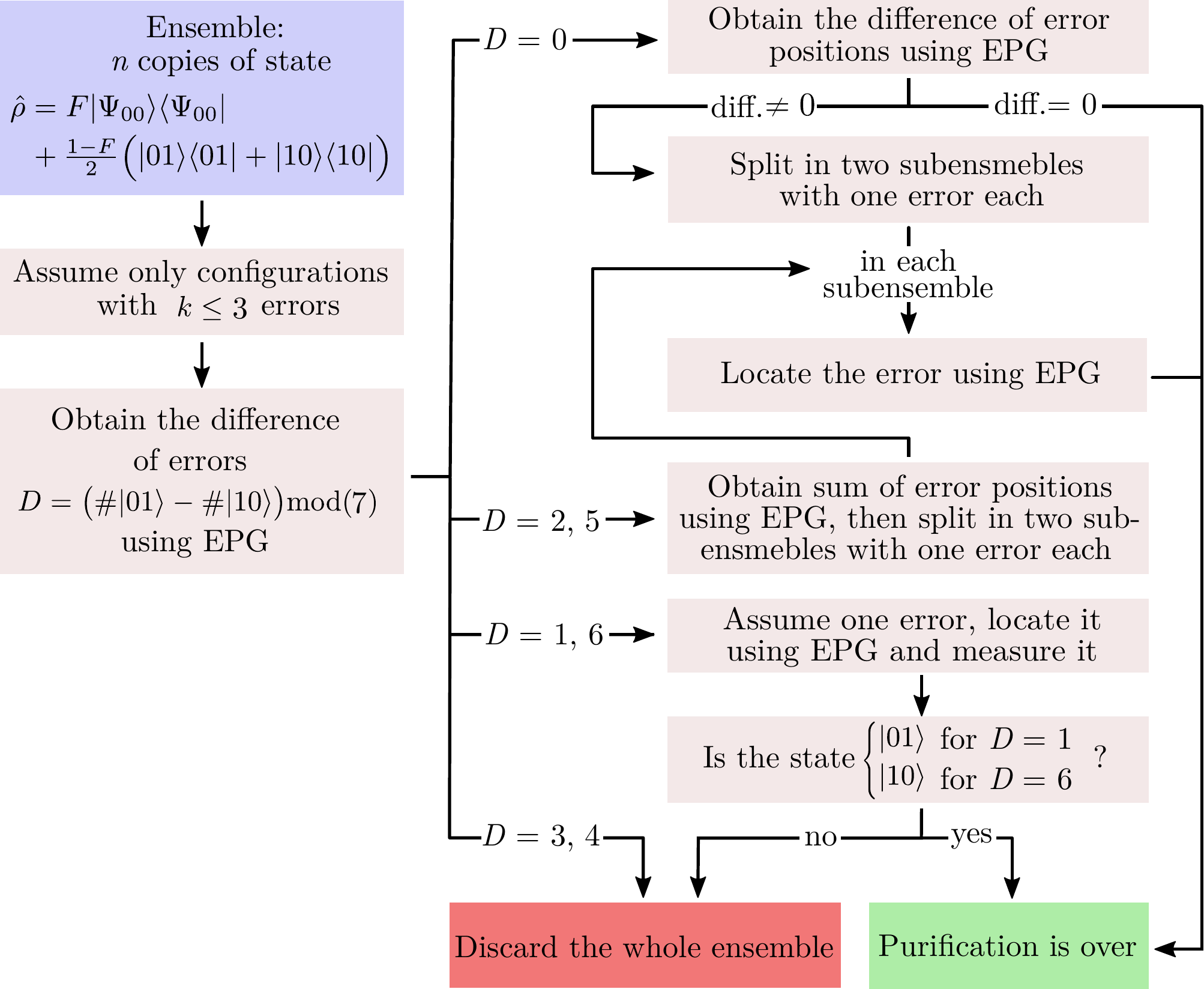} \label{fig:tree:aEIP3}
    \centering
    \caption{Diagram illustrating the steps of the $a$EIP(3). The blue box corresponds to the stating point, the green box to the situation where one end up with a successful purification and the red box to situation where too many errors are found and the procedure aborts.}
    \label{fig:tree_aEIP(3)}
\end{figure}

With three errors we can face up with two different scenarios (three identical errors or one different error). The first difference with the EIP(2) procedure is given in the first step. As the value of $\lambda$ is increased from 2 to 3, now there are seven possible values for the difference $j=(\#01-\#10)$. Obtaining this difference allows us to detect the situations where the ensemble contains three identical errors, revealed by $j=3$ or $4$. On the other hand, we have an ambiguity in the number of errors for $j=1$ or $j=6$, which is not present for the EIP(2) procedure. In these cases, the ensemble can contain either one or three (different) errors. To distinguish the two situations, we first assume that there is one error in the ensemble and locate it with the help of the procedure of Sec.~\ref{sec toy model 1 error}. Then, we locally measure the qubits identified as erroneous in the $Z$ basis. Straightforward combinatorics show that we observe the expected measurement results if and only if the ensemble contains one error indeed (see Appendix~\ref{sec app detect three errors}). In any other measurement results, the ensemble is discarded. In Fig.~\ref{fig:tree_aEIP(3)}, we illustrate the steps of the protocol in a diagram.

These two simple modifications translate into a remarkable increase in the fidelity of the purified ensemble. The yield of the protocol is however slightly reduced for small $n$ and it approaches $Y_\epsilon\to 1-p_{\text{abort}}$ in the asymptotic limit. This improvement can be observed in Figs.~\ref{fig det vs prob Fid} and \ref{fig det vs prob Y}. With some probability $p_{\text{abort}}$, the whole ensemble is discarded. In Appendix~\ref{app sec 3errors}, we plot the expected value of $F'$ and the occurrence probability of each possible branch with given total numbers of errors (\#01,\#10).

\subsubsection{Blocking strategy}

As we have seen, for big ensembles with a large expected number of errors EIP(2) and $a$EIP(3) protocols only provide marginal improvement of fidelity. Naturally, in these situations more complex strategies, taking more errors into account, are required. Nevertheless, there is a simple alternative approach that consists in dividing the ensemble into blocks of fixed size, for which the assumption on a small total number of errors is justified. Each block is then purified independently.

As can be seen from Figs.~\ref{fig det vs prob Fid} and \ref{fig det vs prob Y} there is a certain freedom in the choice of the size of the blocks where a successful purification can be performed. While the fidelity of the purified states decreases with the number of states, the yield increases. We find that there is a maximal yield corresponding to some fixed size of the blocks.

\subsection{Purification of full--rank states}
\label{sec bell diagonal deploraization}

In this section we show how we can purify an ensemble of $n$ copies of a full-rank noisy Bell state, i.e., states of the form of Eq.~\eqref{eq initial state} with $p_\mu>0$ $\forall\,\mu$, by repeating the strategy described in Sec.~\ref{sec rank3 states}, and adding a basis change.

If the initial ensemble is depolarized as described in Sec.~\ref{sec protocol structure}, the states are given by
\begin{equation}\label{eq: full rank bell}
    \begin{aligned}
        \hat{\rho}_{AB}&=F\,|\Psi_{00}\rangle\langle\Psi_{00}|+p_3|\Psi_{10}\rangle\langle\Psi_{10}|\\ &+\frac{1-\widetilde{F}}{2}\Big(\,|01\rangle\langle01|+|10\rangle\langle10|\,\Big),
    \end{aligned}
\end{equation}
where $\widetilde{F}=F+p_3$. Therefore, in this scenario, the ensemble can contain three different kinds of errors. From Eq.~\eqref{eq action bCX}, we see that the counter gate transfers information about the 01 and the 10 errors but does not distinguish between the error $|\Psi_{10}\rangle$ and the target $|\Psi_{00}\rangle$ states. In consequence, to deal with this situation of three kinds of errors in the same ensemble, two purification rounds are needed.

The first purification round consists in locating the 01 and the 10 error states with the EIP (or $a$EIP), described in Sec.~\ref{sec rank3 states}, with an effective fidelity of $\widetilde{F}$. After this first round, the probability of having 01 and 10 error states is decreased and the dominant noise contribution comes from the error states $|\Psi_{10}\rangle$. To purify this kind of errors we can transform them into the 01/10 type, while keeping the target states unchanged. To do so to each pair of qubits we apply the channel $\mathcal{D}_3:\{\frac{1}{\sqrt{2}}\id\otimes\id,\frac{1}{\sqrt{2}}\hat{X}\otimes\hat{X}\}$, followed by the QFT Eq.~\eqref{eq QFT}, and finally the channel $\mathcal{D}_2$, Eq.~\eqref{eq separant Bell states}. In particular, the effect of this operation on the four possible pure states appearing in Eq.~\eqref{eq: full rank bell} reads
\begin{equation*}
    \begin{aligned}
        |\Psi_{00}\rangle\langle\Psi_{00}|\mapsto &\;|\Psi_{00}\rangle\langle\Psi_{00}|, \\|\Psi_{10}\rangle\langle\Psi_{10}|\mapsto &\;\frac{1}{2}\big(\,|01\rangle\langle01|+|10\rangle\langle10|\,\big),\\|01\rangle\langle01|\;\,\,\mapsto &\;\frac{1}{2}|\Psi_{10}\rangle\langle\Psi_{10}|+\frac{1}{4}\big(|01\rangle\langle01|+|10\rangle\langle10|\big), \\ |10\rangle\langle10|\;\,\,\mapsto &\;\frac{1}{2}|\Psi_{10}\rangle\langle\Psi_{10}|+\frac{1}{4}\big(|01\rangle\langle01|+|10\rangle\langle10|\big).
    \end{aligned}
\end{equation*}
This whole operation leaves the Bell state $|\Psi_{00}\rangle$ invariant and transforms the pure error state $|\Psi_{10}\rangle$ into a mixture of 01 or 10 error states, making them detectable with the counter gate. With a 50\% chance the remaining 01 and 10 noise from the first round becomes ``hidden'', but with 50\% probability, it remains detectable. After this operation, a second purification round is performed applying the EIP (or $a$EIP) protocol again, see Sec.~\ref{sec rank3 states}.

\subsection{Error identification protocols vs hashing \& recurrence protocols}

\begin{figure}
    \centering
    \vspace{-0.3cm}
    \subfloat[\centering]{\includegraphics[width=1\columnwidth]{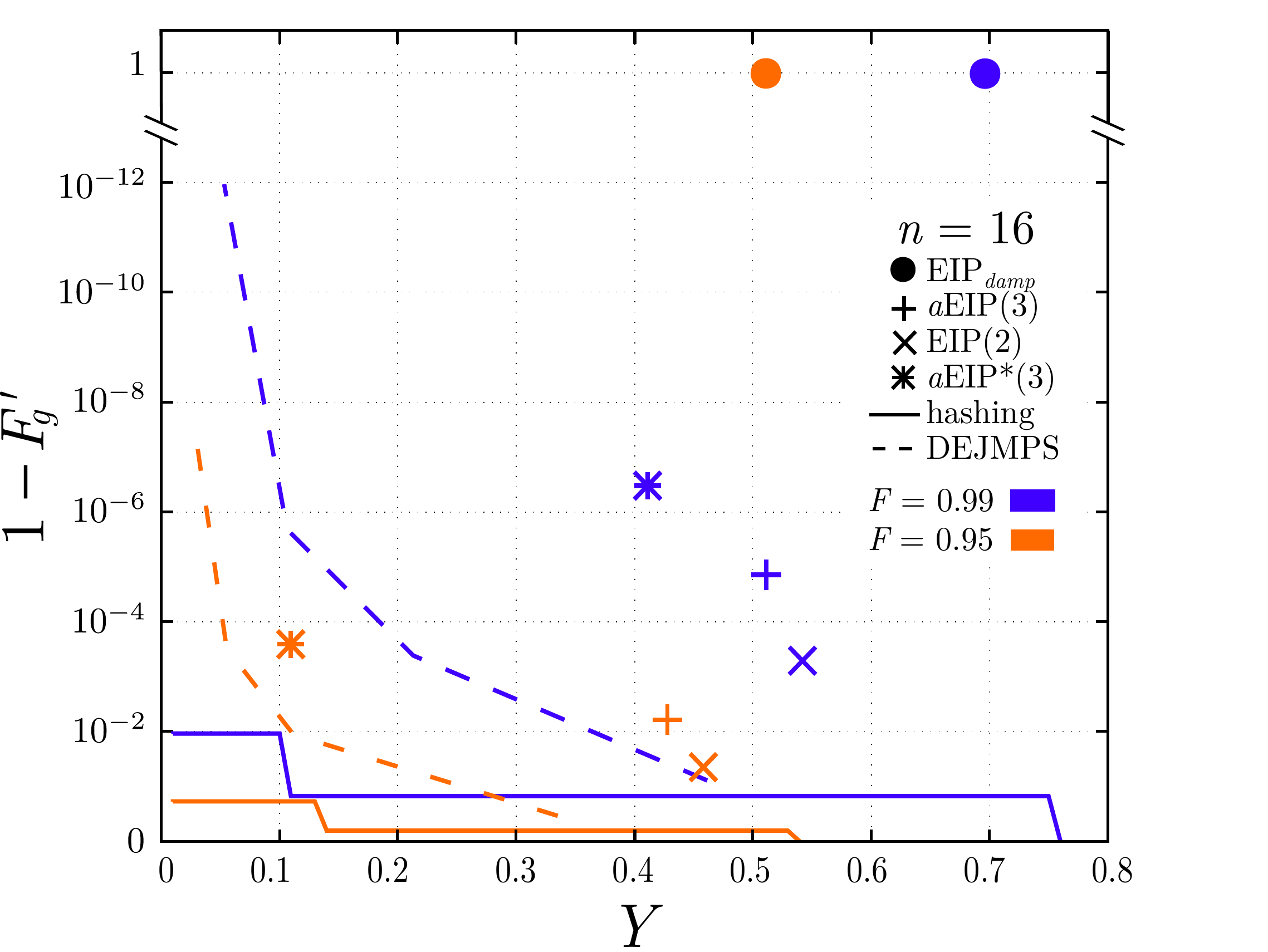}}\hfill\vspace{-0.2cm}
    \subfloat[\centering]{\includegraphics[width=1\columnwidth]{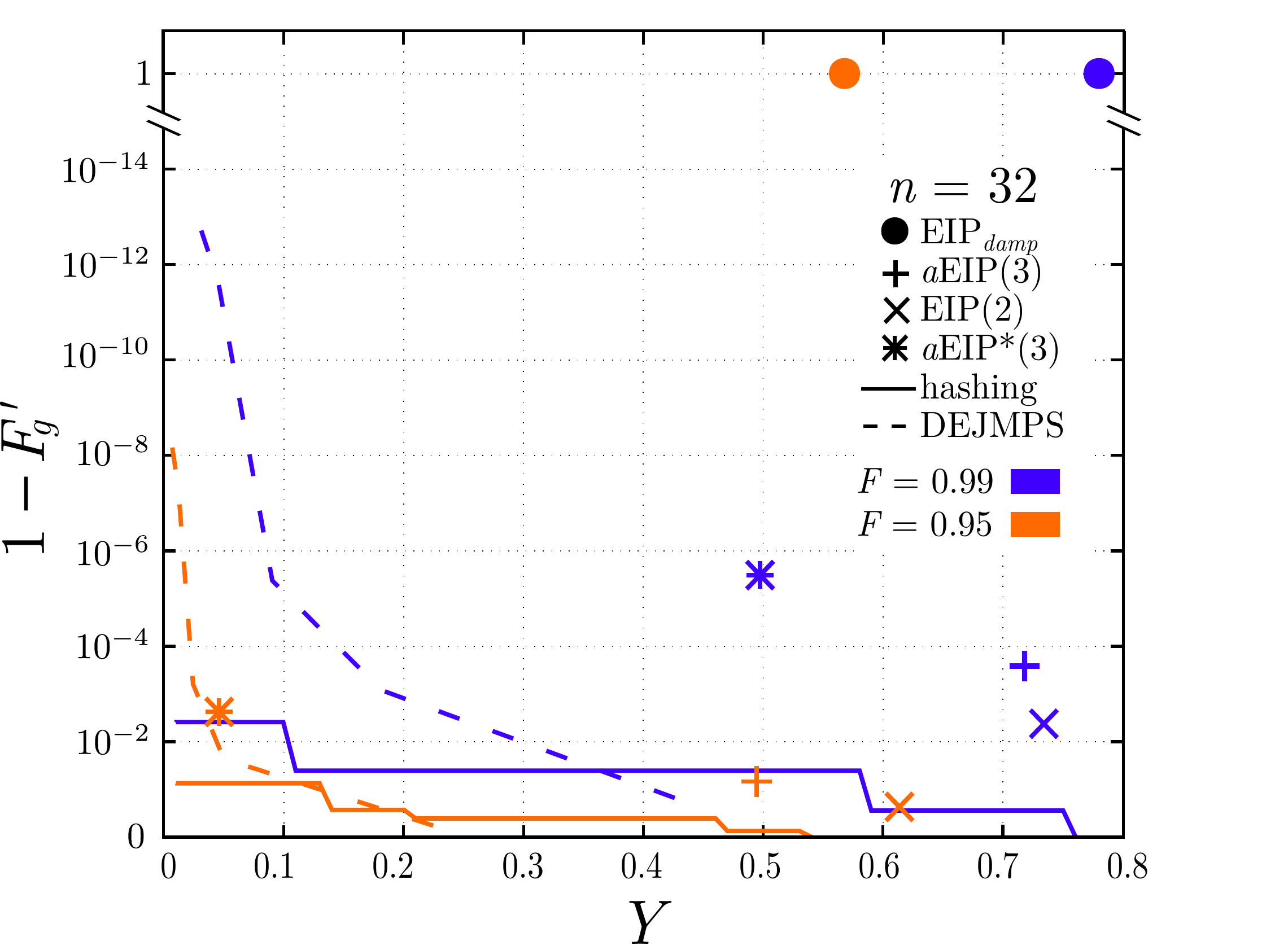}}
    \caption{\label{fig EIPvsHashDEJ}Global fidelity of the purified ensemble as a function of the yield of the purification. Each point corresponds to a different EIP, dashed lines to the DEJMPS recurrence protocol, and solid lines to the finite hashing protocol, where we use the upper bound for the output fidelity derived in Appendix~\ref{app sec upper bound hash}. Each color corresponds to a different initial fidelity. The initial ensemble consists of $n$ copies of the state of Eq.~\eqref{eq rank3 state}, except for EIP$_{damp}$ where we assume copies of the state of Eq.~\eqref{eq rank2}. In (a) the initial ensemble is of $n = 16$ states, and in $b)$ of $n = 32$.}
\end{figure}

In this section, we compare different EIPs with the DEJMPS recurrence protocol \cite{deutsch1996quantum} and the standard hashing protocol in the regime of a finite ensemble \cite{zwerger2018long}.

In Fig.~\ref{fig EIPvsHashDEJ}, we plot the global fidelity of the purified ensemble $F'_g$ as a function of the yield of the purification (see Appendix~\ref{app sec EIPvsDEJMPShash} for local fidelity comparison) for the DEJMPS protocol, the finite hashing protocol, and three different EIPs--$a$EIP$^*(3)$ corresponds to an alternative approach for the $a$EIP(3), where the only case where we do not abort is if no errors are detected. We take as initial ensemble $n$ copies of a rank-3 Bell-diagonal state given in Eq.~\eqref{eq rank3 state}, but for the EIP$_{damp}$--the EIP described in Sec.~\ref{sec toy model}--we assume states of the form of Eq.~\eqref{eq rank2}. Note that we obtain the same yield and output fidelity when we purify these two different ensembles with the recurrence or the hashing protocol. That is because both protocols require an initial depolarization procedure that makes them identical.

From Fig.~\ref{fig EIPvsHashDEJ} we show that EIPs provide much better results than the recurrence and the hashing when the noise is modeled via the damping channel. Whereas EIP allows one to obtain fidelity-one states with a considerable large yield, for recurrence protocols the yield decreases drastically when high fidelity states are required. For the hashing protocol, we show that for $n\leq 32$, the ensemble is too small to provide high fidelity states.

For rank-3 Bell-diagonal ensembles, we show that a priori--as we did not consider the possibility of iterating the EIP--EIP's can not provide states with high fidelity as the recurrence protocol. However, the yield obtained with the EIP is much higher. For instance, for the case of an ensemble of $F=0.99$ and $n=32$, for the same value of output fidelity, the yield of the EIP is five times larger than the yield obtained with the DEJMPS protocol.

\section{Influence of noise and imperfections}
\label{sec Noise model}

In general, with error identification protocols no pure states can be obtained, even though maximally entangled auxiliary states are assumed. In order to do a more realistic analysis, we consider here that the auxiliary states which the protocol makes use of, are noisy. This makes the protocol readily applicable. We also consider noisy operations and decoherence.

\subsection{Amplitude noise}

First, we study the effect of amplitude noise. This noise is modeled by the \textit{amplitude noise channel} defined as
\begin{equation}
    \label{eq channel amplitude}
    \mathcal{X}_{p}\left(\hat{\rho}\right)=p\,\hat{\rho}+\frac{1-p}{d}\sum_{j=0}^{d-1}\hat{X}^j\hat{\rho}\,\hat{X}^{\dagger j}.
\end{equation}
This channel consist of random applications of $\hat{X}$ with probability $1-p$.

\subsubsection{Auxiliary state}

The result of sending one of the two sub-systems of a maximally entangled state of qudits throught $\mathcal{X}_{p}$ Eq.~\eqref{eq channel amplitude} is a mixed state, $\left(\id_{d}\otimes\mathcal{X}_{p}\right)|\Psi^{(d)}_{00}\rangle\langle\Psi^{(d)}_{00}|=\hat{\varrho}^{(d)}_{AB}$, given by
\begin{equation}
    \label{eq noisy1 aux}
    \hat{\varrho}^{(d)}_{AB}=p\,\big|\Psi^{(d)}_{00}\big\rangle\big\langle\Psi^{(d)}_{00}\big|+\frac{1-p}{d}\sum_{j=0}^{d-1}\big|\Psi^{(d)}_{0j}\big\rangle\big\langle\Psi^{(d)}_{0j}\big|,
\end{equation}
where the fidelity of the state is $\mathcal{F}=p+\frac{1-p}{d}$. The amplitude noise channel destroys the information about the amplitude index of the maximally entangled state.

In order to see if auxiliary states of the form $\hat{\varrho}_{AB}^{(d)}$ can be used to implement the protocol, we need to see the action of the noise on the ensemble and $b$CX. It is given by
\begin{equation}
    \begin{aligned}
        b\text{CX}\Big(|\Psi_{00}\rangle\langle \Psi_{00}|\otimes\hat{\Sigma}\Big) b\text{CX}^{\dag} &= |\Psi_{00}\rangle\langle \Psi_{00}|\otimes\hat{\Sigma},\\ b\text{CX}\Big(|m\;\!n\rangle\langle m\;\!n|\otimes\hat{\Sigma}\Big) b\text{CX}^{\dag} &= |m\;\!n\rangle\langle m\;\!n|\otimes\hat{\Sigma},
    \end{aligned}
\end{equation}
where $\hat{\Sigma}\equiv\frac{1}{d}\sum_{j=0}^{d-1}|\Psi^{(d)}_{0j}\rangle\langle\Psi^{(d)}_{0j}|$ describes the noise part. Note that the amplitude noise does not affect the states of the ensemble. Therefore, by linearity, after applying $b$CX and measuring it, the density operator is only modified with a certain probability $p$, i.e., $\hat{\Gamma}\rightarrow p\,\hat{\Gamma}'+(1-p)\hat{\Gamma}$.

This kind of noisy auxiliary states can also be locally obtained from enough copies of the mixed state
\begin{equation}
    \label{eq rank2}
    \hat{\mu}_{AB}=F\,|\Psi_{00}\rangle\langle\Psi_{00}|+\big(1-F\big)|\Psi_{10}\rangle\langle\Psi_{10}|.
\end{equation}
This is done by embedding several copies of the state $\hat{\mu}_{AB}$ (see Appendix~\ref{app sec embedding}). Specifically one can transform $\hat{\mu}_{AB}^{\otimes k}\to \hat{\varrho}^{(2^k)}_{AB}$ with fidelity $\mathcal{F}=F^k$. However, only noisy auxiliary states of $d=2^k$ levels and $k\in\mathbb{N}$ can be obtained.

From an initial ensemble of $n$ copies of a Bell-diagonal rank-2 state, i.e., $\hat{\mu}_{AB}^{\otimes n}$, we can use an auxiliary pool of copies of the same state $\hat{\mu}_{AB}$ to purify the initial ensemble. After depolarizing the $n$ copies of the first ensemble to the form of Eq.~\eqref{eq rank3 state}, we can purify it following the procedure described in Sec.~\ref{sec rank3 states}, but using auxiliary states obtained from the auxiliary pool of copies of $\hat{\mu}_{AB}$. This allows us to directly compare the initial states of the ensemble and the used auxiliary states.

\begin{figure}
    \centering
    \textbf{Noisy EIP(1) for rank-2 Bell-diagonal states}\par\medskip
    \vspace{-0.4cm}
    \subfloat[\centering]{\includegraphics[width=0.5\columnwidth]{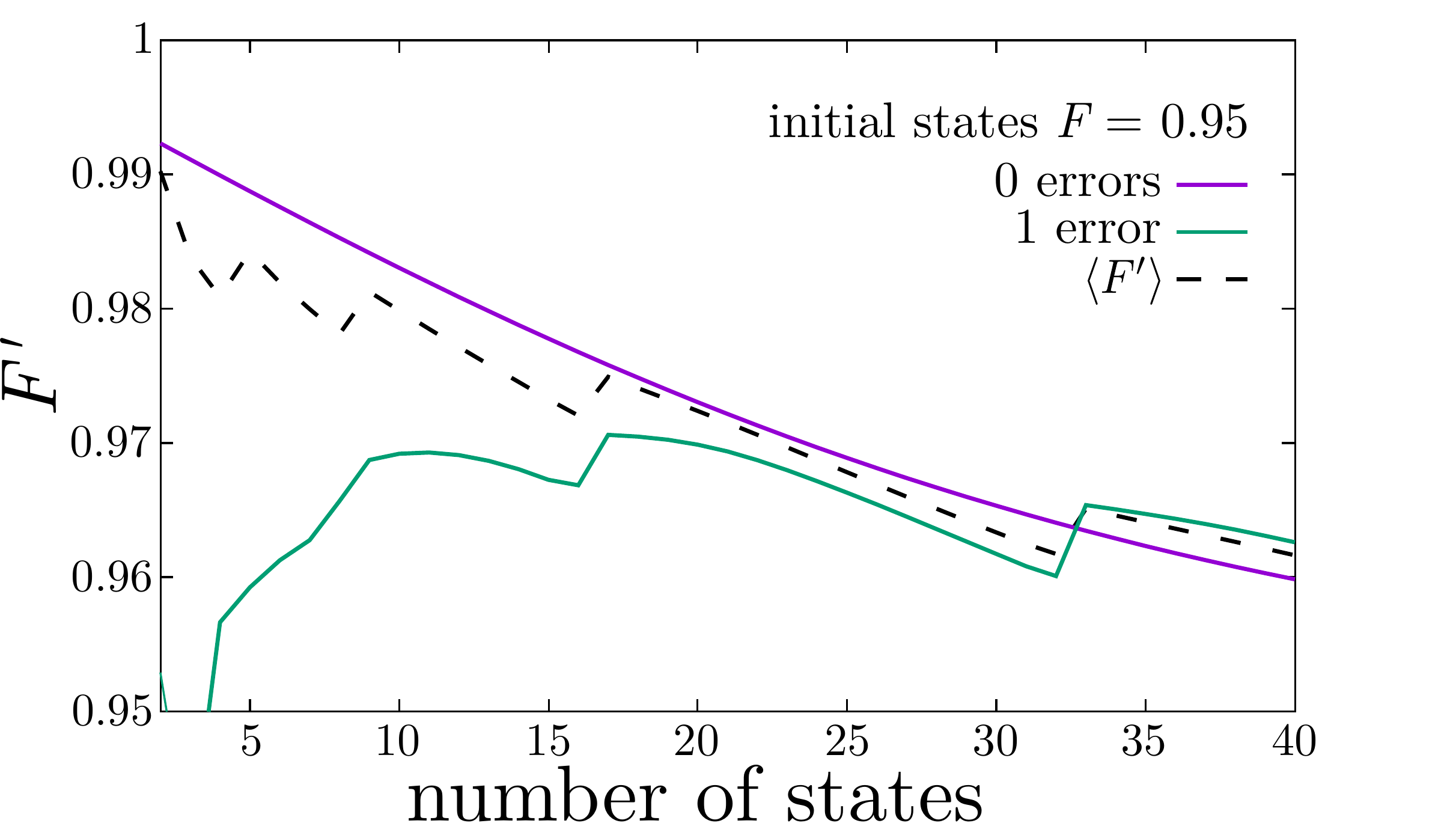} \label{fig hashing 1 error a}}
    \subfloat[\centering]{\includegraphics[width=0.5\columnwidth]{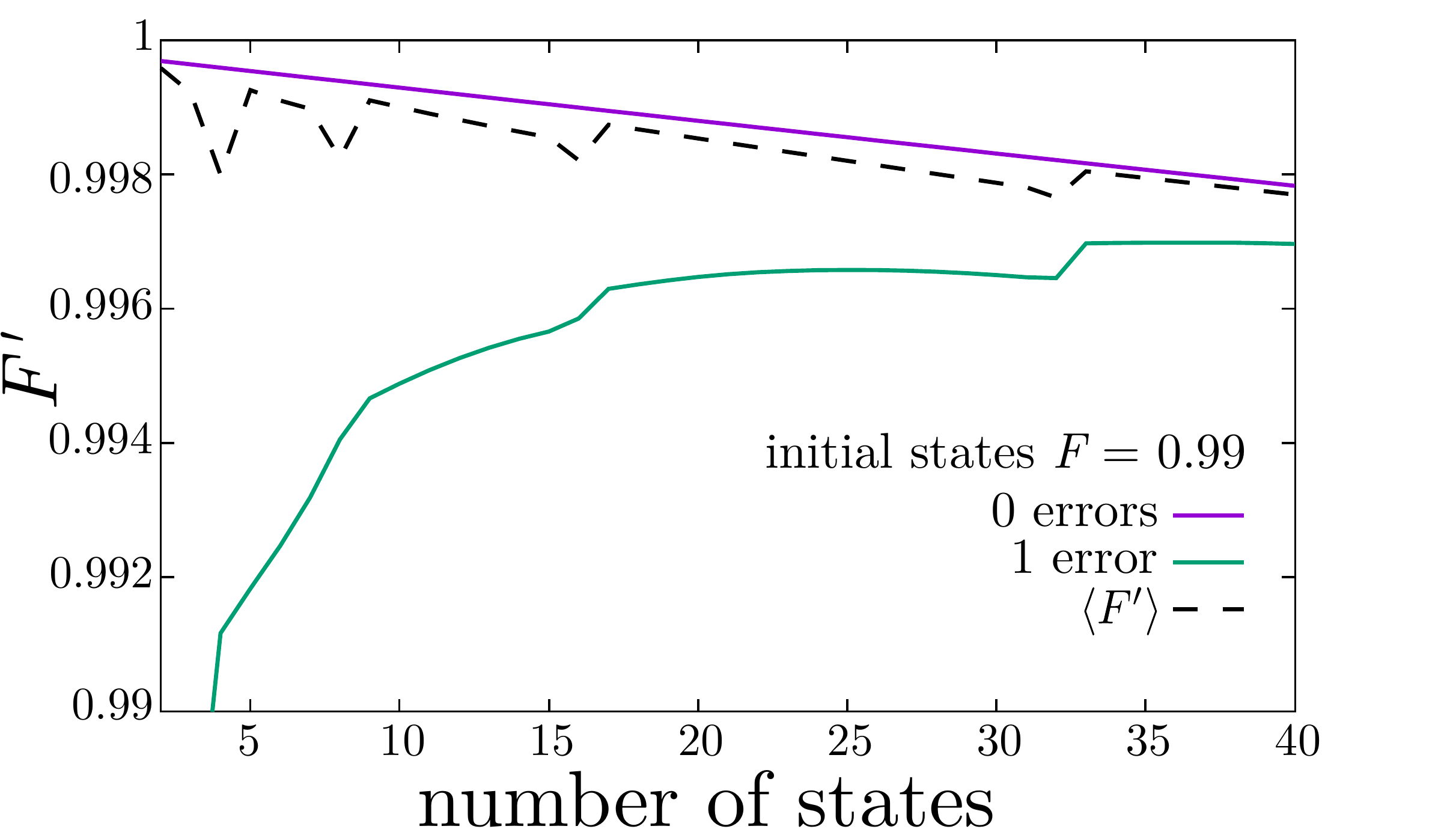} \label{fig hashing 1 error b}}
    \hfill
    \vspace{-0.25cm}
    \subfloat[\centering]{\includegraphics[width=0.5\columnwidth]{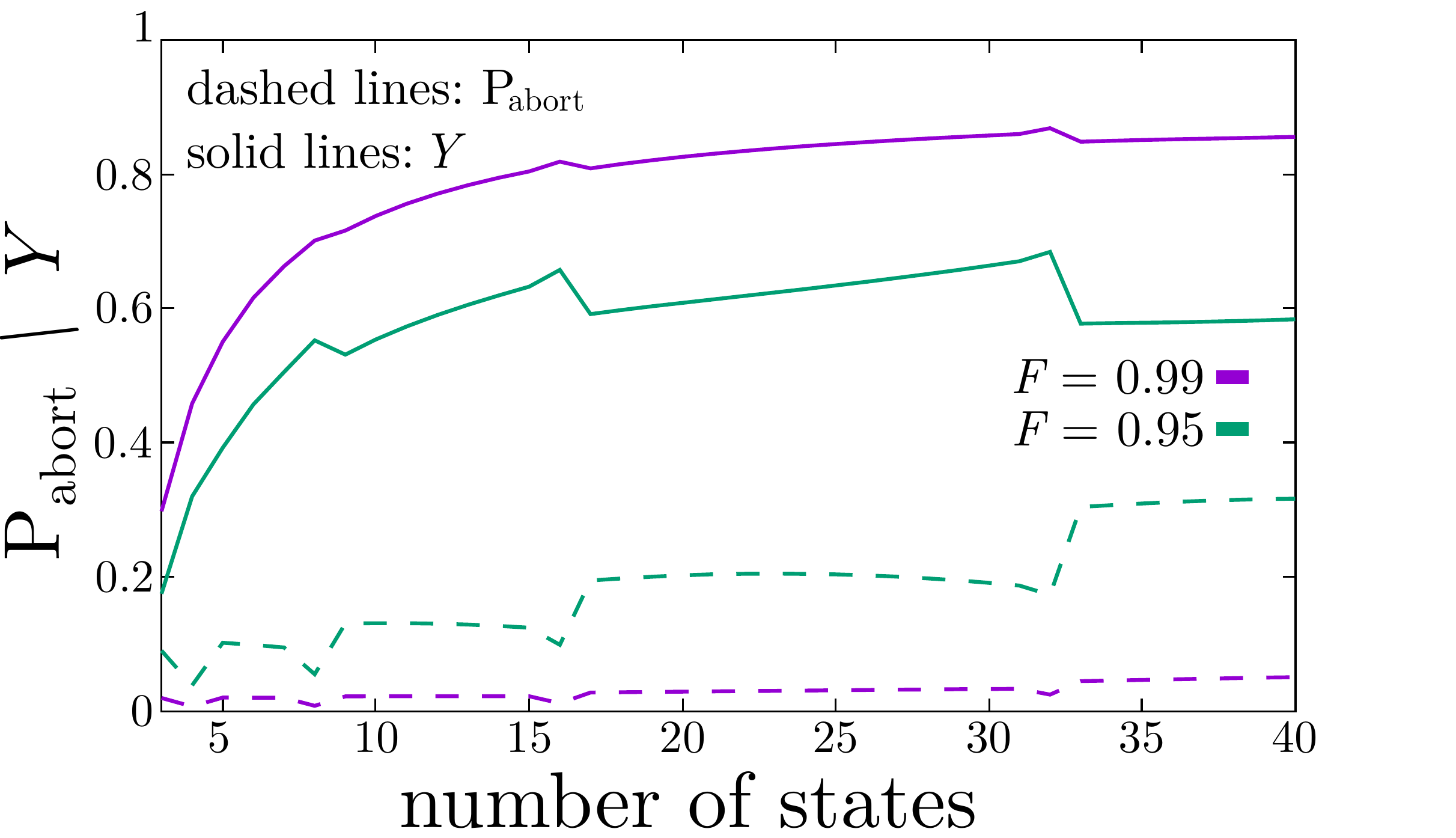} \label{fig hashing 1 error c}}
    \subfloat[\centering]{\includegraphics[width=0.5\columnwidth]{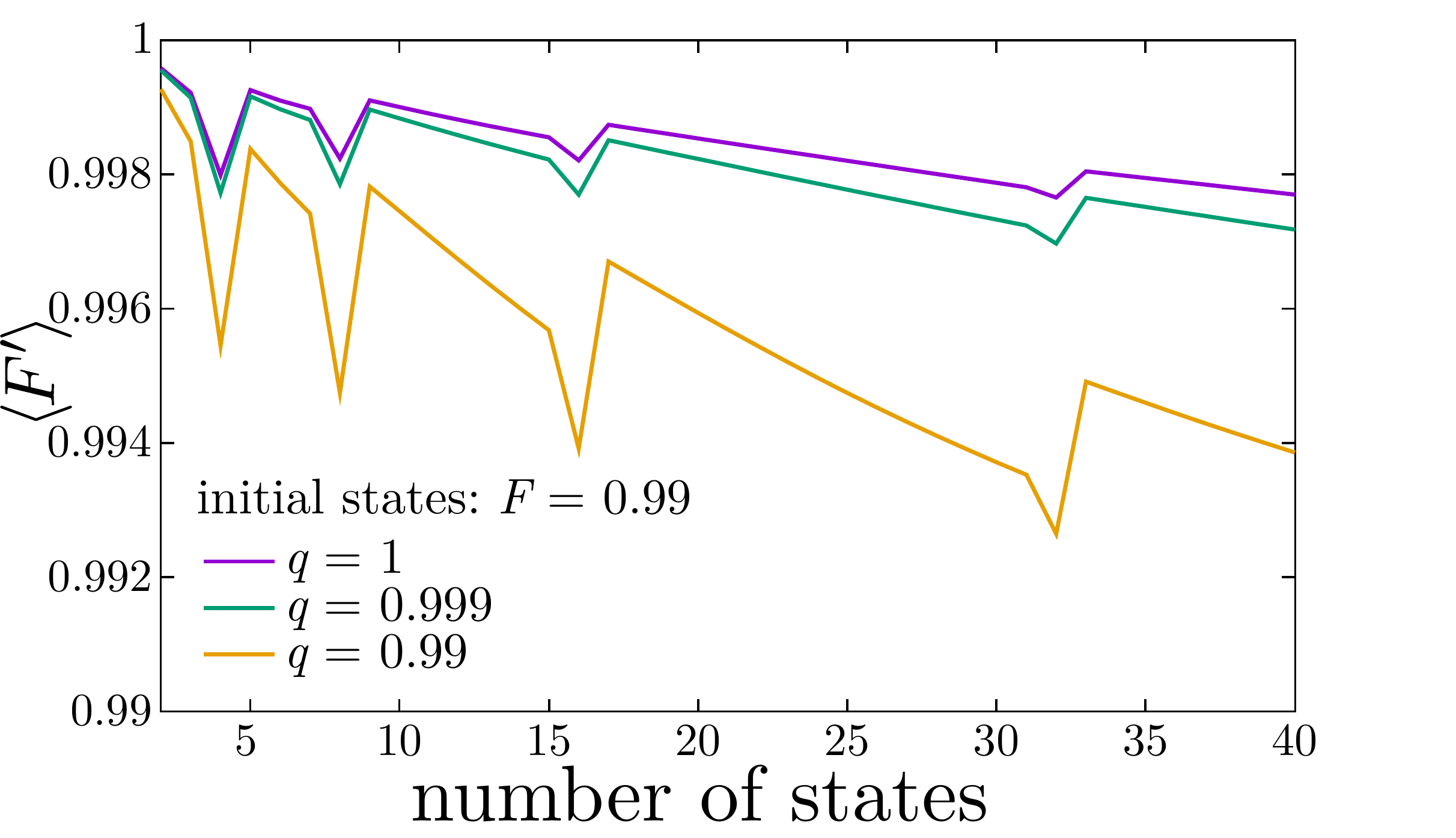} \label{fig hashing 1 error d}}
    \hfill
    \caption{Purification of the ensemble of rank-2 Bell states via EIP(1) using auxiliary states which are obtained from extra copies of the state of the initial ensemble. In (a) and (b) we plot the local fidelity of the purified states as a function of the number of states if no errors or one error is identified. In dashed lines the average. In (c), we plot the probability of aborting and the yield of the purification as a function of the number of states. In (d), we plot the local fidelity of the purified ensemble as a function of the number of states for an ensemble of states with $F=0.99$. We assume noisy operations of the form of Eq.~\eqref{eq noisy C} and each color corresponds to a different value of the parameter $q$. The local minima that the curves present correspond to $n=2^a$ for $a\in \mathbb{N}$ when the failure probability is minimal.}
\label{fig hashing 1 error}
\end{figure}

In Figs.~\ref{fig hashing 1 error a} and \ref{fig hashing 1 error b}, we show $F'$ for the purified states considering that we purify an ensemble of Bell-diagonal rank-2 states up to $\lambda=1$ error, with auxiliary states of the form of Eq.~\eqref{eq noisy1 aux} with $d=2^k$. Note that even using noisy auxiliary states, purification is possible. The main difference with the ideal case is that $F'$ (when one error is detected) increases with the number of states in a regime. This is due to the nonzero probability of detecting one error, but failing in its identification. A second difference comes from the fact that the auxiliary states can only have certain values for $d$. This has two consequences. First, the protocol can not be deterministic. Since in general, we are using auxiliary states of $d$ larger than is required, we can obtain values that are inconsistent with our assumptions. In this case, we abort the protocol. Second, the total expected value of $F'$ shows local minima which corresponds to the power of 2 values of $n$. For these sizes of the ensemble, we can obtain an auxiliary state of the exact dimension required when one error is detected. Therefore here we do not have the possibility of finding an incompatible result. This eliminates the probability of aborting the protocol but also decreases the local fidelity of the purified states as more state configurations are still compatible.

In Fig.~\ref{fig hashing 1 error c}, we plot the yield and the aborting probability for the protocol up to $\lambda=1$ error using noisy auxiliary states. One can see that the yield and the aborting probability also show discontinuities for the same reason as the fidelity.

\subsubsection{Noisy operations}
In this section, we analyze the effect of noisy operations on the protocol. We introduce a simplified model for amplitude noise. It consists in considering that the auxiliary state is affected by $\mathcal{X}_q$ Eq.~\eqref{eq channel amplitude} before each application of $b$CX, i.e.
\begin{equation}
    \label{eq noisy C}
    \begin{aligned}
        b\text{CX}\left[\hat{\rho}\otimes\hat{\varrho}^{(d)}\right]b\text{CX}^\dagger& \\ \rightarrow b\text{CX}\left[\hat{\rho}\otimes\big(\id_d\otimes\mathcal{X}_q\big)\hat{\varrho}^{(d)}\right]b\text{CX}^\dagger&,
    \end{aligned}
\end{equation}
where $q$ is the noisy parameter of the gate. Such a noise makes the standard hashing protocol useless \cite{zwerger2014robustness}. In the standard hashing routine, the noise accumulated into the target state completely destroys the information of the ensemble transferred there.

The application of this kind of noisy operation corresponds to a decrease of the fidelity of the auxiliary state, such that $\mathcal{F}\to\mathcal{F}'=(1-q+qd\mathcal{F})/d$. In Fig.~\ref{fig hashing 1 error d}, we plot $F'$ if we purify the ensemble up to one error with noisy auxiliary states from the initial ensemble and with noisy operations. The noise coming from the operations amplifies the probability of obtaining an incorrect result from the measurement. However, we see that purification is still possible for $q\geq 0.99$.

\subsection{General noise}

Finally, we discuss the effect of more general noise, which is described by a maximally mixed state. This noise is modeled via the \textit{depolarizing channel} defined as
\begin{equation}
    \mathcal{D}_p\left(\hat{\rho}\right)=p\,\hat{\rho}+\frac{1-p}{d}\id_{d}.
\end{equation}
This channel destroys all the information contained in $\hat{\rho}$ with probability $1-p$. If we send a maximally entangled state or one of the two sub-systems through a depolarization channel, the resulting state is the so-called isotropic state $\left(\id_{d}\otimes\mathcal{D}_{p}\right)|\Psi^{(d)}_{00}\rangle\langle\Psi^{(d)}_{00}|=\hat{\rho}_{\text{I}}^{(d)}$, given by
\begin{equation}
    \hat{\rho}_{\text{I}}^{(d)}=p\,\big|\Psi^{(d)}_{00}\big\rangle\big\langle\Psi^{(d)}_{00}\big|+\frac{1-p}{d^2}\id_{d^2}.
\end{equation}
where the fidelity of the state is $\mathcal{F}=p+\frac{1-p}{d^2}$.

In order to check if isotropic auxiliary states can be used to implement the protocol, we need to analyze the action of $b$CX when it is applied between the maximally mixed state $\hat{\rho}=\frac{1}{d^2}\id_{d^2}$, and the state of the ensemble $|\Psi_{00}\rangle\langle\Psi_{00}|$ (see Appendix~\ref{app sec general noise}). We find that the two states become correlated. Because of this, when the auxiliary state is measured, with probability $1-p$ the state of the ensemble becomes separable. This means that the fidelity of the states of the ensemble is decreased after each measurement. In general, this extra noise makes the protocol useless. If one wants to use isotropic states as auxiliary states, high-fidelity $d$-level states are required, in comparison to the states of the ensemble.

\subsubsection{Pre-purification of the auxiliary states}

In order to purify an ensemble using isotropic states obtained from copies of the same state that we want to purify, we have to pre-purify them. In Refs. \cite{jorge,bombin2005entanglement,horodecki1999reduction,alber2001efficient,vollbrecht2003efficient,cheong2007entanglement,sheng2010deterministic}, different protocols are introduced to purify bipartite systems of qudits. We just require to purify the $Z$ noise (or phase noise) of an auxiliary state to make use of it, since we can use states of the form of Eq.~\eqref{eq noisy1 aux} to implement the protocol. One also can purify the qudits systems with the generalized hashing protocol. However, in this case, since the purified states do not only contain amplitude noise, the purified qudits states have to fulfill $pF'>F$, where $p$ is the probability that the state of the ensemble is not affected by the noise, e.g., if only one auxiliary state is used in the purification $p=(\mathcal{F}d^2-1)/(d^2-1)$.

\section{Extension to multipartite systems}
\label{sec multipartite}

There exist different protocols to purify multipartite states \cite{dur2003multiparticle,kruszynska2006entanglement,goyal2006purification,glancy2006entanglement,miyake2005distillation}. In this section we extend our protocol to the multipartite case. In particular, we describe the protocol to purify GHZ states. In analogy to the bipartite case, the idea is to use multipartite maximally entangled states of qudits to transfer information about error states of a certain ensemble of multipartite qubits states. In this section, we first introduce the GHZ state and the different kinds of errors that a noisy GHZ contains. Then we introduce the entanglement purification procedure.

One of the two classes of maximally entangled states of qubits for tripartite systems is given by the GHZ states \cite{dur2000three},
\begin{equation}
    \big|\text{GHZ}\big\rangle=\frac{1}{\sqrt{2}}\Big(\,\big|000\big\rangle+\big|111\big\rangle\,\Big).
\end{equation}
Ensembles of noisy GHZ states of qubits can be purified using auxiliary GHZ states of qudits. These maximally entangled states of qudits form a basis
\begin{equation}
    \begin{aligned}
        \!\big|\Psi^{(d)}_{mnk}\big\rangle_{\!ABC} \!=\!\frac{1}{\sqrt{d}}\sum^{d-1}_{j=0}e^{i\frac{2\pi}{d}mj}\big|j\big\rangle_{\!A}\big|j\ominus n\big\rangle_{\!B}\big|j\ominus k\big\rangle_{\!C},
    \end{aligned}
\end{equation}
where $m$ is the phase index and, $n$ and $k$ are the amplitude indices. The value of the phase index or the amplitude indices can be locally obtained by measuring each qubit on the $X$ basis for the phase index and the $Z$ basis for the amplitude indices and communicating the outcomes to the other parts. Note that the value of all amplitude indices can be obtained simultaneously, but not with the phase index. When the amplitude indices are determined the information of the phase index is destroyed, and vice versa \cite{maneva2002improved}.

For noisy GHZ states, we can obtain an analogous protocol to the bipartite case. We can always depolarize the ensemble into an unknown collection of pure states where each state is either the $|\Psi_{000}\rangle$ state or an error state. However, in the tripartite case, we need to differentiate two classes of errors: separable error states and phase error states.

\subsection{Separable error states}

We call the separable error states those which can be detected with the standard generalization of the counter gate, which we defined as
\begin{equation}
    t\text{CX}_{1\to 2}^{(d)}=\text{CX}_{1\to 2}^{A_1A_2}\otimes\text{CX}_{1\rightarrow 2}^{B_1B_2}\otimes\text{CX}_{1\rightarrow 2}^{C_1C_2}.
\end{equation}
\noindent
If we apply $t$CX between an auxiliary state of qudits with zero phase index $|\Psi^{(d)}_{0ij}\rangle_{\text{aux}}$ and the computational basis state the resulting state reads
\begin{equation}
    \label{eq counter gate GHZ}
    \begin{aligned}
        t\text{CX}_{k\to \text{aux}}\big|\ell mn\big\rangle_k\big|\Psi^{(d)}_{0ij}&\big\rangle_{\text{aux}}\\ = \big|\ell mn\big\rangle_k\Big(\hat{X}^\ell\otimes\hat{X}^m\otimes\hat{X}^n\big|\Psi^{(d)}_{0ij}&\big\rangle_{\text{aux}}\Big) \\ = \big|\ell mn\big\rangle_k\big|\Psi^{(d)}_{0,i\oplus m \ominus \ell,j\oplus n \ominus \ell}&\big\rangle_{\text{aux}}.
    \end{aligned}
\end{equation}
In analogy to the bipartite case, $t$CX leaves the qubits state invariant and modifies the auxiliary state. In this case, the information of the qubits system is transferred in the two amplitude indices of the auxiliary state. Note that, from Eq.~\eqref{eq counter gate GHZ}, there exists a subspace of the qubits system spanned by $|000\rangle$ and $|111\rangle$ which remains invariant. On the other hand, each one of the remaining basis states modifies the amplitude indices in a different way. Meanwhile $t$CX allows one to detect the noise in the sup-space $\text{span}\{|x\rangle\}_{x=1}^6$, where $|x\rangle\equiv|i\rangle|j\rangle|k\rangle$ and $x=4i+2j+k$ for $i,j,k\in\mathbb{Z}_2$, the error state $|\Psi_{100}\rangle$ remains indistinguishable to the $|\Psi_{000}\rangle$ state.

\subsection{Phase error states}

We call the $|\Psi_{100}\rangle$ states \textit{phase error states}. This kind of error is not distinguishable to the $|\Psi_{000}\rangle$ states with the standard application of the counter gate, Eq.~\eqref{eq counter gate GHZ}. For that reason, to deal with them we have to use a different procedure. We apply the counter gate with an auxiliary state of $d=2$ levels, i.e., $|\Psi_{000}\rangle_{\text{aux}}$, as a control and the states of the ensemble as a target. This results in the sum of the phase bit of the auxiliary states and the phase bit of the state of the ensemble, i.e.
\begin{equation}
    \label{eq new C application}
    t\text{CX}^{(2)}_{\text{aux}\to i}|\Psi_{\!m00}\rangle_i|\Psi_{\!n00}\rangle_{\text{aux}}\!=\!|\Psi_{\!m00}\rangle_i|\Psi_{\!n\oplus m,00}\rangle_{\text{aux}}.
\end{equation}
Note that this expression, Eq.~\eqref{eq new C application}, is only fulfilled if the auxiliary state has $d=2$ levels.

This second approach of applying the counter gate $t$CX allows one to obtain one bit of information for measurement. However, when we identify the error states we do not have to discard them as they can be locally transformed to the $|\Psi_{000}\rangle$ state. In addition, we can use as auxiliary states the states of the ensemble, and because the states of the ensemble have zero amplitude bits, there is no backaction to the unmeasured states.

Combining the two different approaches of applying $t$CX, Eq.~\eqref{eq new C application} and Eq.~\eqref{eq counter gate GHZ}, we can detect any kind of error states, and hence purify any mixed state.

\subsection{Purification of general noisy GHZ states}

Consider three parties sharing an ensemble of $n$ copies of a noisy GHZ of the form
\begin{equation}
    \begin{aligned}
        \hat{\rho}_{ABC}=F\,&|\Psi_{000}\rangle\langle\Psi_{000}| \\+\,\alpha\,&|\Psi_{100}\rangle\langle\Psi_{100}|+\beta\sum_{x=1}^{6}p_{x}|x\rangle\langle x|,
    \end{aligned}
\end{equation}
where $F,\,\beta,\,p_x,\,\alpha\geq0$, $\sum_{x=1}^6p_{x}=1$ and $F+\alpha+\beta=1$. This is a completely general scenario as any noisy GHZ state can be depolarized into this form \cite{dur1999separability,dur2000classification}. In this scenario we need two purification rounds one for each kind of error. The initial expected number of amplitude and phase errors are given by $\langle k_a\rangle=n\beta$ and $\langle k_p\rangle=n\alpha$ respectively.

First, we detect the separable error states which are detectable with the standard application of $t$CX, Eq.~\eqref{eq counter gate GHZ}, up to a certain number of errors. Up to one error $\lambda=1$, by applying the counter gate once between each state of the ensemble and an auxiliary state of $d=3$ levels, and by obtaining the new amplitude indices of the auxiliary state, we can verify which kind of amplitude error the ensemble contains, or if otherwise, it does not contain any of these errors, see Eq.~\eqref{eq new C application}. Once the error is known, we can identify it with the same procedure as in the bipartite case, Sec.~\ref{sec toy model 1 error}.

Up to two errors, in a similar way to the bipartite case, one can not distinguish between all possible scenarios from the first measurement. In the tripartite case, the number of possible kinds of errors is larger than in the bipartite case. Therefore, their identification becomes more complex and requires more resources.

When we have identified all the separable error states, the remaining errors in the ensemble are phase error states, i.e., $|\Psi_{100}\rangle$. Then, we proceed with the second purification round. It consists in dividing the ensemble into several subensembles in a way that we can neglect the probability of having more than one error in a subensemble. Then, each subensemble is treated independently. In each, the counter gate is applied with each state and an auxiliary state $|\Psi_{000}\rangle_{\text{aux}}$ as in Eq.~\eqref{eq new C application}. After this operation, the new phase bit of the auxiliary state is 0 if there are no errors in the subensemble, and it is 1 if it contains one phase error. If no errors are detected the protocol is over in that subensemble. On the other hand, if one phase error is detected, the subensemble is divided again and the last step iterated until the error can be identified.

\section{Summary and conclusions}
\label{sec Conclusions}

In this paper, we have introduced a new class of entanglement
purification protocols for noisy qubit states based on the usage of auxiliary high-dimensional entangled states. While they are inspired by asymptotic hashing and breeding schemes, they are optimized for a small and moderate number of copies. The protocols work particularly well if the expected number of errors in the ensemble is small, however, we also provide schemes to deal with a large number of errors and smaller fidelities. The protocols use auxiliary entanglement to access nonlocal information about the ensemble, thereby purifying it. The essential new element as compared to previous approaches is the usage of auxiliary entangled states of high dimension, together with a so-called counter gate. This allows us to directly access the desired information about the ensemble, and thereby to construct efficient schemes to detect the number and position of errors. The protocols can run in two different modes, deterministic and probabilistic, and deal with a pre-defined expected number of errors in the ensemble. There is a trade-off between achievable fidelity and yield. We also have shown that a probabilistic multi-step procedure with an abort option at different points can provide a higher yield. Probabilistic protocols typically provide a higher fidelity state, but may also they have a higher yield in some cases. The abort option is essential in cases where too many errors are found in a first step, as detecting the position of errors would require more resources than can be gained by purifying the remaining ensemble.

The error states that we deal with are product states, which require a different depolarization algorithm as for previously known EPPs. We introduce such depolarization procedures. When dealing with amplitude damping channels -which represent the main source of noise in many relevant situations- only one kind of error state is present. In this case, our approach is simple to understand and apply, and particularly efficient. In the first step, the number of errors in the ensemble is determined by a direct application of the counter gate, and in the second step, the position of the errors is figured out. For a small number of errors, we provide optimized schemes that we construct explicitly and give general methods to deal with a larger number of errors. The situation is slightly more complicated if different kinds of errors are present, as the counter gate counts up for one kind of error, and down for the other. This implies that a no-error situation is in this way not directly distinguishable from cases where there is an equal number of both kinds of errors, and more advanced -and more costly- methods are required, which we also introduce. In this way, we can then deal with situations arising, e.g., from dephasing noise. Full rank noise requires a two-step procedure, where an intermediate basis change transfers errors from an undetectable subspace to the kind our protocol can deal with. This finally allows us to deal with any kind of noise, and purify any kind of noisy entangled states with our method, provided the initial fidelity is sufficiently large.

A direct comparison with finite-size hashing and breeding protocols, as well as with recurrence schemes, shows an improved efficiency and output fidelity of our approach for ensembles of finite size. In particular, for Bell-diagonal rank-2 states, our EIP$_{\text{damp}}$ is the only protocol that provides fidelity one states with a positive yield. For general noisy ensembles of small size, EIPs keeps the higher fidelity-yield trade off as the standard hashing protocol requires an asymptotically large number of copies to justify its applications, and even recurrence protocols are more suitable for such ensembles, they only provide a exponential reduced fraction of the original ensemble. We have also demonstrated that the scheme works in presence of noise and imperfections, in particular, if auxiliary entangled states are noisy, and local operations are imperfect. For dephasing noise, the required auxiliary states can in fact be obtained directly from the noisy ensemble, while for general noise pre-purification of the auxiliary states using other methods is required. Given the fact that entanglement is a crucial resource for many applications in quantum technologies, in particular for long-distance quantum communication and in quantum networks, our protocols should provide a valuable tool to optimize such communication schemes and develop methods for efficient, long-distance quantum communication. However, entanglement is also a crucial resource in distributed quantum metrology, where in particular GHZ states are an important resource. The generalization of our methods to directly purify multipartite GHZ states opens new possibilities in this respect.

Finally, we believe that there is still open the possibility of optimizing our EIPs for ensembles or large number of copies. Even though we provide different ways to purify them, we think that with a different approach of the protocol, a more efficient procedure based on the usage of high dimensional systems can be developed. In addition, the idea of directly accessing nonlocal information of an ensemble using high-dimensional auxiliary entangled states can be useful for other types of problems. For instance, state verification or tomography protocols based on such methods can be designed, which may be more efficient than known protocols that just rely on local measurements of several copies.
\\
\section*{Acknowledgements}

This work was supported by the Austrian Science Fund (FWF) through project No. P30937-N27 and the Swiss National Science Foundation (SNSF) and the NCCR Quantum Science and Technology, through Grant No. PP00P2-179109 in particular.

We thank Julius Walln\"ofer for interesting discussions.

\bibliographystyle{apsrev4-1}
\bibliography{Entanglement_purification_by_counting_and_locating_errors_with_entangling_measurements.bib}
\clearpage
\onecolumngrid
\appendix
\section{Fidelities tradeoff}
\label{app fidelities}

\begin{figure}
    \centering
    \includegraphics[scale=0.2]{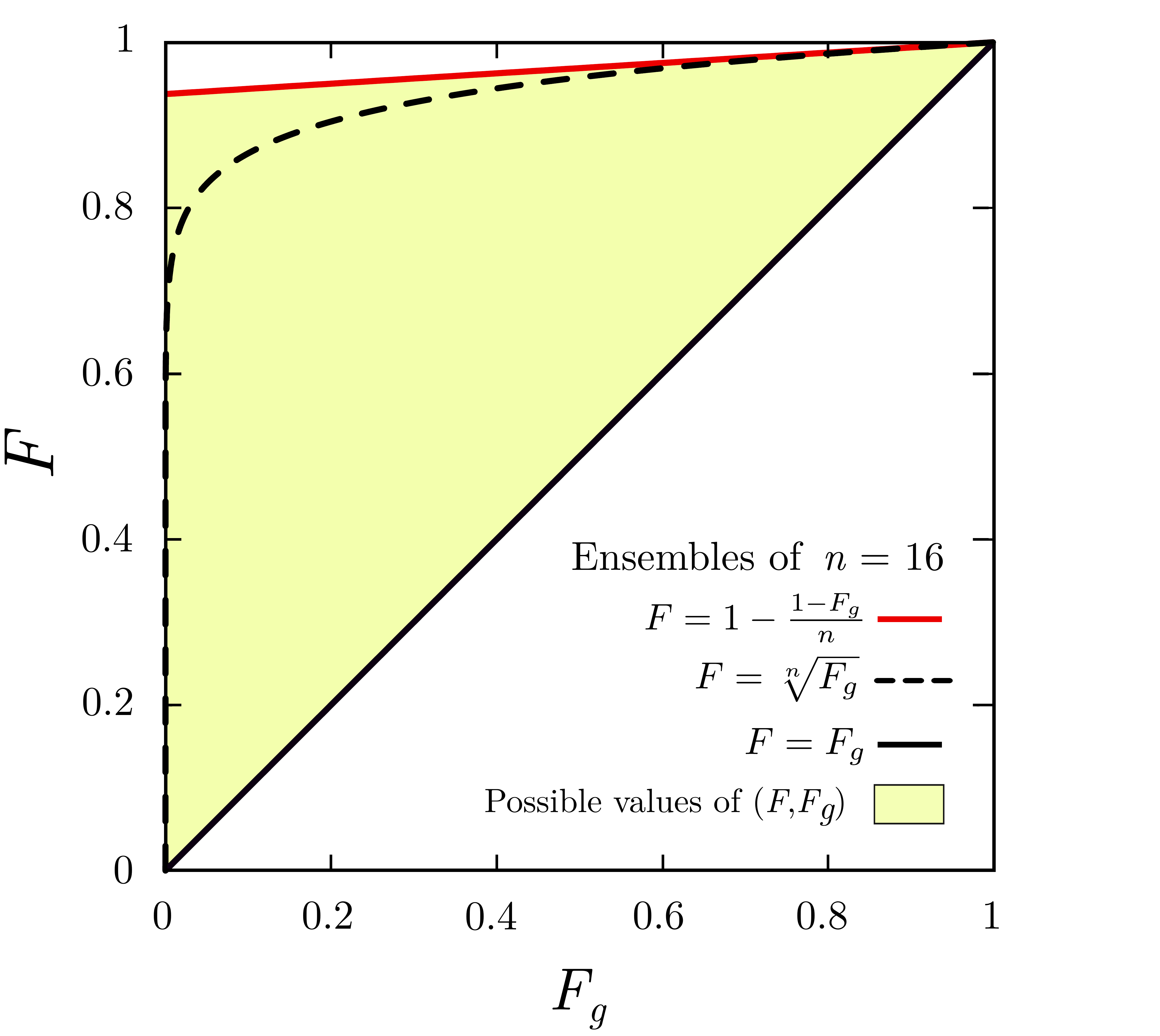}
    \caption{The range of allowed fidelities $(F_g,F)$ for 16 qubit pairs. The top line is attained by an ensemble where the errors anti-bunch, there is at most one error state in the ensemble. The lower limit is attained by states where the errors bunch, either all the qubit pair are in the target state or all give an error. The dashed line gives the relation for product states $F_g = F^{16}$, the errors on each qubit pair are uncorrelated and identically distributed.}
     \label{fig:FgFl}
\end{figure}

In this appendix, we establish the relation between global and local fidelity as quantifiers of the purified ensemble. Precisely, we discuss the range of the possible values $(F_g,F)$ in the $[0,1]^{\times 2}$ square, and what can one learn about the error distribution in the ensemble by looking at the two numbers.
\\ \\
For a given state $\hat{\rho}$ of the $n$ qubit pairs ensemble, the global and (average) local fidelities are defined as
\begin{equation}
    \begin{split}
        F_g & = \text{tr} \left(\hat{\rho} \, \projf{\Psi_{00}}^{\otimes n}\right) , \\
        F & = \text{tr} \left(\hat{\rho} \, \frac{1}{n}\sum_{k = 1}^n \projf{\Psi_{00}}_k \otimes \id_{\neg k}\right),
    \end{split}
\end{equation}
and we also name the Hermitian operators $O_g =\proj{\Psi_{00}}^{\otimes n}$ and $O_\ell =\frac{1}{n}\sum_{i=1}^n |\Psi_{00}\rangle\langle\Psi_{00}|_i\otimes \id_{\neg i}$. Some remarks are in order to facilitate the argumentation. Since both operators are diagonal in the joint Bell basis (in fact joint basis which has $\{\Psi_{00}\}$ as an element of the local basis), it is only the weight of the state in this basis that matters for the two quantities. Thus without loss of generality we can consider states of the form
\begin{equation}
    \label{eq: rho BD}
    \hat{\rho} = \sum_{\boldsymbol{i,j}} p_{\boldsymbol{i j}} \projf{\Psi_{\boldsymbol{ij}}},\quad  \text{with} \quad  \big|\Psi_{\boldsymbol{ij}}\big\rangle = \bigotimes_{k=1}^n \big|\Psi_{i_kj_k}\big\rangle.
\end{equation}
We start with the simple question, given a fixed local fidelity $F$ what is the maximal possible value of the global fidelity. The answer is straightforward, it is $F_g=F$, attained by a state where the errors bunch perfectly. That is a mixture
\begin{equation}
    \hat{\varrho}_{[F_g=F]} = F_g \projf{\Psi_{00}}^{\otimes n} + (1-F_g) \hat{\varrho}_\perp
\end{equation}
of either all qubit pairs in the good states or all qubit pairs in some error states (e.g., $\varrho_\perp = \proj{\Psi_{11}}^{\otimes n}$). Trivially, any higher global fidelity is impossible for a given $F$. We can conclude that $F_g\leq F$.
\\ \\
Next, consider the converse question: what is the minimal global fidelity given a fixed local one $F$. It is less trivial. Yet it is very intuitive to find a state which gives zero global fidelity while keeping the local one high. It is given by any mixture of states where $n-1$ qubit pairs are in the target state but one pair gives an error
\begin{equation}\label{eq:one error}
    \hat{\rho}_{1 \textrm{err}} = \frac{1}{\sum p_\pi} \sum_{\pi} p_\pi\,\pi\Big[ \projf{\Psi_{00}}^{\otimes(n-1)}\otimes \projf{\Psi_\perp}\Big].
\end{equation}
Any such state satisfies $(F_g, F) = (0, 1-\frac{1}{n})$, this allows us to conclude that for $F\leq 1- \frac{1}{n}$ the global fidelity can be zero.
\\ \\
It remains to find the minimal $F_g$ for $F \in \left(1-\frac{1}{n},1\right)$. To do so, consider the contribution of each component $\ket{\Psi_{\boldsymbol{ij}}}$ of the state $\hat{\rho}$ in Eq.~\eqref{eq: rho BD} to the fidelities
\begin{equation}
    \text{tr}\, \Big( O_g \projf{\Psi_{\boldsymbol{ij}}}\Big) = 
    \begin{cases}
        1 & \boldsymbol{ij} = \boldsymbol{00}\\
        0 & \textrm{otherwise}
    \end{cases}
\end{equation}
and
\begin{equation}
    \text{tr}\,\Big(O_\ell \projf{\Psi_{\boldsymbol{ij}}}\Big) = \frac{n_{00}(\boldsymbol{ij})}{n},
\end{equation}
 where $n_{00}(\boldsymbol{ij})$ is the number of $00$ pairs in the bitstrings $\boldsymbol{ij}$. So that
\begin{equation}
    \label{eq: Fids from p}
    F_g = p_{\boldsymbol{00}} \qquad \text{and} \qquad F = \sum_{\boldsymbol{ij}} p_{\boldsymbol{ij}} \frac{n_{00}(\boldsymbol{ij})}{n}.
\end{equation}
Now, we can easily find the states maximizing the local fidelity for a fixed $F_g$. Given $p_{\boldsymbol{00}} = F_g $, the remaining probability $P =1-F_g$ has to be distributed on the state $\ket{\Psi_{\boldsymbol{ij \neq 00}}}$ is such a way as to maximize $F$. This is done by populating the states with only one error, as follows from Eq.~\eqref{eq: Fids from p}. We can now construct the states which maximize $F$ for a fixed $F_g$,
\begin{equation}
    \hat{\rho}_{[F_g\ll F]} = F_g \projf{\Psi_{00}}^{\otimes n} + (1-F_g) \hat{\rho}_{1\text{err}}
\end{equation}
with any ``one error'' state in Eq.~\eqref{eq:one error}. By convexity of quantum states, it also implies the states that minimize $F_g$ for a fixed $F$. The states $\hat{\rho}_{g\ll \ell}$ give a local fidelity of $F = 1- \frac{1}{n}(1-F_g)$, and correspond to the case where the errors maximally anti-bunch, each term in the mixture has one error at most.
\\ \\
In summary, we have identified the region of allowed values $(F_g,F)$, as can be summarized by tight bounds
\begin{equation}
    \boxed{F_g\leq F\leq 1-\frac{1}{n}(1-F_g)}.
\end{equation}
It is interesting to compare the bounds with the relation $F_g = F^{n}$ valid for product states. In plot of the three curves is given in Fig.~\ref{fig:FgFl} for the case $n=16$.
 
\section{Entanglement cost of dichotomic measurements of a Bell state ensemble}
\label{app sec dichotomic measurement}

In this section, we discuss the entanglement cost of elementary measurements on the ensemble on $n$ two-qubit states. Precisely, we consider dichotomic measurements, specified by two completely positive maps $\{\mathcal{E}_+, \mathcal{E}_- \}$ that describe the evolution of the state of the system conditional to the observation of one of the two possible outcomes $+$ or $-$.
\\ \\
Among all possible dichotomic measurements, we consider those that can extract information about the Bell-state configuration of the ensemble. To formalize this, we require that the measurement allows to perfectly distinguish between at least two different sequences of Bell pairs
\begin{equation}
    \begin{aligned}
        \big|\Psi_{\boldsymbol{ij}}\big\rangle &= \bigotimes_{k=1}^n \big|\Psi_{i_kj_k}\big\rangle\\
        \big|\Psi_{\boldsymbol{i'j'}}\big\rangle& = \bigotimes_{k=1}^n \big|\Psi_{i_k'j_k'}\big\rangle,
    \end{aligned}
\end{equation}
without disturbing them. In particular, if the input of the measurement is a mixture of the two sequences only, such a measurement will determine which sequence describes the state of the ensemble without disturbing it. Formally, we then write
\begin{equation}
    \begin{split}
        \mathcal{E}_+ :&\; \big|\Psi_{\boldsymbol{ij}}\big\rangle \mapsto \big|\Psi_{\boldsymbol{ij}}\big\rangle \\
        &\;\big|\Psi_{\boldsymbol{i'j'}}\big\rangle \mapsto 0 \\[10pt]
        \mathcal{E}_- :&\; \big|\Psi_{\boldsymbol{ij}}\big\rangle \mapsto 0\\
        &\;\big|\Psi_{\boldsymbol{i'j'}}\big\rangle \mapsto \big|\Psi_{\boldsymbol{i'j'}}\big\rangle,
    \end{split}
\end{equation}
and note that the positivity of the map also implies that any coherence between the states is destroyed $\mathcal{E}_\pm:|\Psi_{\boldsymbol{i'j'}}\rangle\langle\Psi_{\boldsymbol{ij}}|\mapsto 0$. We will assume that the sequences are different for the first $m$ qubit pairs and identical for the remaining pairs  (otherwise we rearrange the pairs accordingly). Furthermore without loss of generality, we can assume that on the nonidentical pairs the sequences are $(i_k,j_k)=(0,0)$ and $(i_k',j_k')= (0,1)$ for $k=1$ to $m$, such that the two states are
\begin{equation}
    \begin{split}
        \big|\Psi_{\boldsymbol{ij}}\big\rangle &= \big|\Psi_{00}\big\rangle^{\otimes m}\otimes \big|\xi\big\rangle, \\
        \big|\Psi_{\boldsymbol{i'j'}}\big\rangle&= \big|\Psi_{01}\big\rangle^{\otimes m}\otimes \big|\xi\big\rangle.
    \end{split}
\end{equation}
Here $\ket{\xi} = \bigotimes_{k=m+1}^n \ket{\Psi_{i_k,j_k}}$
specifies the irrelevant states of the identical pairs.
\\ \\
Now consider the following initial state of the ensemble plus an auxiliary pair of qubits:
\begin{equation}
    \begin{split}
        \big|\Theta_0\big\rangle &= \frac{1}{\sqrt 2} \Big( \big|\Psi_{00}\big\rangle\otimes\big|\Psi_{\boldsymbol{ij}}\big\rangle + \big|\Psi_{01}\big\rangle\otimes\ket{\Psi_{\boldsymbol{i'j'}}}\Big)\\
        & =  \frac{1}{\sqrt 2} \left(\big|\Psi_{00}\big\rangle\otimes \big|\Psi_{00}\big\rangle^{\otimes m}+ \big|\Psi_{01}\big\rangle\otimes \big|\Psi_{01}\big\rangle^{\otimes m}\right) \otimes \big|\xi\big\rangle.
    \end{split}
\end{equation}
By assumption, for the input state 
$\ket{\Theta_0}$ both outcomes are equiprobably $p_+=p_-=\frac{1}{2}$, and the two final states read
\begin{equation}
    \begin{split}
        \mathcal{E}_+: \big|\Theta_0\big\rangle&\mapsto \big|\Theta_+\big\rangle= \big|\Psi_{00}\big\rangle^{\otimes m+1} \otimes \big|\xi\big\rangle,\\[2pt]
        \mathcal{E}_-: \big|\Theta_0\big\rangle&\mapsto \big|\Theta_-\big\rangle=\big|\Psi_{01}\big\rangle^{\otimes m+1} \otimes \big|\xi\big\rangle.
    \end{split}
\end{equation}
Let us now compare the entanglement of the three states $\ket{\Theta_0},\ket{\Theta_+}$ and $\ket{\Theta_-}$. As the last $n-m$ pairs remains in the state $\ket{\xi}$ product to the rest, we only consider the partial states of the other $m+1$ pairs, i.e $\ket{\Theta_0}\mapsto \frac{1}{\sqrt 2} \big( \ket{\Psi_{00}}^{\otimes m+1} + \ket{\Psi_{01}}^{\otimes m+1}\big)$, $\ket{\Theta_+}\mapsto\ket{\Psi_{00}}^{\otimes m+1}$ and $\ket{\Theta_-}\mapsto\ket{\Psi_{01}}^{\otimes m+1}$. The two possible final states are LU equivalent
\begin{equation}
    \big|\Psi_{00}\big\rangle^{\otimes m+1} \simeq 
    \big|\Psi_{01}\big\rangle^{\otimes m+1},
\end{equation}
and contain $m+1$ Bell pairs.
To find the Schmidt vector for the initial state we use
\begin{equation}
    \begin{split}
        \big|\Psi_{00}\big\rangle^{\otimes m+1} &= \frac{1}{2^{(m+1)/2}} \sum_{\boldsymbol{k}} \big|\boldsymbol{k}\big\rangle\big|\boldsymbol{k}\big\rangle,\\
        \big|\Psi_{01}\big\rangle^{\otimes m+1} &= \frac{1}{2^{(m+1)/2}} \sum_{\boldsymbol{k}} (-1)^{|\boldsymbol{k}|}\big|\boldsymbol{k}\big\rangle\big|\boldsymbol{k}\big\rangle,
    \end{split}
\end{equation}
where $\boldsymbol{k}$ runs through all bitstrings of length $m+1$ and $|\boldsymbol{k}|$ is the number of ones in the bitstring. Then, by computing the Schmidt decomposition of the initial state explicitly
\begin{equation}
    \begin{split}
        \frac{1}{\sqrt 2} &\left( \big|\Psi_{00}\big\rangle^{\otimes m+1} + \big|\Psi_{01}\big\rangle^{\otimes m+1}\right)
        = \frac{1}{2^{(m+2)/2}} \sum_{\boldsymbol{k}} \left(1+(-1)^{|\boldsymbol{k}|}\right)\big|\boldsymbol{k}\big\rangle\big|\boldsymbol{k}\big\rangle
        = \frac{1}{2^{m/2}} \sum_{\boldsymbol{k} \, \text{even}} \big|\boldsymbol{k}\big\rangle\big|\boldsymbol{k}\big\rangle  \simeq \big|\Psi_{00}\big\rangle^{\otimes m},
    \end{split}
\end{equation}
we find that it is LU equivalent to only $m$ Bell states. We have just shown that any measurement satisfying our requirement has created an additional Bell pair for both measurement outcomes. Obviously, this is not possible with only LOCC, and requires consuming at least one auxiliary ebit (Bell pair).
\\ \\
It is worth mentioning that for the example we just presented, it is possible to distinguish between any two sequences $\ket{\Psi_{\boldsymbol{ij}}}$ and $|\Psi_{\boldsymbol{i'j'}}\rangle$ by consuming one ebit. This is because without loss of generality we can assume that for the sequences the states of the first pair of qubits are different and given by $\ket{\Psi_{00}}$ and $\ket{\Psi_{01}}$. The two can be distinguished by acting with a bilateral CNOT and an auxiliary Bell pair.
\\ \\
The next natural question is whether any dichotomic measurement on the ensemble can be implemented by consuming at most one ebit on average. The answer to this question is no. The simplest counter-example is given by the following measurement for one pair of qubits
\begin{equation}\label{eq: mes counter}
    \begin{split}
        \mathcal{E}_+ :\bullet &\mapsto \Pi_+\bullet \Pi_+,\qquad \Pi_+ =\big|\Psi_{00}\big\rangle\big\langle\Psi_{00}\big|,\\[4pt]
        \mathcal{E}_-: \bullet &\mapsto \Pi_-\bullet \Pi_-,\qquad \Pi_-  =\big|\Psi_{01}\big\rangle\big\langle\Psi_{01}\big| + \big|\Psi_{10}\big\rangle\big\langle\Psi_{10}\big|+\big|\Psi_{11}\big\rangle\big\langle\Psi_{11}\big|.
    \end{split}
\end{equation}
This is easy to see by applying it on the qubit pair $AB$ of the state $\ket{\Psi_{00}}_{AA'}|\Psi_{00}\rangle_{BB'}$, product with respect to the $AA'|BB'$ partition, and computing the Schmidt decomposition of the two final states. Precisely, the first outcome happens with probability $\frac{1}{4}$ and generates two Bell pairs between $AA'$ and  $BB'$. The second outcome happens with probability $\frac{3}{4}$ and generates a state LU equivalent to
\begin{equation}
    \big|\theta_-\big\rangle= \frac{\sqrt{3}}{2}\big|0|0\big\rangle + \frac{1}{2 \sqrt{3}} \big|1|1\big\rangle+ \frac{1}{2 \sqrt{3}} \big|2|2\big\rangle+ \frac{1}{2 \sqrt{3}} \big|3|3\big\rangle.
\end{equation}
In particular, the entropy of entanglement of this state $S(\text{tr}_{BB'} \proj{\theta_-})\approx 1.21$  exceeds the entanglement entropy of a Bell pair (equal to $1$). Thus, for both branches the entanglement entropy of the final states exceeds $1$ ebit. Hence, it is impossible to transform a single Bell pair to the post-measurement state  $\rho=\frac{1}{4} \proj{\Psi_{00}}^{\otimes 2}\oplus \frac{3}{4}\proj{\theta_-}$  using LOCC~\cite{horodecki2009quantum}. We conclude that the measurement in Eq.~\eqref{eq: mes counter} can not be implemented with only one auxiliary Bell pair.  


\section{Protocol for a general number of errors for amplitude damping channel}
\label{app:sec:amp:general}

In the following we minimize the dimensions $d_{i}$ of the auxiliary states $|\Psi^{(d_{i})}_{00}\rangle_{\text{aux}}$ necessary for step \ref{enum:toy:general:2} of the EIP$_{\text{damp}}$ of Sec.~\ref{sec:toy:general:protocol}, compute the number of resources $R_F(k)$ for identifying the error configuration $(k_1,\ldots,k_a)$ for $k$ errors, and the number of resources $R_L(k)$ to locate them.
\\ \\
For that purpose, recall that the number of errors $k_i$ for $1\leq i\leq a$ distributed over $a$ blocks of size $n/a$ each follows a hypergeometric distribution
\begin{equation}
    k_i\sim \mathrm{Hyp}\left(n-\frac{(i-1)n}{a},k-\sum\limits^{i-1}_{j=1}k_j, \frac{n}{a}\right).
\end{equation}
Observe that for the first block the number of errors is hypergeometrically distributed with $k_1 \sim \mathrm{Hyp}\left(n,k,\frac{n}{a}\right)$. Therefore, we can find at most $\min(k,n/a)$ errors in the first block. In other words, the most errors we can find is either $k$ (all errors in the first block) or the block size $n/a$. For the second block, given that we have found $k_1$ errors in the first block, we can find at most $\min(k-k_1,\, n/a)$. Hence, for block $i$ the number of errors we can find at most is given by $\min(k-\sum^{i-1}_{j=1}k_j,\, n/a)$, which implies for the dimension $d_i$ of the auxiliary state that
\begin{equation}
    d_i=\log_2 \!\left[\min\left(k-\sum^{i-1}_{j=1}k_j, \frac{n}{a} \right)\right].
\end{equation}
Next we investigate the number of resources which we need to identify an error distribution $(k_1,\dots,k_a)$ over all $a$ blocks for a given value of $k$, i.e., $R_F(k)$. Observe that there are in principle many configurations of errors $(k_1,\dots,k_a)$ which give rise to $k$ errors in total, i.e., such that $\sum_i k_i = k$. The total number of resources for a given value of $k$ such that the protocol is able to distinguish all possible configurations $(k_1,\dots,k_a)$ is therefore
\begin{equation} \label{eq:toy:general:resources:find}
    R_{F}(k)=\log_2\left(\min(k,n/a)\right)+\sum_{\substack{k_1,\ldots,k_a \\k_1+\ldots+k_a=k}}\sum\limits^{a-1}_{i=2} \left(\prod\limits^{i-1}_{j=1} P(k_j)\right)\log_2\! \left[\min\left(k-\sum^{i-1}_{j=1} k_j, \frac{n}{a} \right) \right].
\end{equation}
The first term corresponds to the resources necessary to identify $k_1$. Observe that in order to determine $k_1$ we always have to use $\log_2(\min(k, n/a))$. The first sum takes into account all different error configurations of $(k_1, \ldots, k_a)$ which give rise to $k$ errors in total, and the second sum iterates over the individual blocks $i$ we have split the ensemble into. The second sum starts with $i=2$ since the first block is considered by the term $\log_2(\min(k, n/a))$. Further, the second sum ends with $a-1$, because $k_a = k - \sum^{a-1}_{i = 1} k_i$. Observe that the probability of finding the error configuration $(k_1, \ldots, k_{i})$ is given by the product over the individual probabilities of the random variables $k_i$.
\\ \\
Now we investigate the number of resources necessary to locate the errors for a given error configuration $(k_1, \ldots, k_a)$, i.e., $R_{L}(k)$. As outlined in the main text, we pursue two strategies to locate them.
\begin{itemize}
 	\item $k_i \leq 2$:In this case, the resources for locating two or less errors have to be used, see Sec.~\ref{sec toy model 1 error} (one error) and Sec.~\ref{sec toy model 2 errors} (two errors).
 	\item $k_i > 2$: In this case, the $i-$th block is again divided into $a$ blocks, and the procedure is applied recursively to block $i$, until only one or two errors are left.
\end{itemize}
By denoting the number of resources for both cases by $R_{L,i}(k_i)$ we find that the number of resources for locating $k$ errors is given by
\begin{equation}
    R_{L}(k) = \sum_{\substack{k_1, \ldots, k_a \\ k_1 + \cdots + k_a = k}} P(k_1) \cdots P(k_a) \left(\sum\limits^{a}_{i = 1} R_{L,i}(k_i) \right).
\end{equation}

\section{upper bound for the standard finite hashing protocol}
\label{app sec upper bound hash}

\begin{figure}
    \centering
    \subfloat[\centering]{\includegraphics[width=0.43\columnwidth]{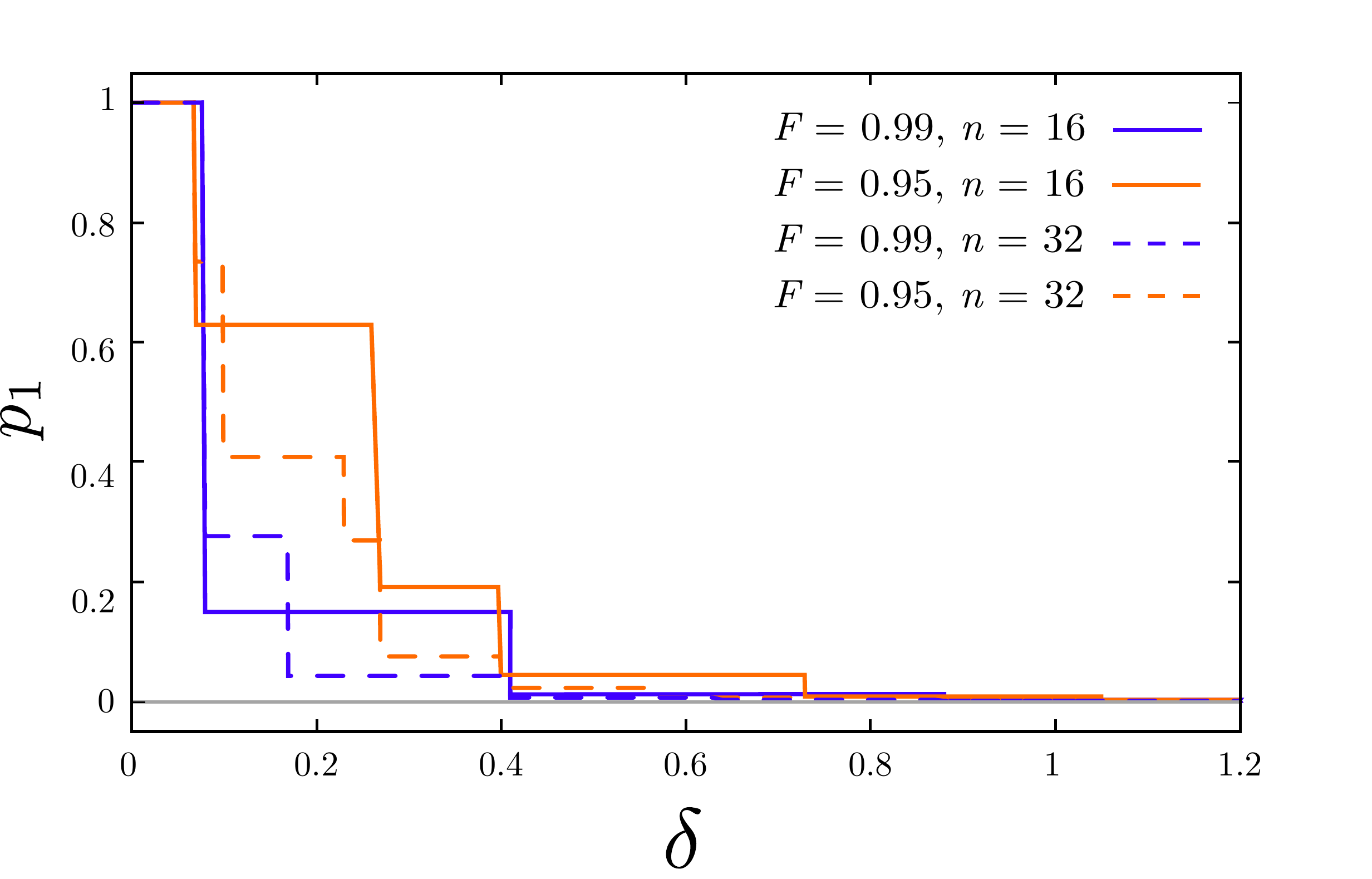} \label{app fig fgvsd}} \subfloat[\centering]{\includegraphics[width=0.43\columnwidth]{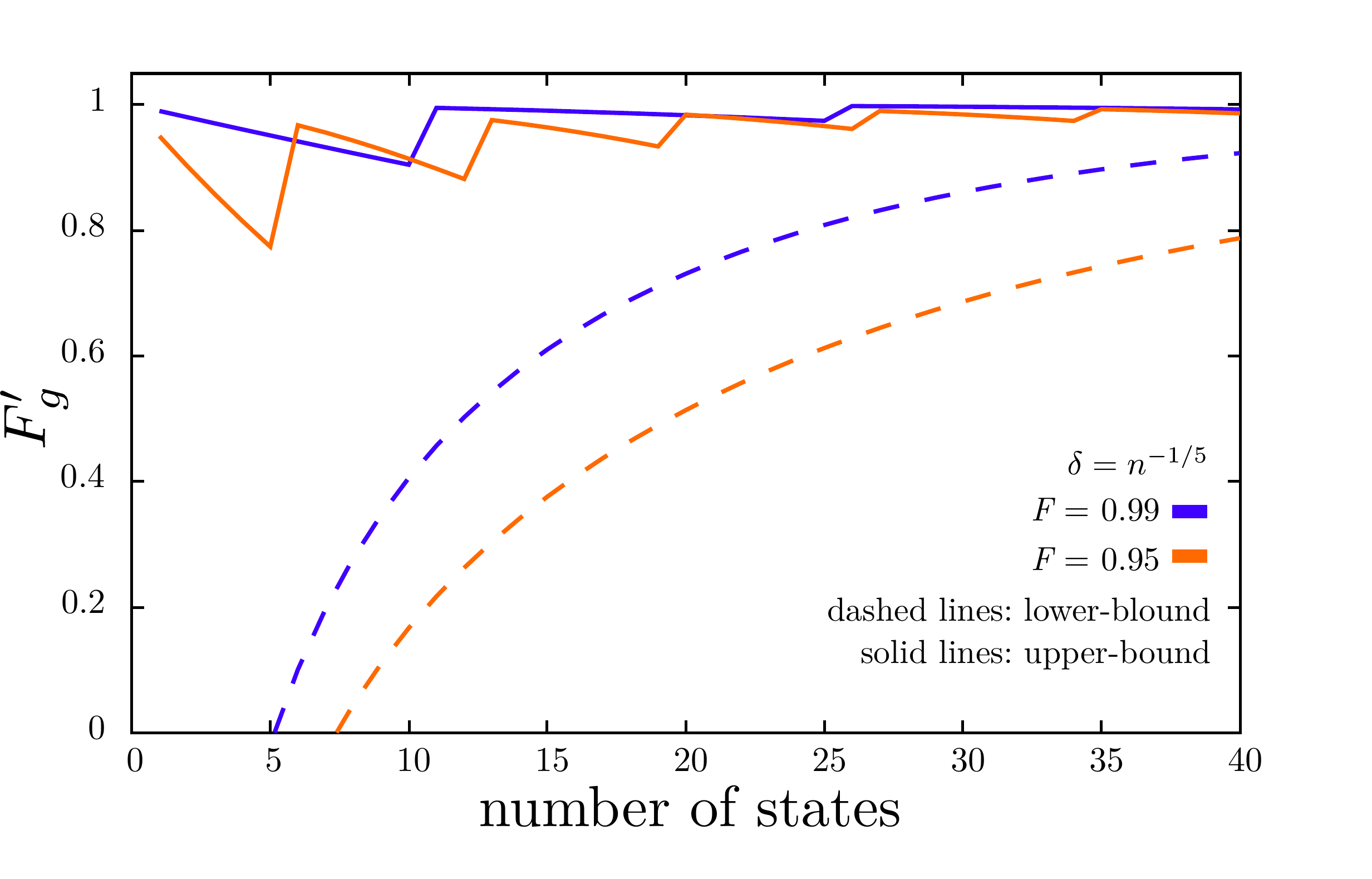} \label{app fig upper fg}} 
    \caption{In (a), we plot $p_1$ as a function of $\delta$ for an ensemble of copies of state Eq.~\eqref{eq 2 erro}. Each color corresponds to a different initial fidelity. Solid lines correspond to an initial ensemble of $n=16$ states and dashed lines to ensembles with $n=32$. In (b), we plot upper and lower bounds for the global fidelity of the purified ensemble of $n$ copies of state Eq.~\eqref{eq 2 erro} with the hashing protocol as a function of the number of states, for $\delta = n^{-1/5}$. Each color corresponds to a different value of the initial fidelity. Solid lines correspond to the upper bound, Eq.~\eqref{eq upperbound}, and dashed lines to the lower bound derived in \cite{zwerger2014robustness}.}
\end{figure}

The hashing protocol is designed to work in the asymptotic limit, i.e., when the number of copies approaches $n\to\infty$. In this limit, the protocol provides pure states with $Y=n(1-S)$, where $S$ denotes the von Neumann entropy of the initial Bell pairs. In any real scenario with a finite $n$ the yield and the output fidelity are reduced. In this situation, the computation of the exact output fidelity is complex. For that reason in order to compare the finite hashing protocol with the EIP's, we derive an upper bound to the output fidelity.
\\ \\
The output global fidelity $F_g'$ of the purified ensemble via the hashing protocol is given by
\begin{equation}
    F_g'=(1-p_1)(1-p_2),
\end{equation}
where $p_1$ is the probability that the classical bit-string describing the configuration of the Bell pairs of the initial ensemble falls outside of the likely subspace $\mathcal{L}$, and $p_2$ is the probability that two strings $\boldsymbol{s}$ and $\boldsymbol{s}'$ remain distinct while having agreed on all $r$ subset parity measurements. To obtain an upper bound of $F'_g$ we assume $p_2=0$ and we compute the exact value of $p_1$.
\\ \\
The probability $p_1$ is the sum of the probability of all strings outside of the likely subset, i.e.
\begin{equation}
    p_1 =\sum_{\boldsymbol{s}\notin \mathcal{L}} \text{Prob}(\boldsymbol{s}).
\end{equation}
The strings that belong to the likely set, $\boldsymbol{s}\in\mathcal{L}$, are those which probability fulfills
\begin{equation}
    2^{-n(S(F)+\delta)}\leq \text{Prob}(\boldsymbol{s}) \leq 2^{-n(S(F)-\delta)},
\end{equation}
where $\delta\geq 0$ determine the size of the likely set. Therefore for an ensemble of $n$ copies of a Bell-diagonal rank-3 state of the form
\begin{equation}
    \label{eq 2 erro}
    \hat{\rho}=F\big|\Psi_{00}\big\rangle\big\langle\Psi_{00}\big|+\frac{1-F}{2}\Big(\big|\Psi_{01}\big\rangle\big\langle\Psi_{01}\big|+\big|\Psi_{11}\big\rangle\big\langle\Psi_{11}\big|\Big),    
\end{equation}
the probability of a random string falling outside of the likely set is given by
\begin{equation}
    \label{eq upperbound}
    p_1=\sum_{\substack{i,j,k=0\\i+j+k=n}}^n\frac{n!}{i!j!k!}F^{i}\left(\frac{1-F}{2}\right)^{j+k}g_{ijk},
\end{equation}
where
\begin{equation}
    g_{ijk} = 
    \begin{cases}
        0 & \text{ if }\quad 2^{-n(S(F)+\delta)}\leq F^{i}\left(\frac{1-F}{2}\right)^{j+k} \leq2^{-n(S(F)-\delta)} \\ 
        1 & \text{ otherwise }.
    \end{cases}
\end{equation}
In Fig.~\ref{app fig fgvsd}, we plot the probability of a random configuration falling inside the likely sequence, i.e., $1-p_1$, as a function of the value of $\delta$. Observe that this probability increases in discrete jumps with $\delta$, as the number of different probabilities, Prob$(\boldsymbol{s})$, is discrete. In Fig.~\ref{app fig upper fg}, we plot the upper bound and the lower bound derived in \cite{zwerger2018long} of $F'_g$, for a fixed value of $\delta=n^{-1/5}$. Note that, for small values of $n$ the two bounds differ considerably, but for larger values, both approach the same value.
\\ \\
The yield of the hashing protocol is just given by $Y=(n-r)/n$, where $r$ is the number of measurements that are performed. In order to make $p_2$ to approach 0, the number of measurements is chosen to $r=n(S(F)+2\delta)$. Then the yield is given by
\begin{equation}
    Y=1-S(F)-2\delta.
\end{equation}
Note that the output fidelity increases with $\delta$, On the other hand, the yield decreases with $\delta$ as more measurements are needed to distinguish a string in the likely set.

\section{Identification of two identical errors}
\label{sec app ident 2 identical errors}

In this section, we introduce the detailed procedure to identify two identical 01 errors at unknown positions in an ensemble of $n$ states, where the remaining $n-2$ states are $|\Psi_{00}\rangle$. The idea of the procedure is to isolate each error state in a subensemble. Then the errors can be identified with the same procedure as in the one error situation, Sec.~\ref{sec toy model 1 error}. To isolate the errors we need information about their positions. In particular, we transfer the sum of their positions into an auxiliary state. Since the positions range from 1 to $n$, there are $2n-3$ possible values. Therefore, we need an auxiliary state with $d=2n-3$ in order to distinguish between all possibilities. To encode the sum of the position to the auxiliary state, we apply the EPG, i.e.
\begin{equation}
    \label{eq C gate position 2}
    \prod_{i=1}^n\,(b\text{CX}_{i\to \text{aux}})^{i}\big|\Phi_{\boldsymbol{\mu}}\big\rangle\big|\Psi^{(2n-3)}_{00}\big\rangle_{\text{aux}} = \big|\Phi_{\boldsymbol{\mu}}\big\rangle\big|\Psi^{(2n-3)}_{0j}\big\rangle_{\text{aux}}.
\end{equation}
The auxiliary state is only modified when the counter gate $b$CX is applied with an error state. Therefore, from Eq.~\eqref{eq action bCX}, if the errors are 01 type, the new amplitude index is given by $j=(r+t)\text{mod}(2n-3)$ where $r$ and $t$ are the position the errors, i.e., $|\phi_{\mu_r}\rangle=|\phi_{\mu_t}\rangle=|01\rangle$. The value of $j$ is the sum of the position of the two errors modulo $2n-3$ from what we can directly obtain the sum of positions $j'=r+t$. From $j'$ we can find two subensembles where there is one error in each.
\\ \\
For example: if we obtain $j'=7$, there are three possible pairs of states: (1+6), (2+5), and (4+3). Therefore, in this example, one error must be in position 1, 2, or 3 and the other in 5, 4, or 6. 
\\ \\
Given a sum of positions $j'$, the number of state in each subensemble, $n'$, is given by
\begin{equation}
    \label{eq size of subsets}
    n'=
    \begin{cases}
        \left\lfloor\frac{j'-1}{2}\right\rfloor & \text{ if }\quad 3\leq j'\leq n+1 \\[8pt] \left\lceil\frac{2n-j'}{2}\right\rceil & \text{ if }\quad n+1< j'\leq 2n-1
\end{cases}
\end{equation}
Note that if $j'=3$ or $j'=2n-1$, the two errors are already identified as the subensembles are just one state.
\\ \\
As the sum of positions of the states is known, when one state is identified, the other is directly determined. Therefore, one only needs to apply the one error procedure in one subensemble. As the size of the two subensembles is the same, no matter which we choose.
\\ \\
Identify the two errors corresponds to distinguish between $\frac{1}{2}n(n-1)$ possible distributions. Therefore, the minimum amount of required information to identify two identical errors is $\log_{2}\left(\frac{1}{2}n(n-1)\right)$ bits. The procedure is not optimal as more entanglement is used. Specifically, the expected value of used resources is given by
\begin{equation}
    R(2)=\log_2\left(2n-3\right)+R_j,
\end{equation}
where $R_j$ are the resources required to locate one error once the sum of the positions $j$ is obtained--the position of the other error is directly inferred once the first error is located. Therefore, $R_j=\langle\,\log_2(n')\rangle$, and it depends on the position of the errors as it is the size of the subsets where we isolate each error. The expected value of $\langle \log_2\left(n'\right)\rangle$ is given by
\begin{equation}
    R_j=\left\langle\, \log_2\left(n'\right)\,\right\rangle=2\sum_{j'=3}^{n}\text{P}(j')\log_2\left(\left\lfloor \frac{j'-1}{2}\right\rfloor\right)+\text{P}(n+1)\log_2\left(\left\lfloor \frac{n}{2}\right\rfloor\right).
\end{equation}
We average over all possible values of $j'$. With Eq.~\eqref{eq size of subsets}, we relate the number of states in each subensemble with the obtained value of $j'$. To obtain the expected value we weigh each term with the probability of obtaining $j'$. Since all configurations are equally probable, the probability of obtaining $j'$ is given by the number of configurations with two errors that the sum of the two positions is $j'$, i.e.
\begin{equation}
    \begin{aligned}
        \text{P}(j')&= \frac{2}{n(n-1)} \sum_{l_1,\dots,l_n=-1}^1 \delta_{\sum_i l_i,2}\,\delta_{\sum_k |l_k|,2}\,\delta_{\sum_s s\cdot l_s\text{mod}(2n-3),j'} \\&= 
        \begin{cases}
            \frac{2}{n(n-1)}\left\lfloor\frac{j'-1}{2}\right\rfloor &\text{ if}\quad 3\leq j'\leq n+1 \\[8pt] \frac{2}{n(n - 1)}\left\lceil\frac{2n - j'}{2}\right\rceil &\text{ if}\quad n+1<j'\leq 2n - 1
        \end{cases}
    \end{aligned}
\end{equation}
\begin{figure}
    \centering
    \subfloat[\centering]{\includegraphics[width=0.4\columnwidth]{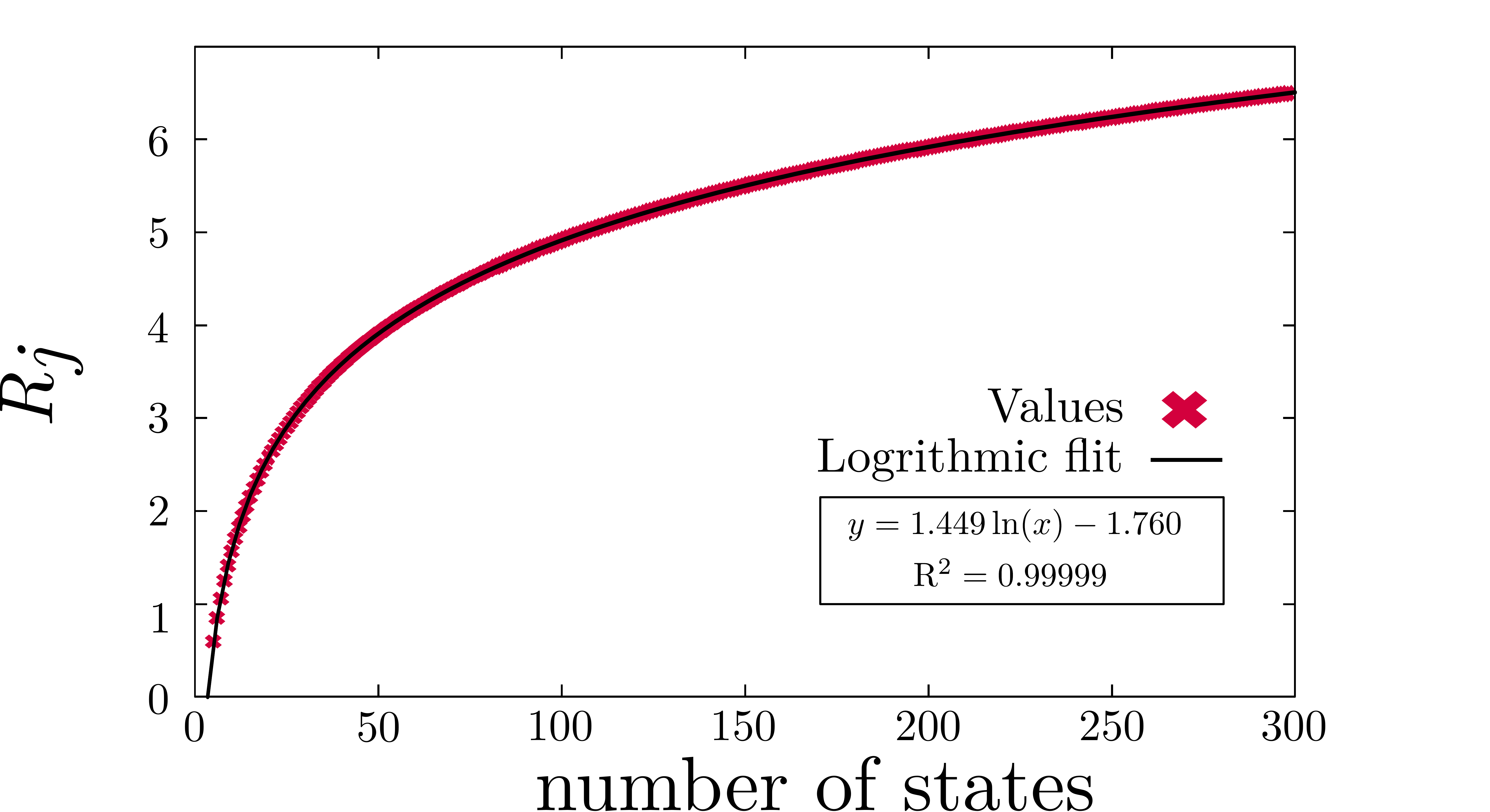} \label{fig log2 a}}
    \subfloat[\centering]{\includegraphics[width=0.4\columnwidth]{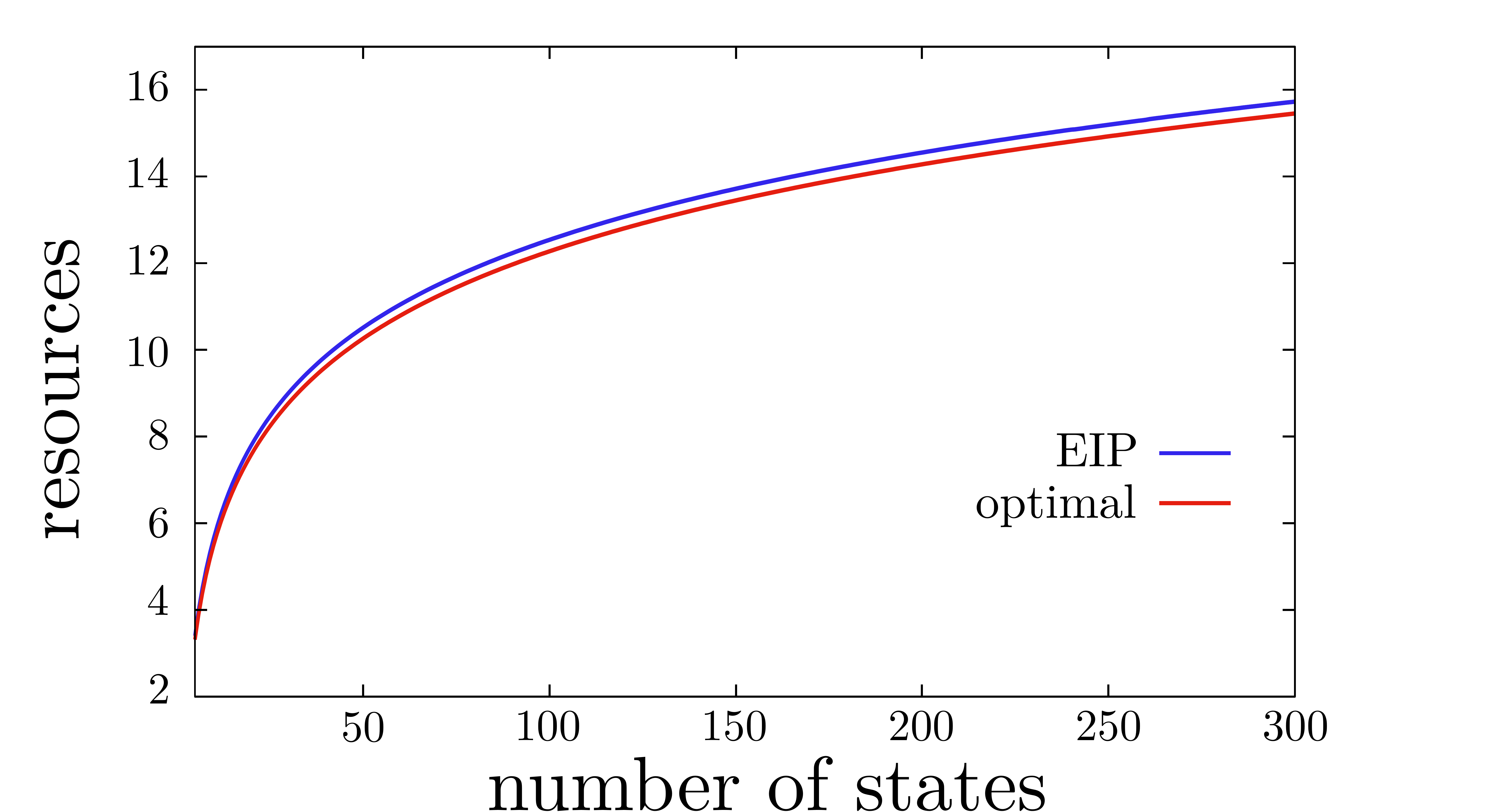} \label{fig log2 b}}
    \centering
    \caption{In (a), we plot $\langle \, \log_2 n' \rangle$ for different ensembles of $n$ states with two identical errors. $R^2$ is the coefficient determined by the logarithmic fit. In (b), we plot the resources needed to identify two identical errors as a function of the number of states of the ensemble. The blue curve corresponds to the procedure of EIP. The red curve corresponds to the optimal (minimum information) required to identify the position of the two errors under the assumption that the information is fully nonlocal and one e-bit per bit is required.}
\end{figure}
\noindent
We compute $\langle\log_2 d_3 \rangle$ (Fig.~\ref{fig log2 a}) for $n < 300$ and we obtain a logarithmic fit to approximate it up to $n$. We obtain that it is approximately given by
\begin{equation}
    R_j \approx 1.449 \ln(n) - 1.760\,.
\end{equation}
In Fig. \ref{fig log2 b}, we can see the amount of resources used by the procedure of EIPs do not differ considerable than the optimal amount of resources. More specifically, the ratio $r = (\text{EIP resources}) / (\text{optimal resources})$ is $r < 1.03$ for $n < 300$.

\section{Identification of two different errors}
\label{sec app 2 different}

\begin{figure*}
    \centering
    \subfloat[\centering]{\includegraphics[width=0.4\columnwidth]{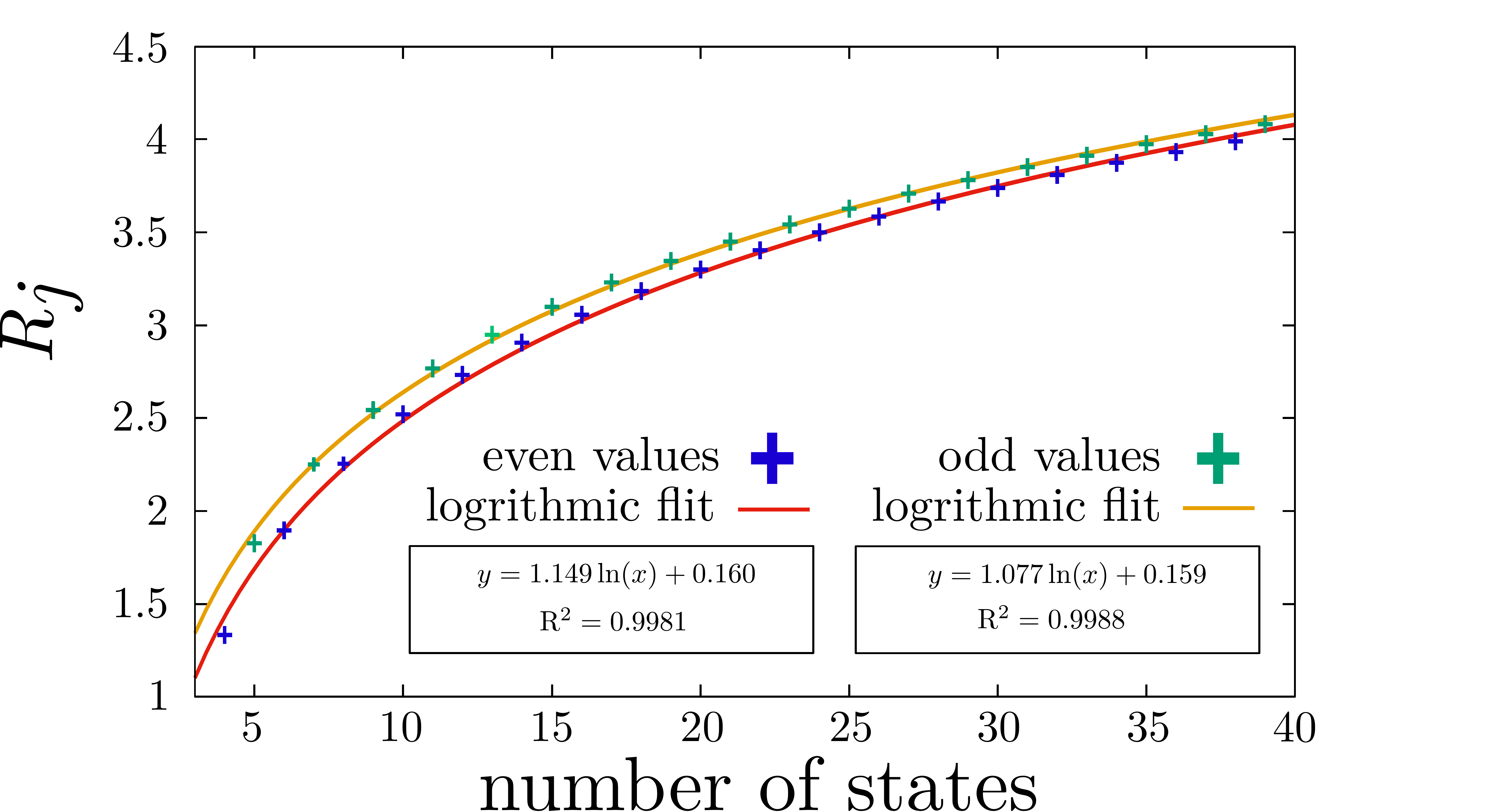} \label{fig log0d a}}
    \subfloat[\centering]{\includegraphics[width=0.4\columnwidth]{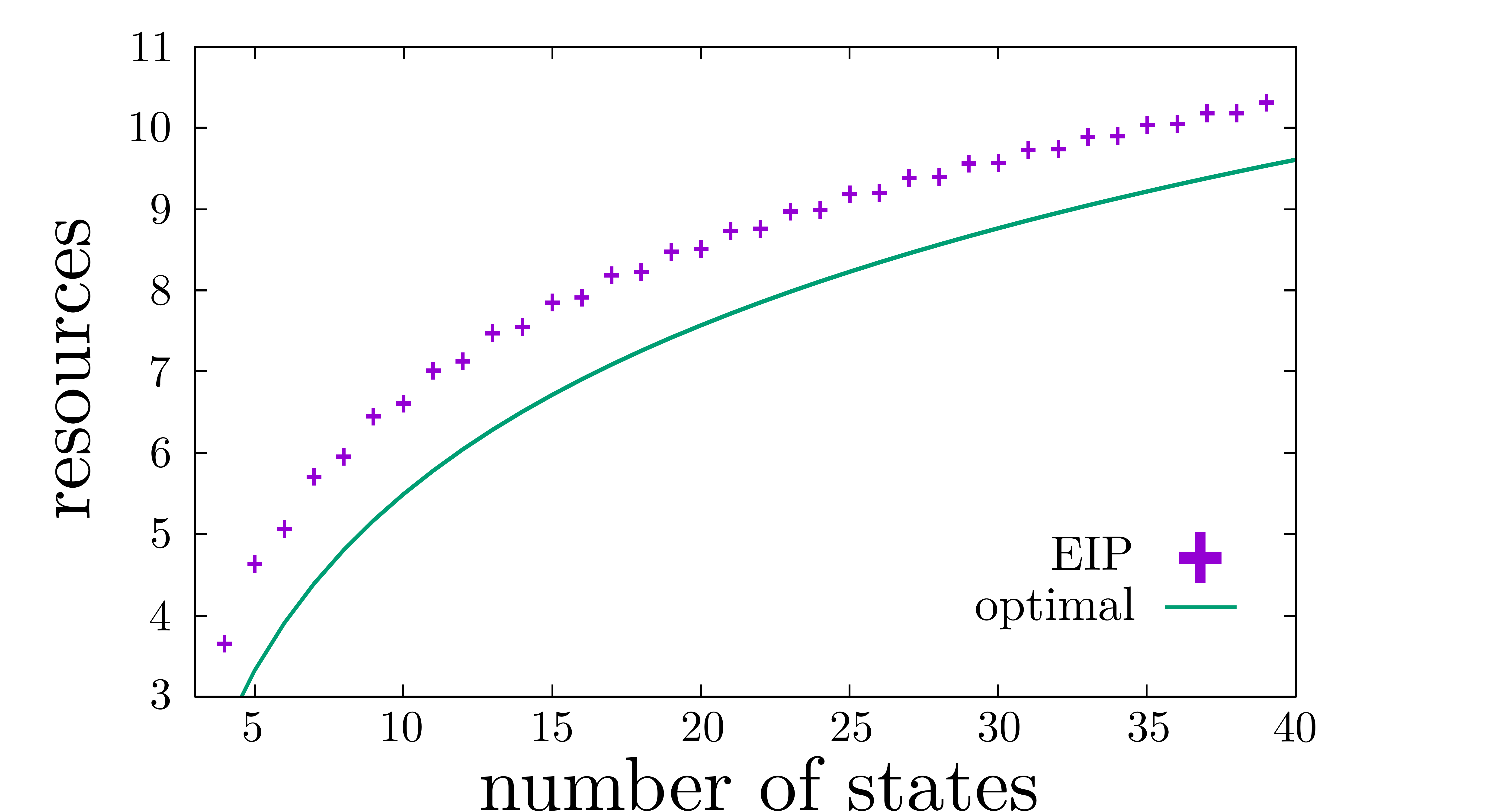} \label{fig log0d b}}
    \centering
    \caption{In (a), we plot of $R_j$ for different ensembles of $n$ states if the ensemble contains two identical errors. $R^2$ is the coefficient determined by the logarithmic fit. In (b), we plot the resources needed to identify two different errors as a function of the number of the states of the ensemble. The green and the purple curves correspond to the procedure proposed in EIP. The black curve corresponds to the optimal (minimum information) required to identify the position of the two errors.}
\end{figure*}

In this section, we introduce the procedure to identify two different errors in an ensemble of $|\Psi_{00}\rangle$ states. We also consider the possibility that there are no errors in the ensemble. This is the scenario when we purify an ensemble of copies of the state Eq.~\eqref{eq rank3 state} with the EIP(2) procedure and in the first step, $( \# 01 - \# 10 ) \text{mod} (2\lambda + 1) = 0$ is obtained.
\\ \\
Consider an ensemble that contains no errors, or two different errors (one 01 and one 10 at unknown positions). In this case, the minimum amount of information (and of resources) is the same as in the two identical errors case $\log_2(\frac{1}{2} n( n - 1 ))$, as we just have to identify the position of the two errors. The procedure that we propose does not reach this optimal value but it follows the same tendency, see Fig. (\ref{fig log0d b}). More specifically, the ratio $r=(\text{EIP resources})/(\text{optimal resources})$, is $r<1.3$ for $n<40$. To identify the positions of the two possible errors, we first isolate each error in a subensemble. For that purpose and to verify if there are no errors in the ensemble, we apply the EPG using an auxiliary state of $d=2n-1$ levels, i.e.
\begin{equation}
    \prod_{i=1}^n\,(b\text{CX}_{i\to \text{aux}})^i\big|\Phi_{\boldsymbol{\mu}}\big\rangle\big|\Psi^{(2n-1)}_{00}\big\rangle_{\text{aux}} = \big|\Phi_{\boldsymbol{\mu}}\big\rangle\big|\Psi^{(2n-1)}_{0j}\big\rangle_{\text{aux}}.
\end{equation}
From Eq.~\eqref{eq action bCX}, if the ensemble contains two different errors at positions $|\phi_{\mu_r}\rangle = |01\rangle$ and $|\phi_{\mu_t}\rangle = |10\rangle$ the new amplitude index of the auxiliary state is $j = (r - t) \text{mod}(2n - 1)$. On the other hand, we have $j = 0$ if there are no errors. In this point we determine the number of errors in the ensemble, as only if the ensemble contains two errors $j\neq 0$. The difference of two different numbers can not be 0, and as $r$, $t < n$, $|r - t|\neq |2n - 1|$.
\\ \\
The value $j = (r - t) \text{mod}(2n - 1)$ can be understood as a distance between the two error states. In addition, from this value, we can determine which error is in a higher position. Specifically
\begin{equation}
    \begin{cases}
        \, 1 \leq j < n \quad & \Longrightarrow \quad t < r \\ n \leq j \leq 2n \quad & \Longrightarrow \quad r < t.
    \end{cases}
\end{equation}
With this ``distance'' we can isolate each error in a subensemble, as we shall see, and then locate them with the EPG. We differentiate three possible situations:
\renewcommand{\labelitemi}{$\bullet$}
\begin{enumerate}
    \item The distance $r - t$ is an odd number. If the distance of two natural numbers is odd, then one of them is an odd number and the other is even. Therefore one subset is the states in even positions and the other set is the states in odd positions. In this case, we can isolate each error in a subensemble, but we do not know which error is in each.
    \item The distance $r - t$ is an even number. In this case, it means that both errors are in even positions or both are in odd positions. We reorder the states in a way that, the distance becomes an odd number. Specifically, we first put the odd states followed by the even keeping the initial order inside the odd and the even numbers. With this new order, the distance is reduced by half $d\rightarrow d/2$.
    \\ \\
    To illustrate this, we consider the following example: Assume an ensemble of 8 states with a ``distance'' between the errors $j = 2$. In this case the possibilities are: (1,3), (2,4), (3,5), (4,6), (5,7) and (6,8). If we reorder the ensemble the new position label of the states looks as follows:
    \begin{equation}
        \begin{pmatrix}
            \textcolor{red}{1} \\ \textcolor{blue}{2} \\ \textcolor{red}{3} \\ \textcolor{blue}{4} \\ \textcolor{red}{5} \\ \textcolor{blue}{6} \\ \textcolor{red}{7} \\ \textcolor{blue}{8}
        \end{pmatrix}
    \longrightarrow
    \begin{pmatrix}
        \textcolor{red}{1} \\ \textcolor{red}{3} \\ \textcolor{red}{5} \\ \textcolor{red}{7} \\ \textcolor{blue}{2} \\ \textcolor{blue}{4} \\ \textcolor{blue}{6} \\ \textcolor{blue}{8}
    \end{pmatrix}
    =
    \begin{pmatrix}
        \textcolor{red}{1'} \\ \textcolor{red}{2'} \\ \textcolor{red}{3'} \\ \textcolor{red}{4'} \\ \textcolor{blue}{5'} \\ \textcolor{blue}{6'} \\ \textcolor{blue}{7'} \\ \textcolor{blue}{8'}
    \end{pmatrix}.
    \end{equation}
    Note that in the new order ``$k'$'', the possible positions of the errors change: $(1, 3) \rightarrow (1', 2')$, $(2, 4) \rightarrow (5', 6')$, $(3, 5)\rightarrow (2',3')$, $(4, 6) \rightarrow (6',7')$, $(5,7)\rightarrow (3',4')$ and $(6, 8) \rightarrow (7',8')$. Therefore, the distance is reduced to 1. In this new order one error is in an even position and the second in an odd one.
    \\ \\
    We can iterate this step until the distance between the errors is an odd number. Note that this step does not require additional resources, it is just a relabeling. At this point we are in the $1^{\text{st}}$ case.
    \item Up to here, we have seen that one can always find two subsets of size $n/2$ where there is one unknown error in each. However, in some cases these subsets can be smaller, and also the kind of error in each can be known. We can find two subsets smaller than $n/2$ if the distance is bigger than $n/2$. For example, consider an ensemble of 8 states and we obtain the distance $j = - 5$. Then the subsets are:
    \begin{equation}
        \text{distance} \; j = - 5 \; \text{mod 15} = 10 \quad \Rightarrow \quad
            \begin{pmatrix}
                \textbf{1} \\ \textbf{2} \\ \textbf{3} \\ 4 \\ 5 \\ \textbf{6} \\ \textbf{7} \\ \textbf{8}
            \end{pmatrix}
        \begin{matrix}
            \Bigg\} |01\rangle \\ \\ \\ \\ \Bigg\}|10\rangle 
        \end{matrix}.
    \end{equation}
    In this example, the 01 error must be in position 1, 2, or 3, and the 10 error in positions 6, 7, or 8. Note that the position of one error determines the position of the second. The error in one of the subsets can be detected as explained in Sec.~\ref{sec toy model 1 error}.
\end{enumerate}
The number of resources needed is this process is given by
\begin{equation}
    R=\log_2(2n-1)+R_j
\end{equation}
where $R_j$ are the resources needed to locate one error after obtaining a ``distance'' $j$--the position of the other error is directly inferred once the first error is located. $R_j$ depends on the positions of the error states, and hence, on the obtained value of $j$. Therefore we need to compute the expected value of the resources used after obtaining $j$, what is given by
\begin{equation}
    R_j = 2 \left( \sum_{j = 1}^{\lceil n/2 \rceil - 1} \text{P}(j) \log_2\left( n \right) + \sum_{i = \lceil n/2 \rceil}^{n - 1} \text{P}(i) \log_2 \left( n - i \right) \right),
\end{equation}
where if a distance $j < n/2$ is obtained we isolate each error in a subensemble of $n / 2$ states and we need an auxiliary state od $d = n$ levels, as we do not know which error is in each subensemble. On the other hand, if the distance obtained is $j \geq n/2$ we isolate each error in a subensemble of $n - j$ states. In this case, we need an auxiliary state of $d = n - j$ levels, as we know which error is in each subensemble. As all configurations are equally probable, the probability of obtaining a distance $j$ is given by
\begin{equation}
    \label{def function com}
    \text{P}(j) = \frac{1}{n (n - 1)} \sum_{l_1, \dots, l_n = - 1}^1 \delta_{\sum_i l_i, 0} \, \delta_{\sum_k |l_k|, 2} \, \delta_{\sum_s s \cdot l_s \text{mod}( 2n - 1 ), j}
\end{equation}
what is obtained by dividing configuration with two different errors with a distance $j$ over all possible configurations with two different errors. We compute $R_j$ for $n<40$ and we obtain the logarithmic fit (Fig.~\ref{fig log0d a}).
\noindent
We obtain that for odd values of $n$ the expected value can be approximated by
\begin{equation}
    R_j \approx 1.077 \, \ln(n) + 0.159 \;.
\end{equation}
and the even by
\begin{equation}
    R_j \approx 1.149 \, \ln(n) + 0.160 \;.
\end{equation}

\section{Resources of EIP(2)}
\label{app sec Y2}

In this section, we provide the expected number of errors detected and the resources required for the EIP(2) protocol, if it is used to purify an ensemble of copies of a Bell-diagonal rank-3 state, Eq.~\eqref{eq rank3 state}.
\\ \\
The expected number of identified errors is given by
\begin{equation}
    \label{app eq ktilde}
    \begin{aligned}
       & \sum_{j = - \lambda}^{\lambda} k(j) \, p_\lambda(j) \\ & = 0 \cdot p_2(j'=0,j=0)+1\cdot\Big[p_2(j=1)+p_2(j=4)\Big]+2\cdot\Big[ p_2(j = 2) + p_2(j = 3) + 2 \cdot p_2 (j'\neq 0, j = 0) \Big],
    \end{aligned}
\end{equation}
where $p_2 (j)$ is the probability of obtaining a difference $j$ with $\lambda = 2$, and $p_\lambda (j', j = 0)$ is the probability of measuring a ``distance'' $j'$ between the error states if in the first step $j = 0$ was obtained. The expressions for $p_2 (j)$ and $p_2 ( j', j )$ are calculated in Appendix~\ref{sec app FL}.
\\ \\
The number of resources $R_{t}$ is given by
\begin{equation}
    R_{t} = \log_2 ( 2 \lambda + 1 ) + \sum_{ j = - \lambda}^{ \lambda} p(j) R_{\lambda} (j).
\end{equation}
For the case of $\lambda = 2$, $R_{\lambda}(j)$ in each case is given by
\begin{equation}
    \label{app eq yields 2 errors}
    \begin{aligned}
        & R_{2} (0) \approx\log_2 ( 2n - 1 ) p (j' = 0 | j = 0) + \Big[ \log_2 ( 2n - 1 ) + 1.077 \ln(n) + 0.159 \Big] p (j' \neq 0 | j = 0 ) 
        \\ & R_{2}(1) = \log_2 \left( n \right) 
        \\ [3pt] & R_{2} (2) \approx \log_2 ( 2n - 3 ) + 1.449 \ln(n) - 1.760 
        \\ [3pt] & R_{2}(3) = R_{2} (2) 
        \\[3pt] & R_{2} (4) = R_{2} (1)
    \end{aligned}
\end{equation}

\section{Computation of local fidelity: Bell-diagonal rank-3 states}
\label{sec app FL}

In order to obtain the local fidelity of each purified state and the probability of each possible outcome, first, we have to see that the density matrix describing the initial ensemble $\hat{\Gamma}$ of $n$ copies of state Eq.~\eqref{eq rank3 state} is given by
\begin{equation}
    \begin{aligned}
        \hat{\rho}^{\otimes n} & = \left[F \Pi_{\Psi}+\frac{ 1 - F }{2}\big( \Pi_{01} + \Pi_{10} \big) \right]^{\otimes n} = \sum_{l_1, \dots, l_n = - 1 }^1 \bigotimes_{ i = 1 }^n F^{\delta_{l_i, 0}} \left( \frac{ 1 - F }{2}\right)^{ 1 - \delta_{l_i, 0}} \Pi_{\Psi}^{\delta_{l_i, 0} } \Pi_{01}^{\delta_{l_i, 1}} \Pi_{10}^{\delta_{l_i, - 1 }}.
    \end{aligned}
\end{equation}
where $\Pi_{\Psi} = |\Psi_{00}\rangle \langle\Psi_{00} |$, $\Pi_{01} = | 01 \rangle \langle 01|$ and $\Pi_{10} = |10 \rangle \langle 10|$. The tensor product over index $i$ is due to the $n$ copies, and the $n$ sums are due to that each system can be in one of three states: $l_i = 0 \rightarrow \Pi_{\Psi}$, $l_i = 1 \rightarrow \Pi_{01}$ and $l_i = - 1 \rightarrow \Pi_{10}$. The value of $l_i$ assigned to each possible states is given due to the action that the state induces on the amplitude index of the auxiliary state. The operator $\hat{\Gamma}$ contains the $3^n$ possible distributions of states with each corresponding probability.
\\ \\
To obtain the density matrix of the ensemble after each measurement we just have to cancel all configurations that are not compatible with the outcome and re-normalize the distribution in $\hat{\Gamma}$. The density operator after obtaining the difference of error $j_1$ is given by
\begin{equation}
    \hat{\Gamma}_{j_1} = \frac{1}{ p_\lambda( j_1 ) } \sum_{ l_1, \dots, l_n = - 1 }^1 \bigotimes^n_{ i = 1 } F^{ \delta_{l_i, 0} } \left( \frac{ 1 - F }{2} \right)^{ 1 - \delta_{l_i, 0}} \Pi_{\Psi}^{\delta_{l_i, 0}} \Pi_{01}^{\delta_{l_i, 1}} \Pi_{10}^{\delta_{l_i, - 1}} \delta_{\left( \sum_k l_k \right) \text{mod}(d), j_1 },
\end{equation}
where $d = 2\lambda + 1$ is the number of levels of the used auxiliary state and $p_\lambda(j)$ the probability of measuring $j$, given by
\begin{equation}
    \label{eq app pj}
        p_\lambda( j ) = \sum_{ \substack{ \alpha, \beta, \gamma = 0 \\ \alpha + \beta + \gamma = n } }^{n} \Bigg[ \; \frac{n!}{  \alpha ! \; \beta ! \; \gamma! } \, F^{\alpha} \left( \frac{ 1 - F }{2} \right)^{ \beta + \gamma } \delta_{ ( \beta - \gamma ) \text{mod}(d), j} \Bigg],
\end{equation}
Consider that in the first measurement one 01 error is detected, $j_1=1$ and then identified at position $j_2$. Then the density operator of the ensemble is given by
\begin{equation}
    \hat{ \Gamma }_{1, j_2} = \frac{1}{ p_\lambda ( 1, j_2 ) } \sum_{ l_1, \dots, l_n = - 1 }^1 \bigotimes^{n}_{ i = 1 } F^{\delta_{l_i, 0}} \left( \frac{ 1 - F }{2} \right)^{ 1 - \delta_{l_i, 0}} \Pi_{\Psi}^{\delta_{l_i, 0}} \Pi_{01}^{\delta_{l_i, 1} } \Pi_{10}^{\delta_{ l_i, - 1}} \delta_{ \left( \sum_k l_k \right) \text{mod}(d), 1 } \delta_{\left(\sum_k k \cdot l_k \right) \text{mod}(n), j_2}.
\end{equation}
The local fidelity of the $k^{\text{th}}$ state after obtaining $j_2$ corresponds to
\begin{equation}
    F'^{(k)} = \frac{1}{p_\lambda(1, j_2)} \sum_{l_1, \dots, l_n = - 1 }^1 \prod_{ i = 1 }^n F^{\delta_{l_i, 0}} \left(\frac{ 1 - F }{2} \right)^{1 - \delta_{l_i,0}} \delta_{ \left(\sum_{r}^n l_r \right) \text{mod}(d), 1} \; \delta_{ \left( \sum_{s}^n s \cdot l_s \right) \text{mod}(n), j_2} \; \delta_{l_k, 0}.
\end{equation}
where the probability of identifying the error state at position $j_2$ is given by
\begin{equation}
    p_\lambda( j_1 = 1, j_2 ) = \sum_{l_1, \dots, l_n = - 1}^1 \prod_{ i = 1 }^n F^{\delta_{ l_i, 0 } } \left(\frac{ 1 - F }{2} \right)^{ 1 - \delta_{l_i, 0} } \delta_{\left(\sum_{r}^n l_r \right) \text{mod}(d), j} \; \delta_{\left( \sum_{s}^n s \cdot l_s \right) \text{mod}(n), j_2}.
\end{equation}

\section{Up to one error \texorpdfstring{$\lambda= 1$}{TEXT}: Additional details}
\label{app sec one error}

\begin{figure}[h]
    \centering
    \textbf{EIP(1) for Bell-diagonal rank-3 states}\par\medskip
    \subfloat[\centering]{\includegraphics[width=0.3\columnwidth]{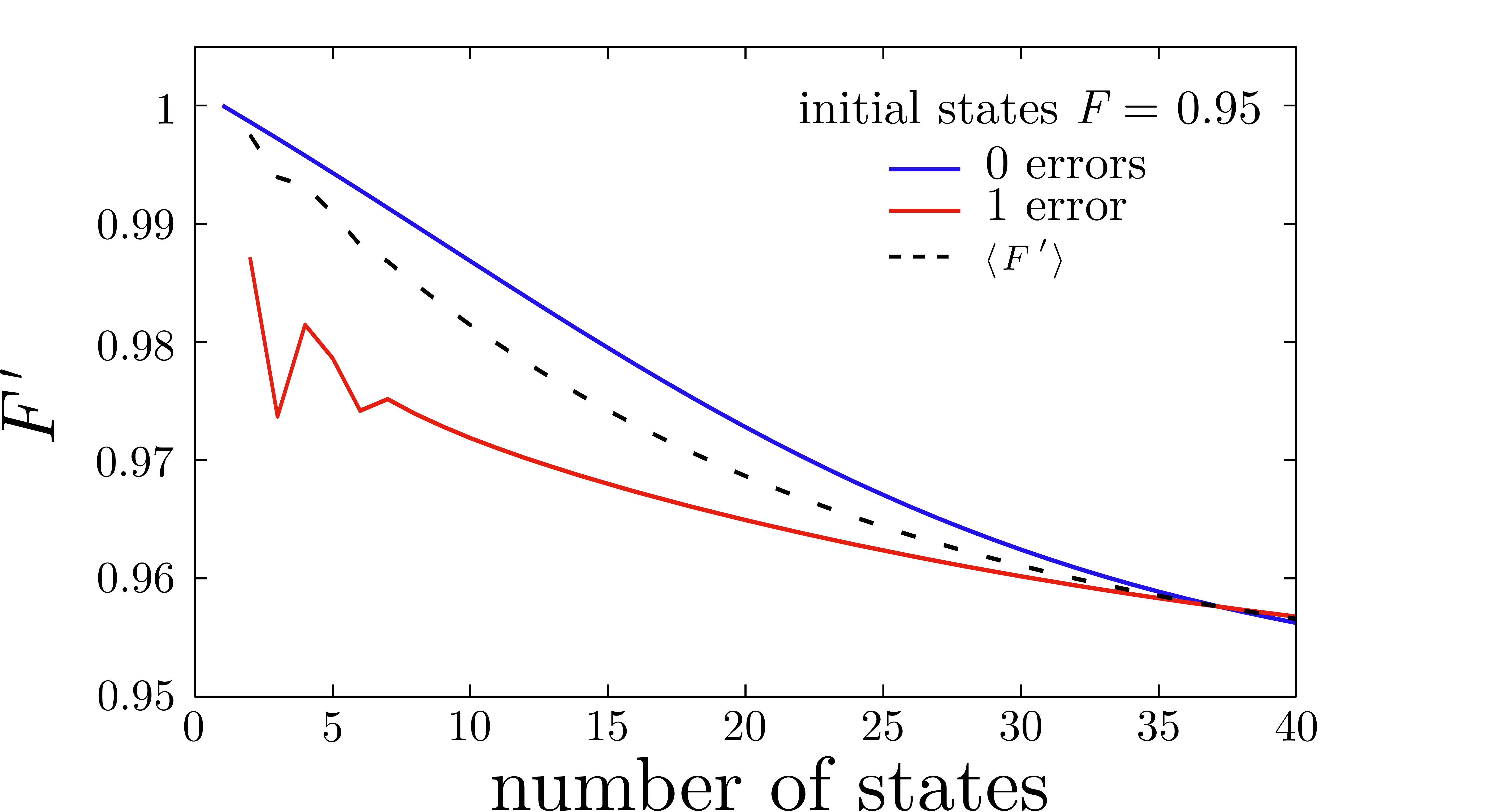} \label{fig R3a}}
    \subfloat[\centering]{\includegraphics[width=0.3\columnwidth]{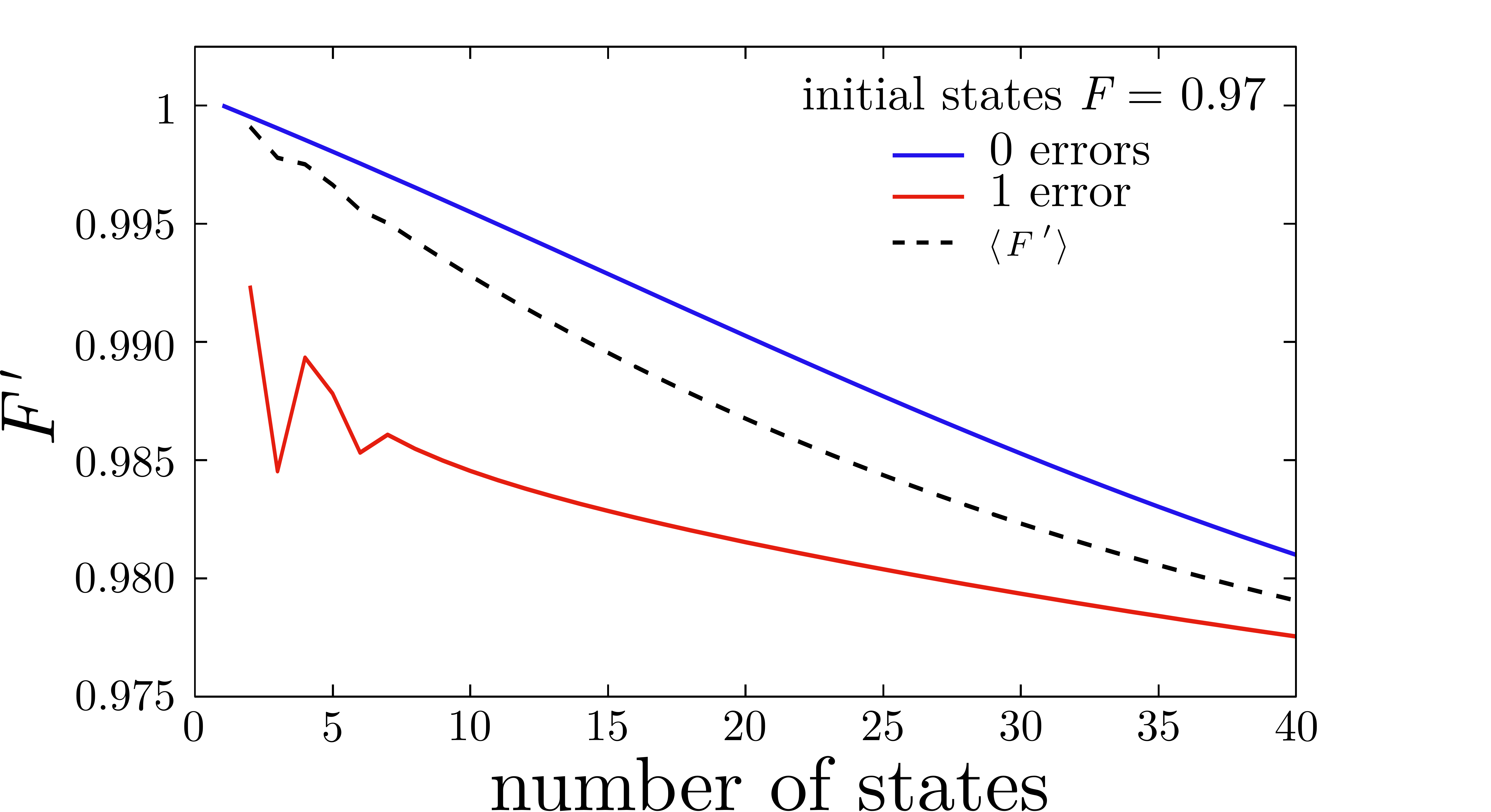} \label{fig R3b}}
    \subfloat[\centering]{\includegraphics[width=0.3\columnwidth]{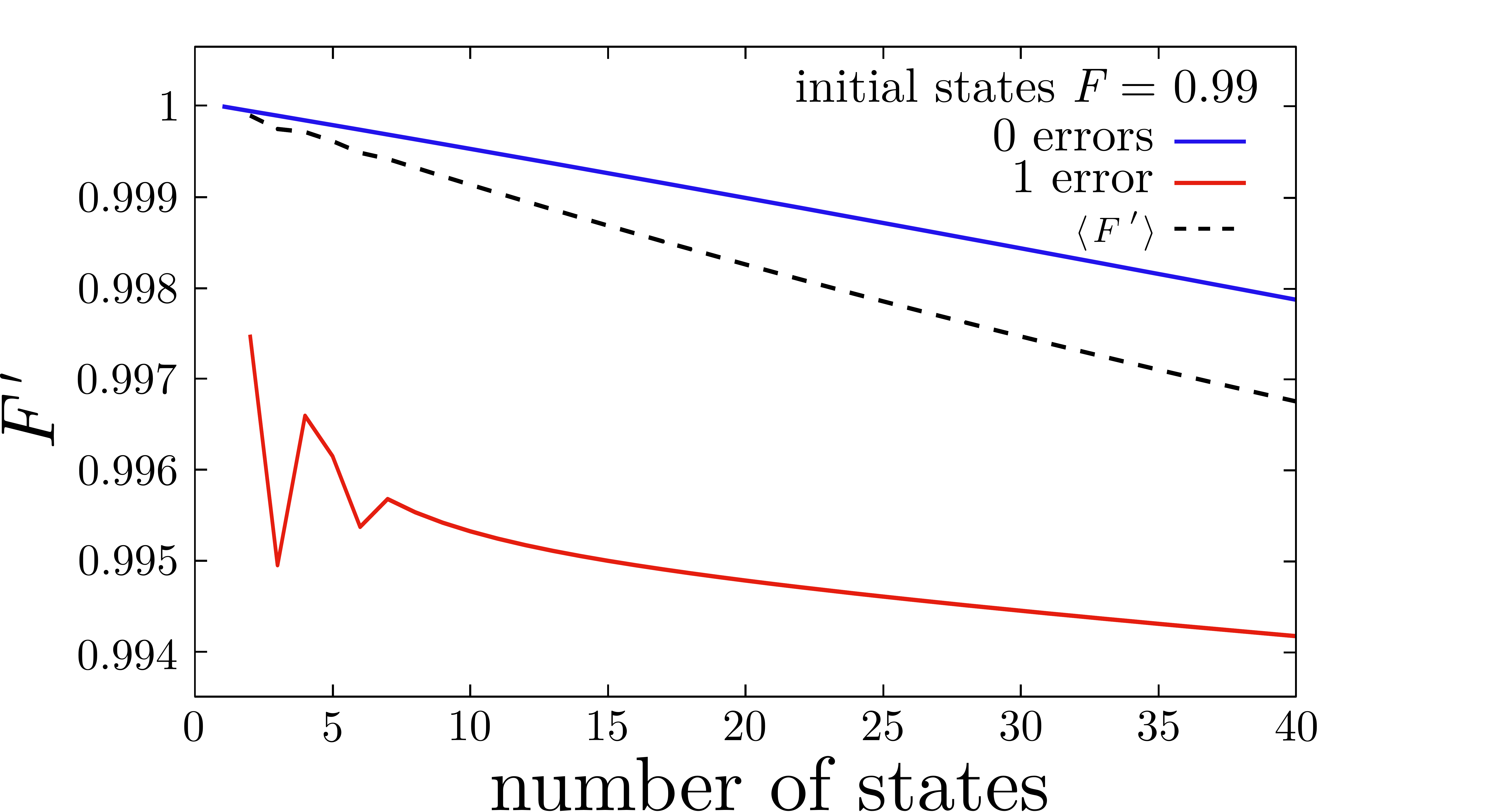} \label{fig R3c}} \hfill
    \subfloat[\centering]{\includegraphics[width=0.3\columnwidth]{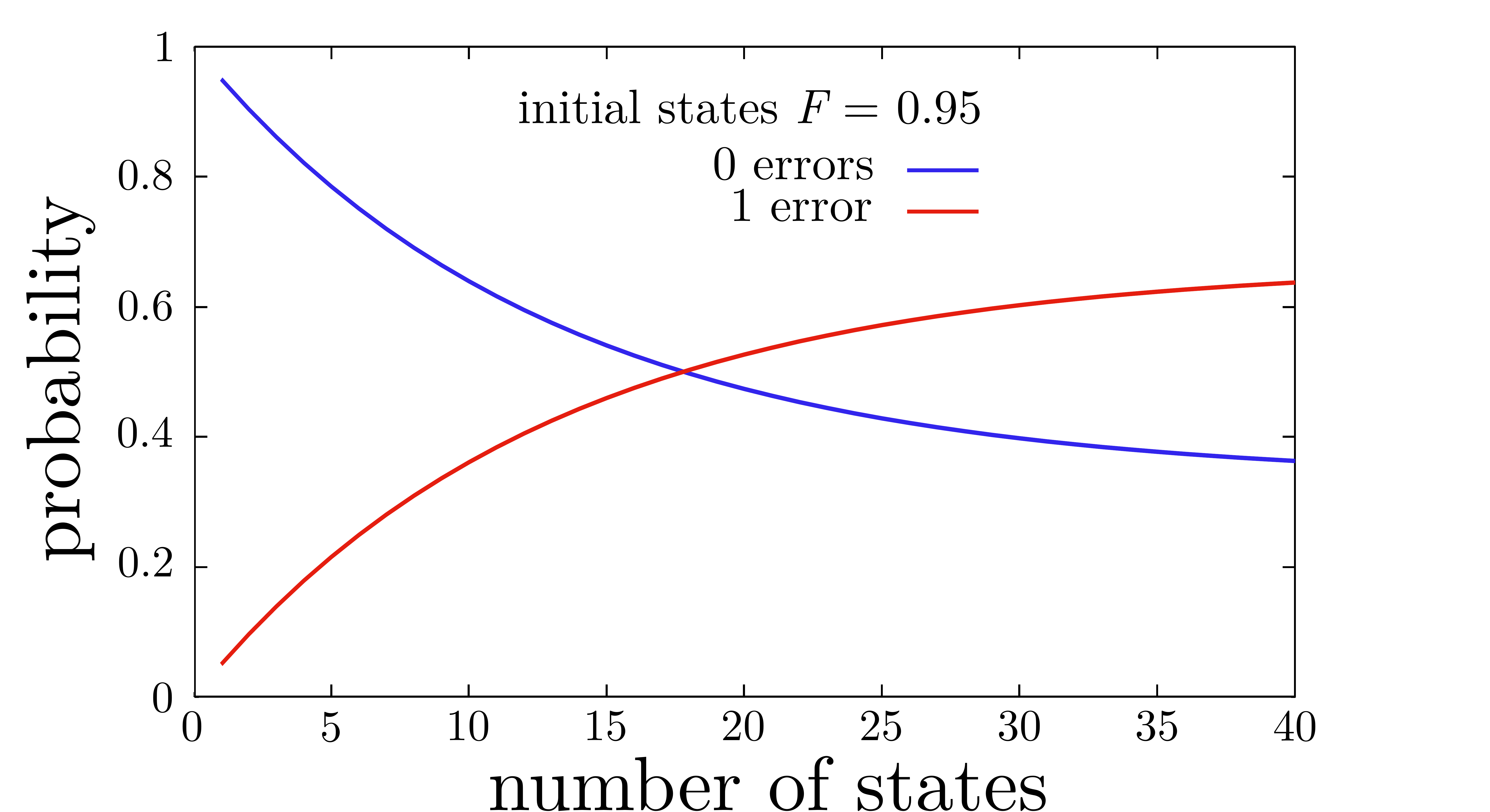} \label{fig R3d}}
    \subfloat[\centering]{\includegraphics[width=0.3\columnwidth]{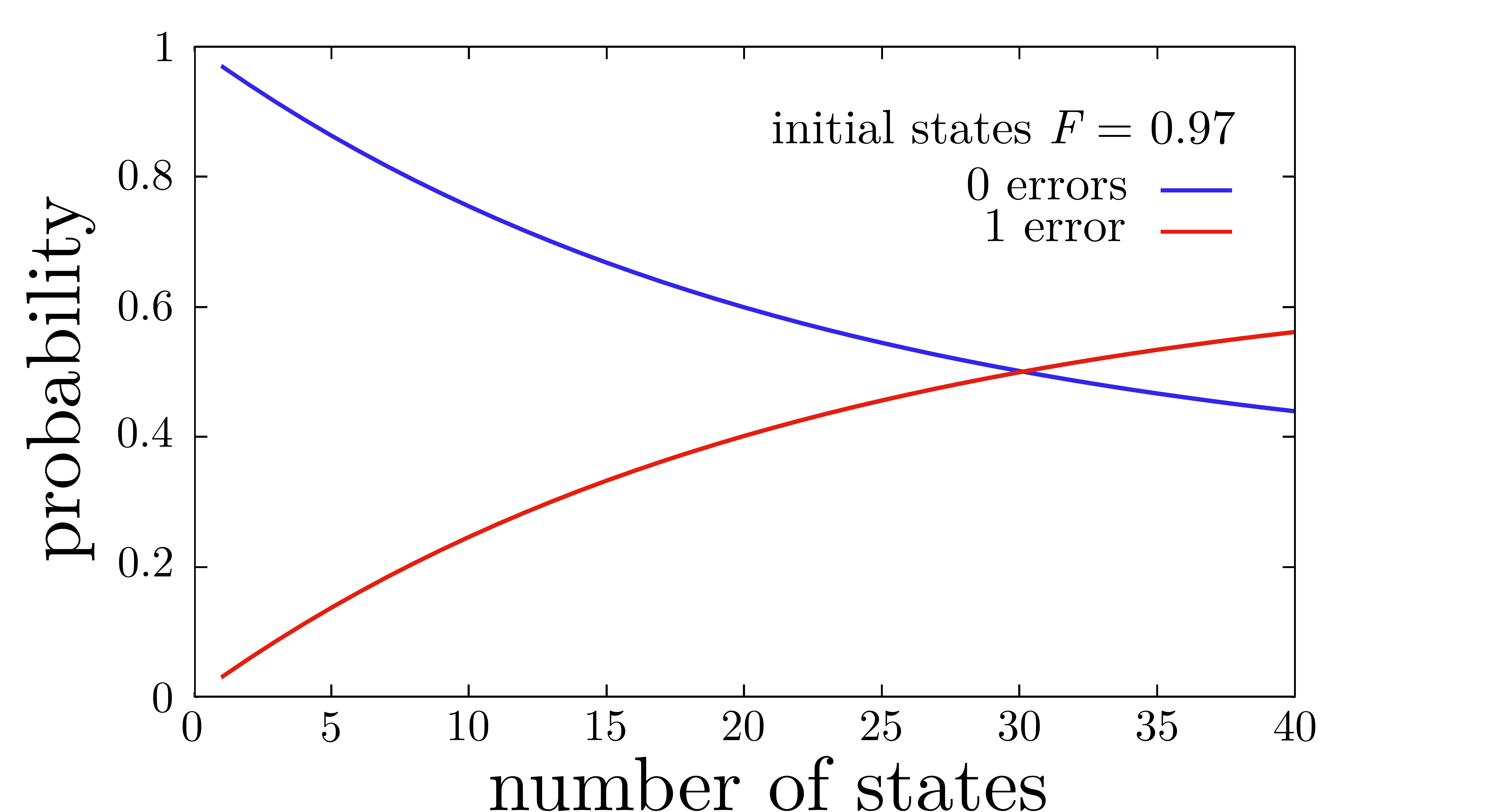} \label{fig R3e}}
    \subfloat[\centering]{\includegraphics[width=0.3\columnwidth]{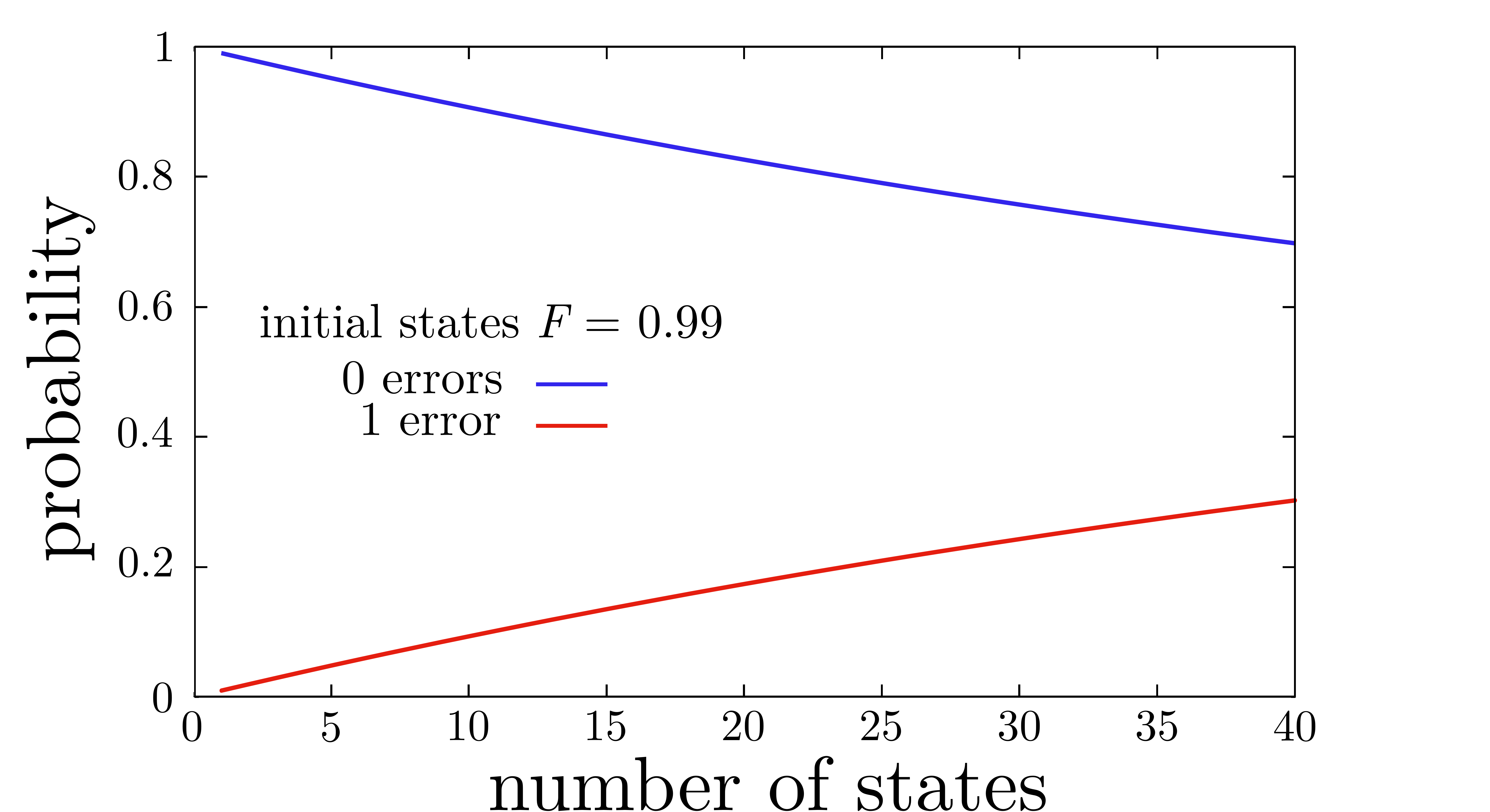} \label{fig R3f}}
    \caption{Purification of an ensemble of copies of a rank-3 Bell-diagonal state, Eq.~\eqref{eq rank3 state}, with the EIP(1). Each color corresponds to a different scenario given by the number and kind of errors identified (solid lines). Dashed curves correspond to the average of the different scenarios. In (a), (b) and (c), we plot the local fidelity as a function of the number of states for different ensembles. In (d), (e) and (f), we plot the probability of detecting one error and the probability of detecting no errors as a function of the number of states for different ensembles. In (a) and (d), the initial local fidelity is $F = 0.90$. In (b) and (e), the initial local fidelity is $F = 0.97$. In (c) and (f), the initial local fidelity is $F = 0.99$.}
    \label{fig Fl one error}
\end{figure}

\begin{figure}[h]
    \centering
    \textbf{EIP(1) for Bell-diagonal rank-3 states}\par\medskip
    \subfloat[\centering]{\includegraphics[width=0.5\columnwidth]{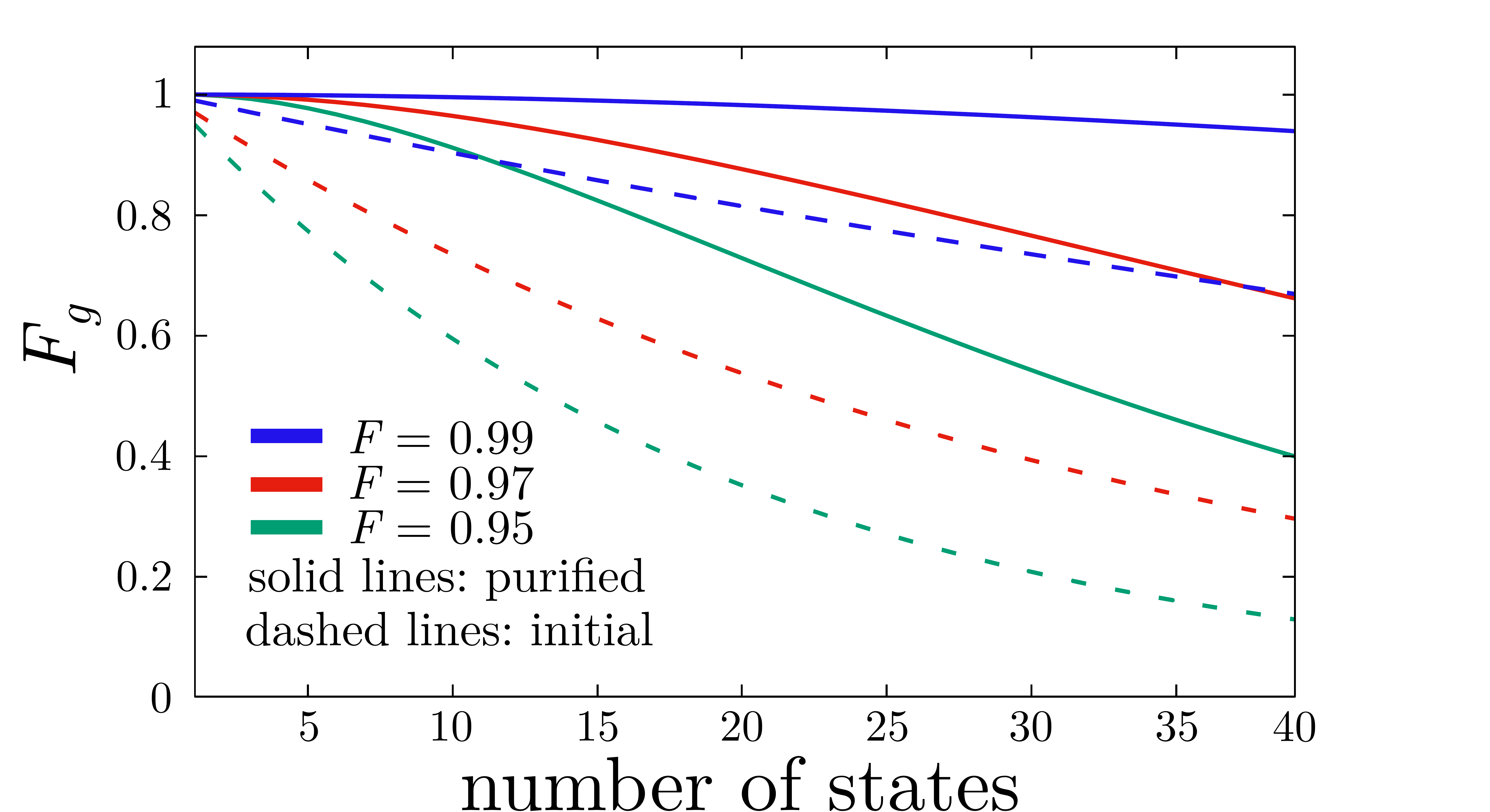} \label{fig FgYone error a}}
    \subfloat[\centering]{\includegraphics[width=0.5\columnwidth]{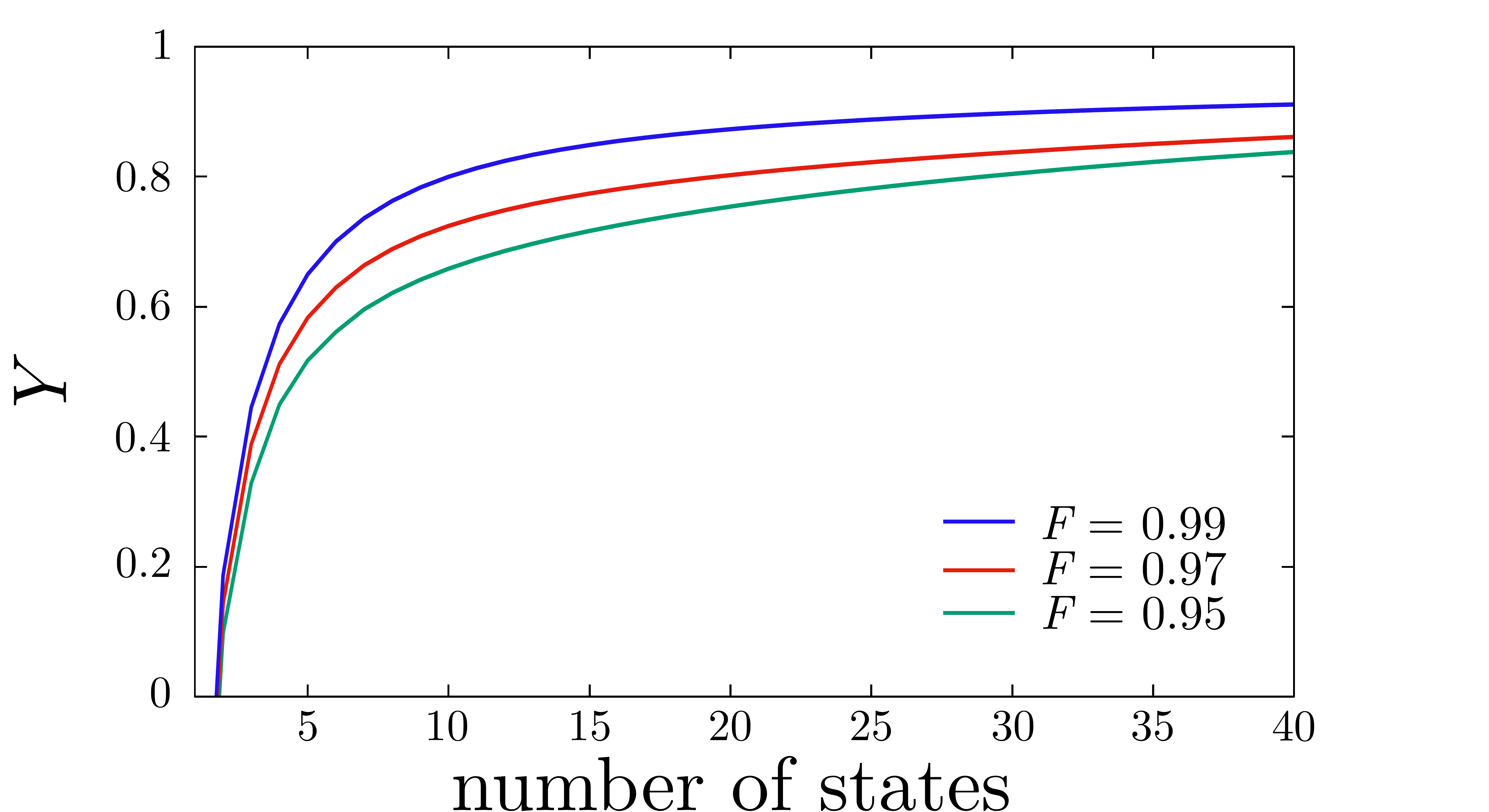} \label{fig FgYone error b}}
    \caption{Purification of an ensemble of copies of a rank-3 Bell-diagonal state, Eq.~\eqref{eq rank3 state}, with the EIP(1). Different colors correspond to different initial fidelities (0.99, 0.97 and 0.95). In (a), we plot the global fidelity as a function of the number of states for different ensembles. Dashed lines correspond to the initial ensemble and solid lines the purified ensemble. In (b), we plot the yield of the purification as a function of the number of states for different ensembles.}
    \label{fig FgYone error}
\end{figure}

In this section, we analyze the fidelity and the yield of the purification of an ensemble of $n$ copies of a Bell-diagonal rank-3 state, Eq.~\eqref{eq rank3 state}, using the EIP(1). In Figs.~\ref{fig R3a}, \ref{fig R3b} and \ref{fig R3c}, we plot the local fidelity $F'$ of the purified states as a function of the number of copies for different ensembles. We plot $F'$ for the cases where we detect one error if we detect no errors and the average. These plots show that we can successfully purify ensembles of $F > 0.95$ and $n < 40$ assuming at most one error. Note that the expected value of the local fidelity is higher if no errors are detected. However, this difference decreases as the expected number of errors increase. Note also that the curve for $F'$ with one error displays local minima for $n = 3$ and $n = 6$. This fact is due to combinatorial reasons. For these two values, the probability of some non-considered configurations gains weight.
\\ \\
In Figs.~\ref{fig R3d}, \ref{fig R3e} and \ref{fig R3f}, we plot the probability of detecting one error or no errors. For a small expected number of errors, it is more probable to detect that the ensemble contains no errors. As one expects, when the number of states $n$ increases the probability of detecting one error increase, and the probability to detect no errors decreases.
In Fig.~\ref{fig FgYone error}a, we plot the global fidelity of the initial and the purified ensemble as a function of the number of states for different initial ensembles. Like the behavior of $F'$, the global fidelity increases for ensembles with $F > 0.95$ and $n < 40$.
\\ \\
In Fig.~\ref{fig FgYone error b}, we plot the yield of the purification as a function of the number of states. We recall that for computing the yield we consider the target fidelity plotted in Fig.~\ref{fig Fl one error}. Note that we always obtain a positive yield.

\section{EIP(2): Additional material}
\label{sec app 2 errors}

\begin{figure}
    \centering
    \textbf{EIP(2) for Bell-diagonal rank-3 states}\par\medskip
    \subfloat[\centering]{\includegraphics[width=0.25\columnwidth]{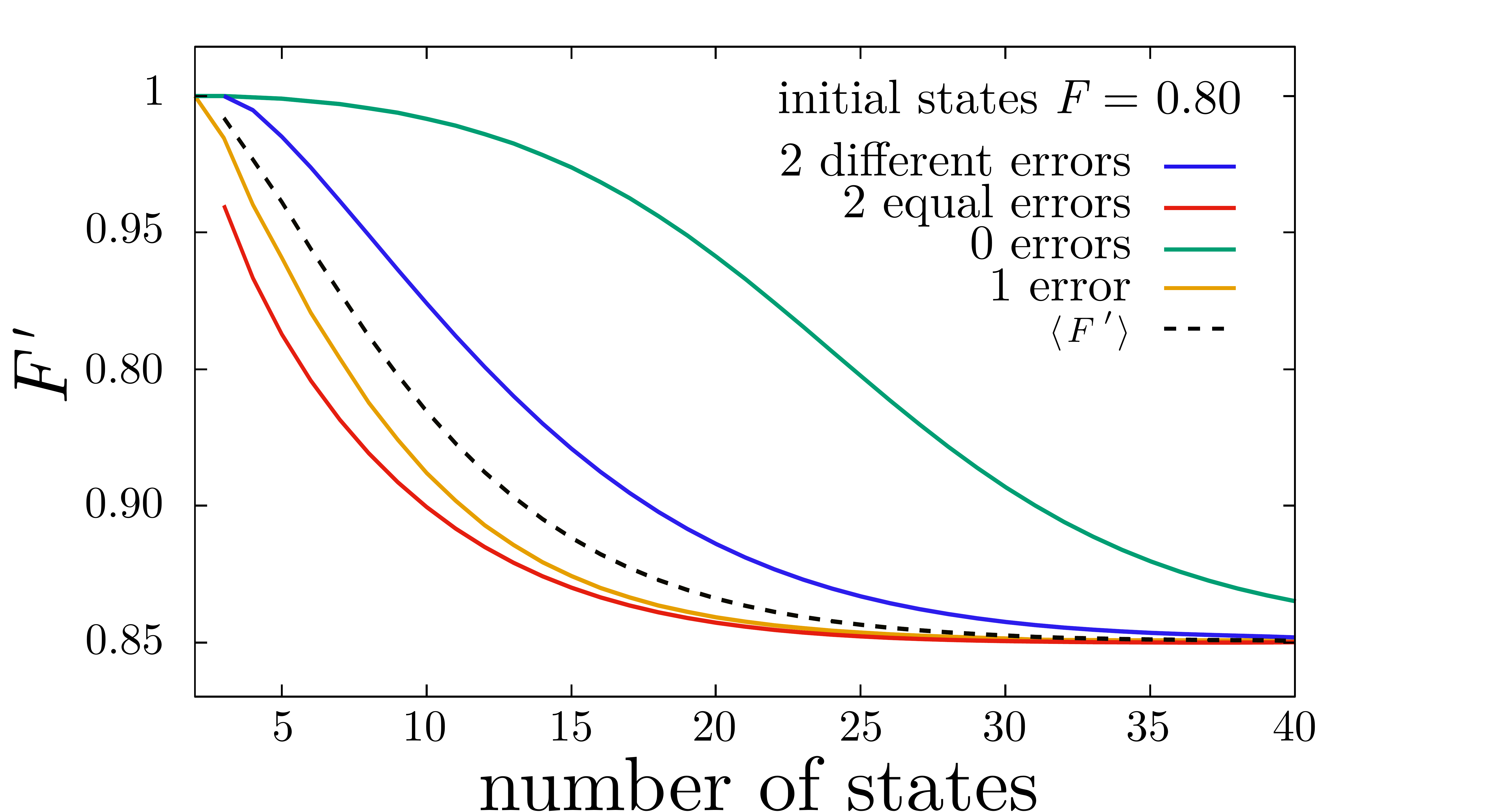} \label{fig r5 FiPa}}
    \subfloat[\centering]{\includegraphics[width=0.25\columnwidth]{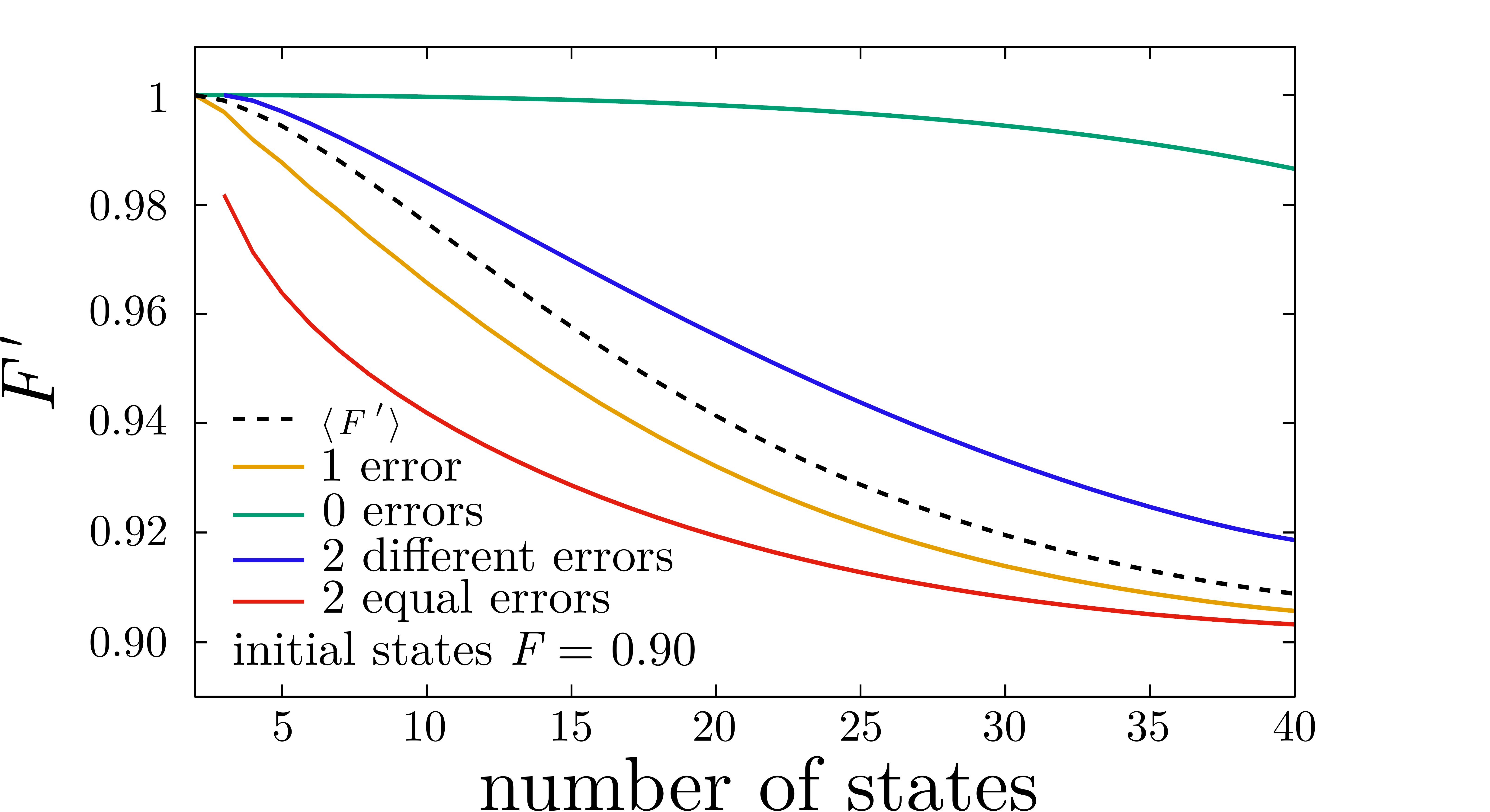} \label{fig r5 FiPb}}
    \subfloat[\centering]{\includegraphics[width=0.25\columnwidth]{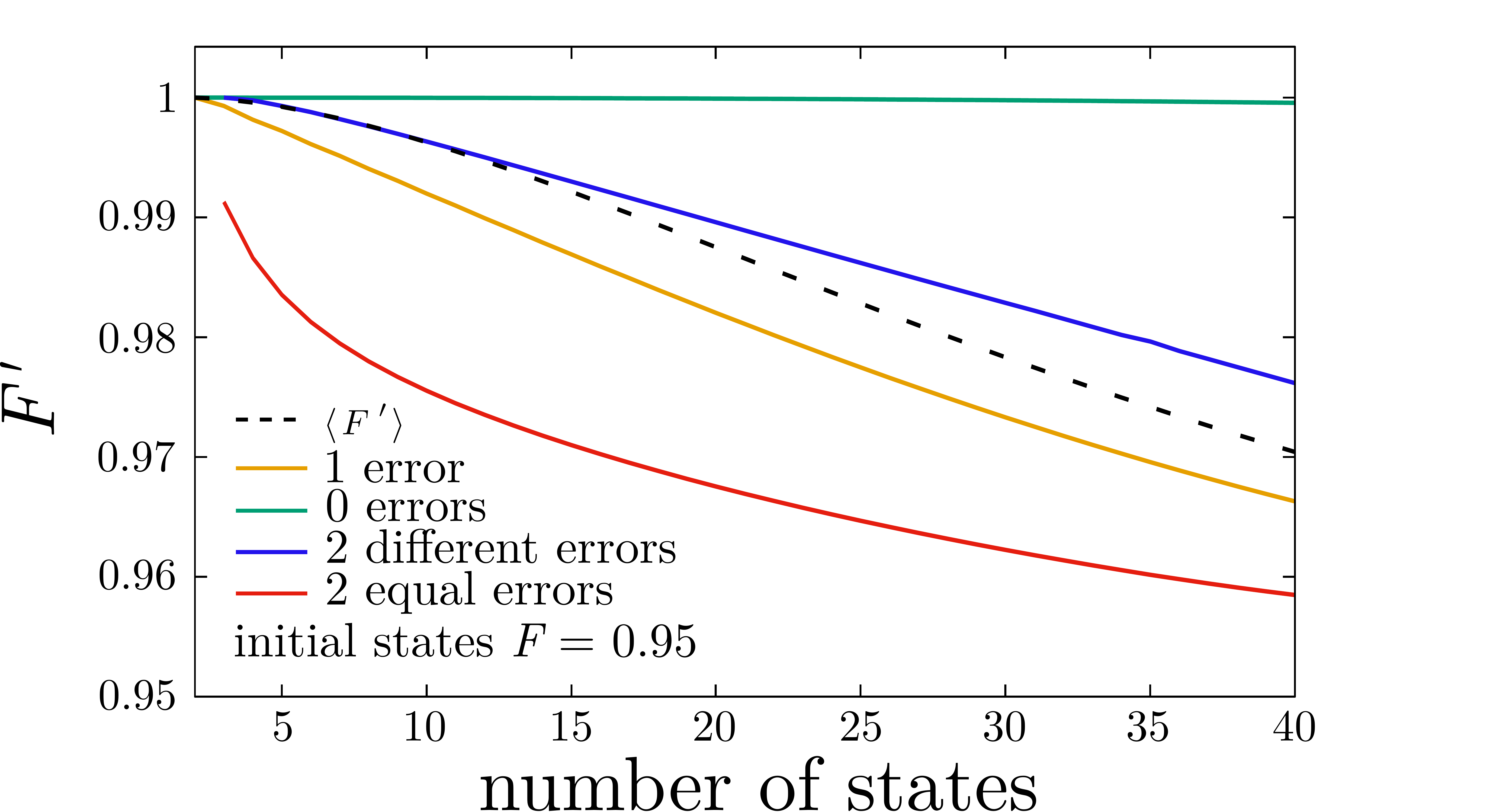} \label{fig r5 FiPc}}
    \subfloat[\centering]{\includegraphics[width=0.25\columnwidth]{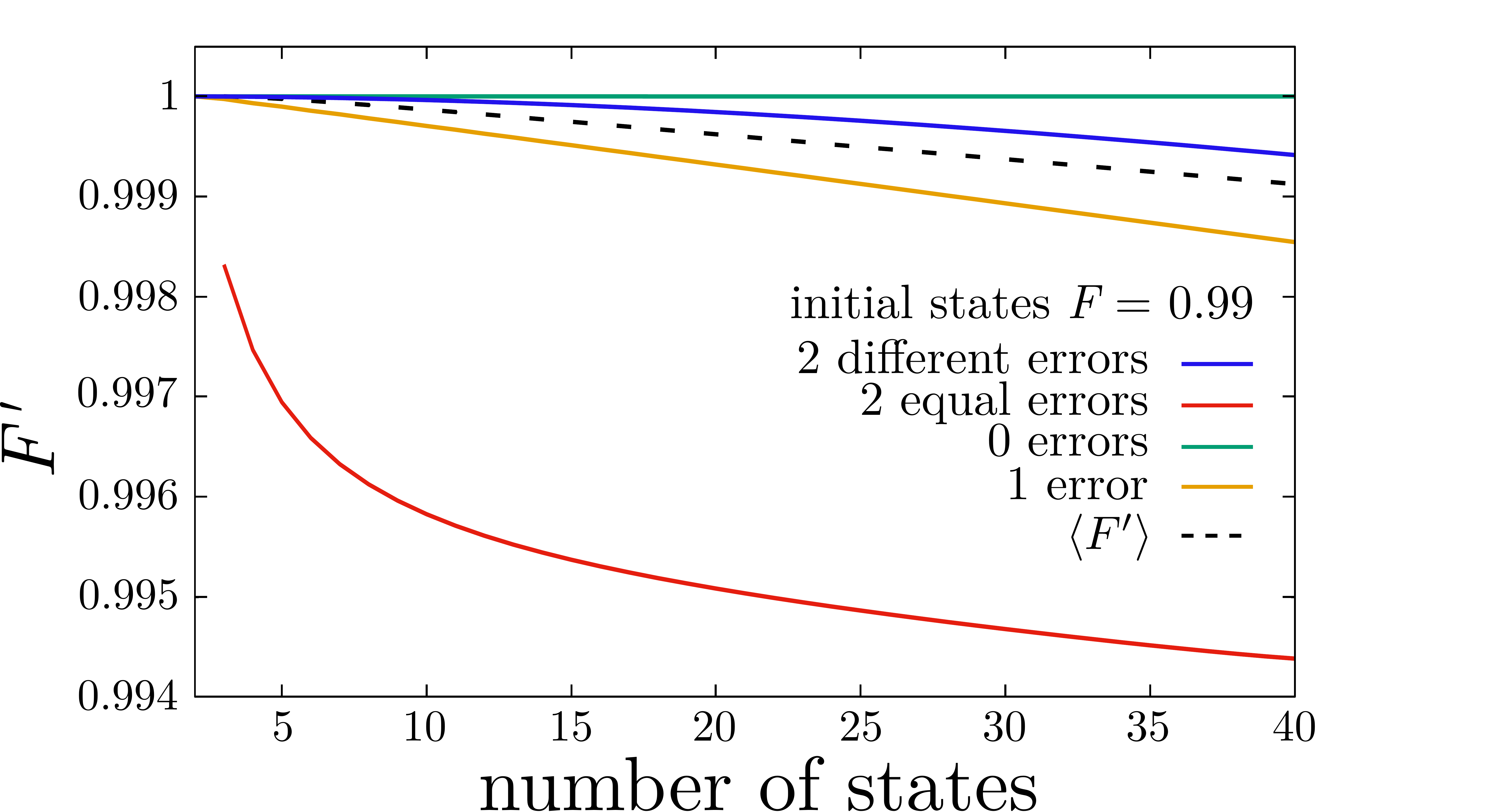} \label{fig r5 FiPd}}\hfill
    \subfloat[\centering]{\includegraphics[width=0.25\columnwidth]{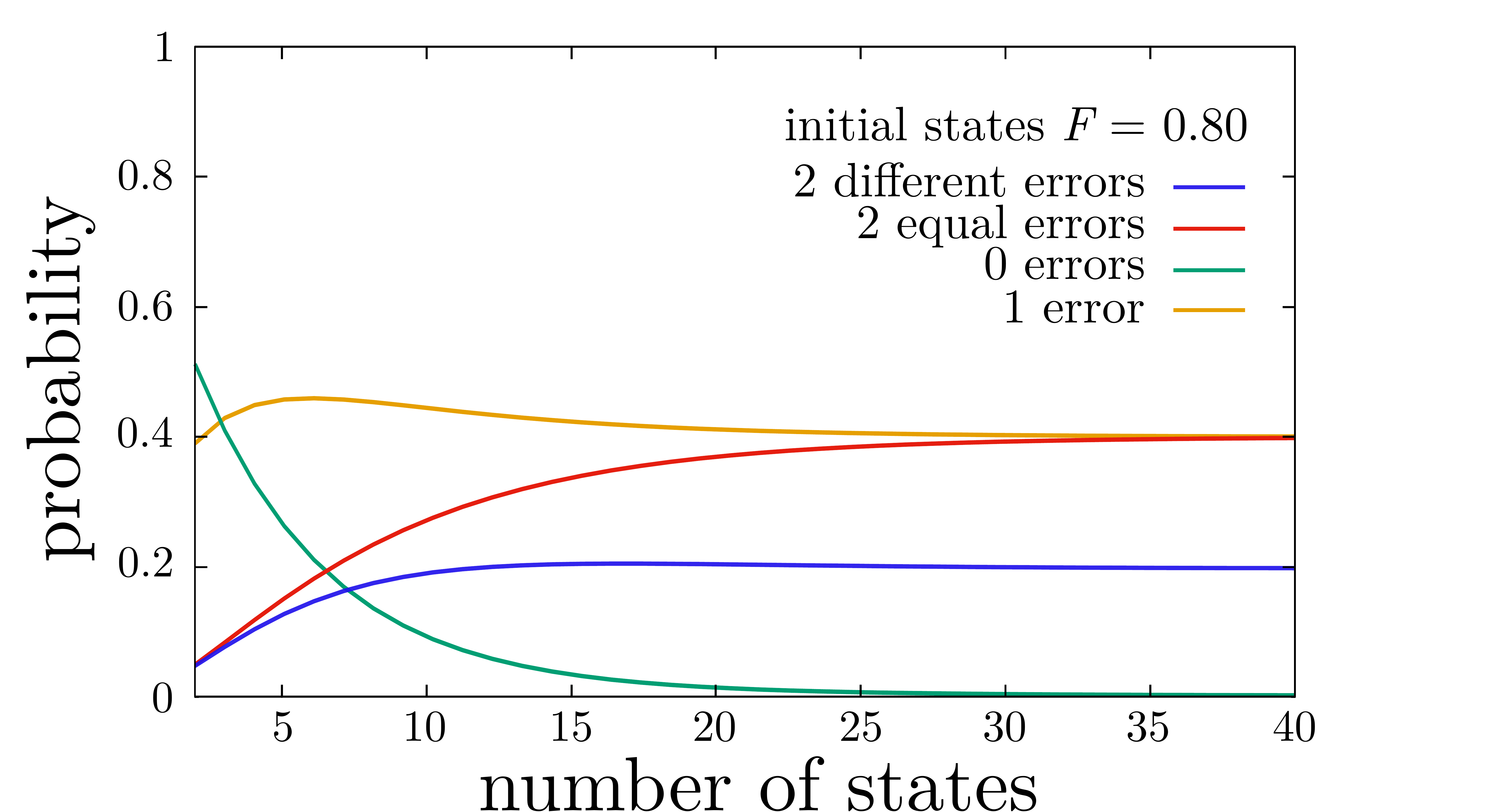} \label{fig r5 FiPe}}
    \subfloat[\centering]{\includegraphics[width=0.25\columnwidth]{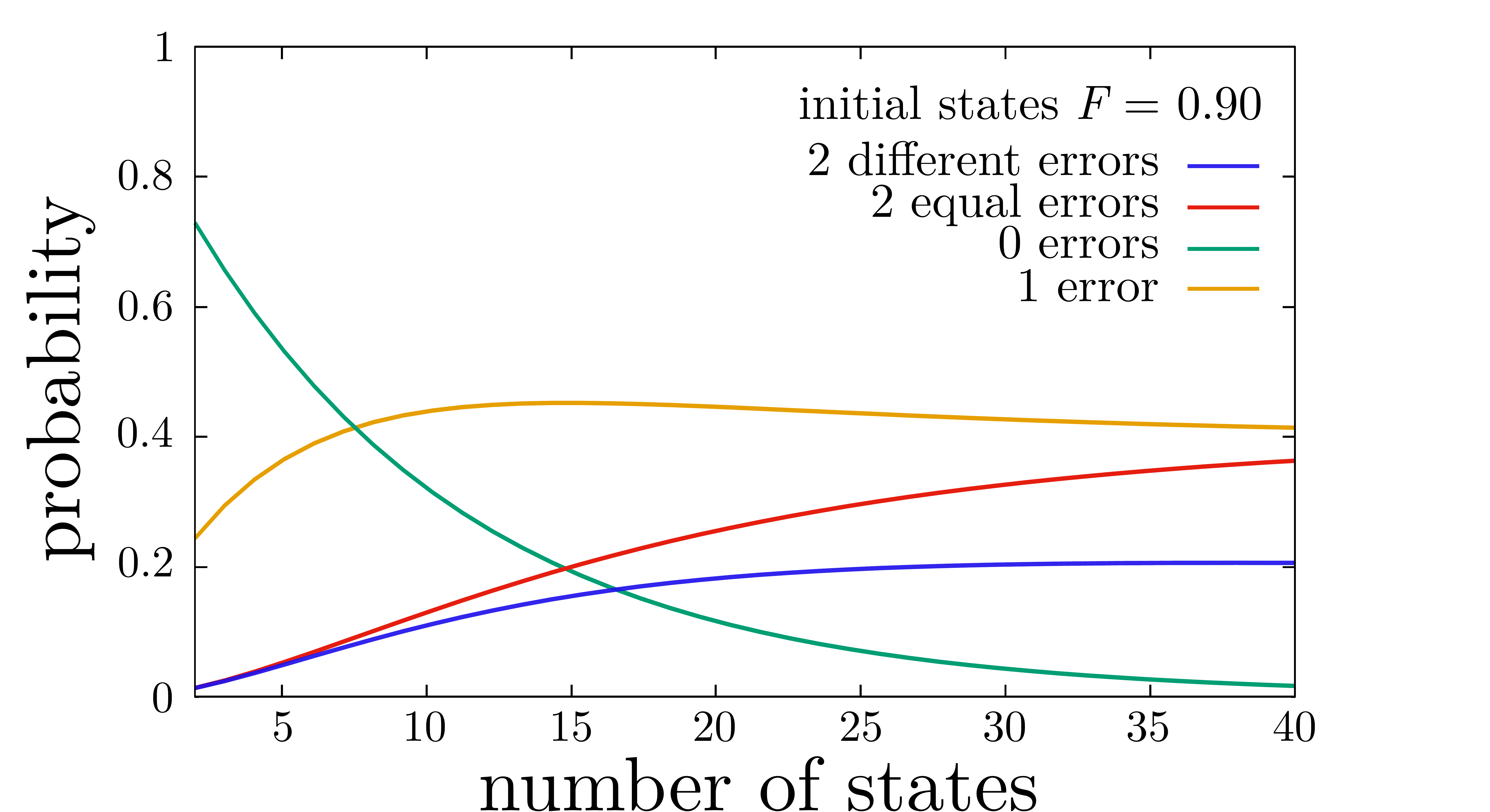} \label{fig r5 FiPf}}
    \subfloat[\centering]{\includegraphics[width=0.25\columnwidth]{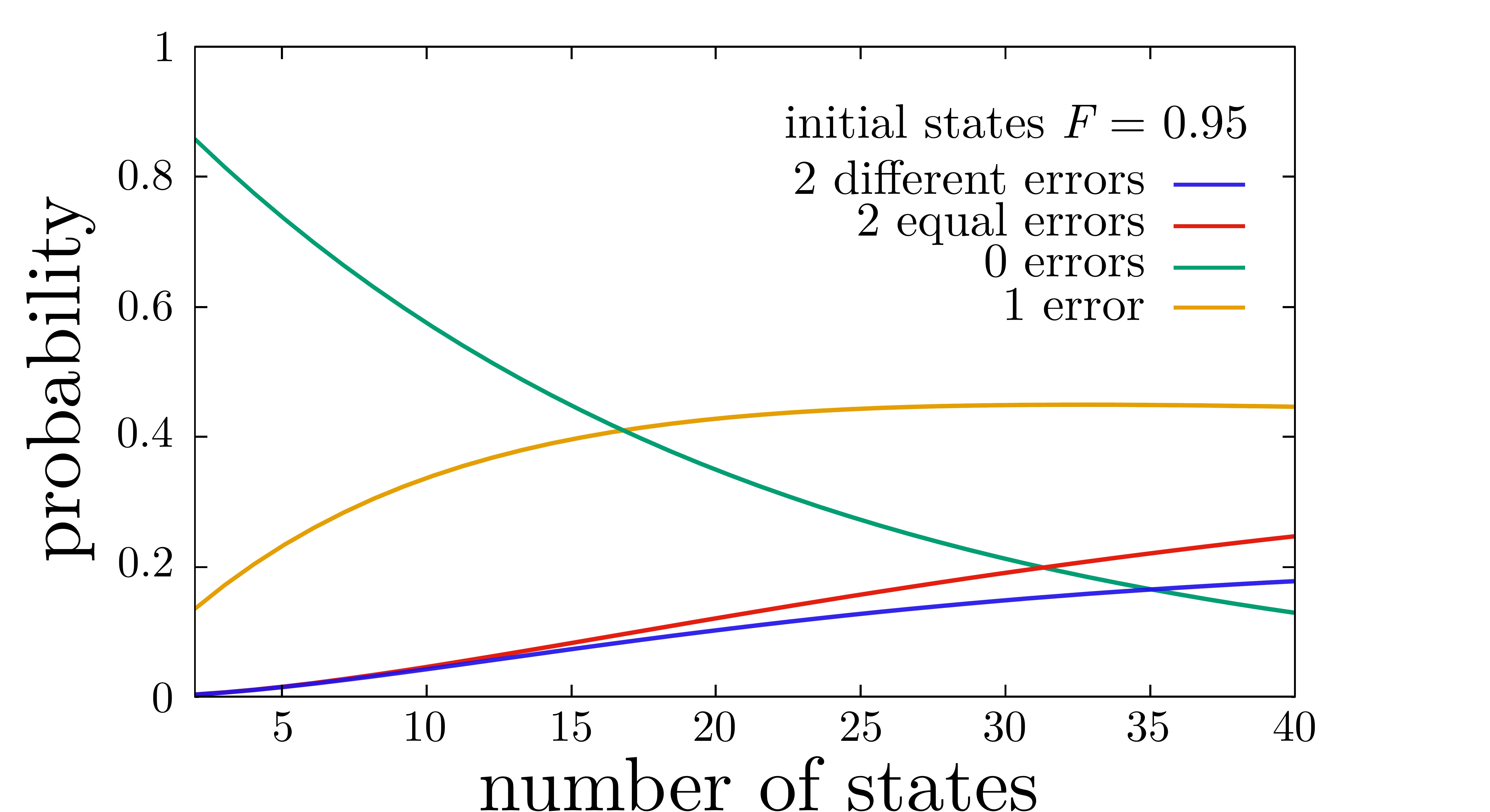} \label{fig r5 FiPg}}
    \subfloat[\centering]{\includegraphics[width=0.25\columnwidth]{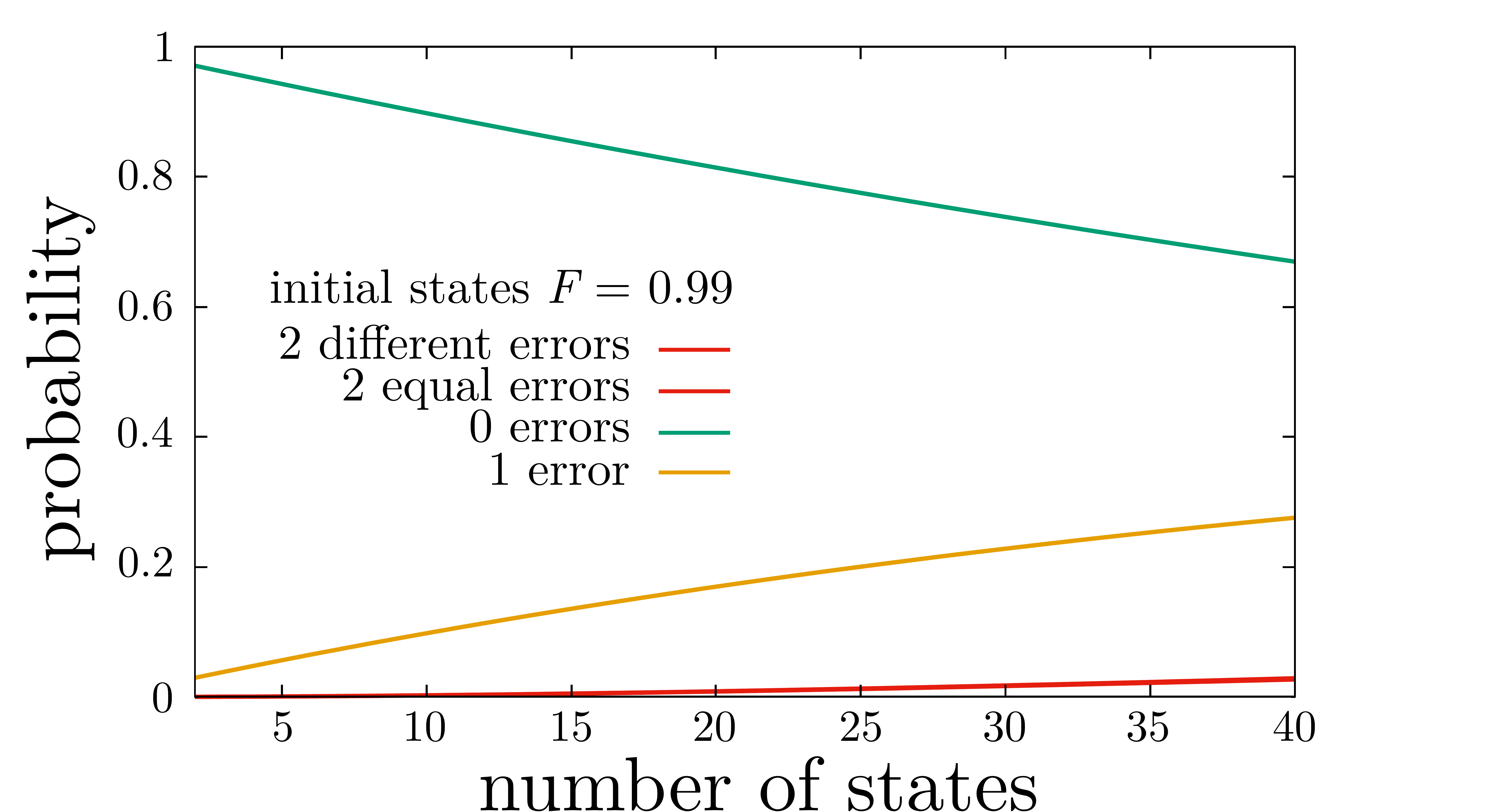} \label{fig r5 FiPh}}
    \caption{Purification of an ensemble of copies of a Bell-diagonal rank-3 state, Eq.~\eqref{eq rank3 state}, with the EIP(2). Each different color corresponds to a different scenario given by the number and kind of errors identified (solid lines). Dashed curves correspond to the average of the different scenarios. In (a), (b), (c) and (d), we plot the local fidelity of the purified ensemble as a function of the number of states of the initial ensemble. The number of states of the purified ensemble depends on each possible scenario. Each different color corresponds to a different scenario given by the number and kind of errors identified (solid lines). Dashed curves correspond to the average of the different scenarios. In (e), (f), (g) and (h), we plot the probability of detecting each of the four possible scenarios as a function of the number of states of the initial ensemble. In a,e) the initial local fidelity is $F = 0.80$. In (b) and (f), the initial local fidelity is $F = 0.90$. In (c) and (g), Initial local fidelity $F = 0.95$. In (d) and (h), the initial local fidelity is $F = 0.99$.}
    \label{fig r5 FiP}
\end{figure}

In Fig.~\ref{fig r5 FiP}, we plot the local fidelity of the purified states as a function of the number of states if a different number of errors are identified. The initial ensemble consists of $n$ copies of a Bell-diagonal rank-3 state and the purification is done via the EIP(2) protocol. In Fig.~\ref{fig r5 FiP}, we also plot the probability of detecting each possible scenario.
\\ \\
We recall that EIP(2) does not provide fidelity one states due to the possibility of the ensemble containing more than $2$ errors. The probability of the ensemble containing more then 2 errors increases with the number of states, the local fidelity of the purified ensemble for each scenario--of a certain number of errors of each kind--decreases as $n$ increases, being $F' = 1$ for $n = 2$. The most relevant feature of the curves is the differences between each scenario. The higher fidelity is archived when no errors are detected, followed by the two different error situations, one error and finally two identical errors. These differences come from the non-considered distributions of states (scenarios with $k > 2$ errors) that are compatible with the output of the performed measurements. The information to distinguish between the distributions with less than three errors indirectly discards other configurations with more errors. For instance, if in the first step we find that $( \# 01 -\# 10) \text{mod} \, 5 = 0$, configurations with 3 errors are no longer compatible. The different slopes of the different curves are due to the most probable configuration with $k > 2$ errors which are still possible after the measurement. When one error is detected, the ensemble could contain also four errors (two 01 + two 10). On the other hand, when two identical errors are detected the ensemble could contain three errors. Since the probability distribution is renormalized after each measurement the probability of the ensemble to contain three errors if two are detected, is much higher than the ensemble containing four if no errors are detected.
\\ \\
The probabilities follow a binomial distribution when our assumption of the number of errors is valid. However, when $n$ increases the outcome of each measurement is random, and the probability of each branch is given by the number of favorable cases over the total number of cases.

\section{\textit{a}EIP(3): Additional details}
\label{sec app detect three errors}

In this section, we introduce the procedure to distinguish between an ensemble with one 01 error and one ensemble with two 01 errors and one 10 error, i.e., $|01\rangle$ or $|01\rangle|01\rangle|10\rangle$. This procedure belongs to the $a$EIP(3).
\\ \\
In this situations, we proceed exactly in the same way as it is described in Sec.~\ref{sec toy model 1 error}, applying the counter gate with each state a number of times corresponding to its position with an auxiliary state of $d=n$ levels, i.e.
\begin{equation}
    \prod_{ i = 1 }^{n} ( b\text{CX}_{i\to \text{aux}} )^i \; \big| \Phi_{\boldsymbol{\mu}} \big\rangle \big| \Psi^{(n)}_{00} \big\rangle_{\text{aux}} = \big| \Phi_{\boldsymbol{\mu}} \big\rangle \big| \Psi^{(n)}_{0j} \big\rangle_{\text{aux}}.
\end{equation}
Measuring the auxiliary state, we determine the new amplitude index $j$ which corresponds to the position of the error in case the ensemble contains only one 01 error, as it is explained in Sec.~\ref{sec toy model 1 error}. In order to distinguish between the three errors situation, we measure the two qubits of the state $|\phi_{\mu_j}\rangle$ in the $Z$ basis. Then, communicating the result to the other, we can verify if the state was a 01 error, a 10 error, or the $|\Psi_{00}\rangle$ state. Here we differentiate two situations: \begin{enumerate}[label=\roman*)]
    \item $| \phi_{\mu_j} \rangle \neq |01\rangle$. If the measured state is not a 01 error, it entails that the ensemble contains more errors. As we have seen in Sec.~\ref{sec toy model 1 error}, if there is one error it must be the state $| \phi_{\mu_j} \rangle$.
    \item $| \phi_{\mu_j} \rangle = |01\rangle$. If the measured state is the $|01\rangle$ state we verify that there are not three errors. This is because configurations with three errors (two 01 plus one 10 error states) are not compatible with this situation. Which we see as follows:
    \\ \\
    At this point of the procedure, there is only one possible scenario of the ensemble containing three errors: $n - 3$ Bell states $|\Psi_{00}\rangle$, two $|01\rangle$ and one $|10\rangle$ error states. The separable states are in the positions $|\phi_{\mu_i} \rangle = |01\rangle$, $|\phi_{\mu_\ell} \rangle = |01\rangle$ and $|\phi_{\mu_k} \rangle = |10\rangle$. Therefore, when we proceed with $b$CX and the auxiliary state $|\Psi^{(n)}_{00}\rangle$ as we described, what we obtain is $j = (i + \ell - k) \, \text{mod} (n)$. Then, if we measure the $|\phi_{\mu_j}\rangle$ state and we obtain that it is the error state $|10\rangle$, it means $j=i$ and therefore $(\ell - k) \, \text{mod} (n) = 0$. However, it is not consistent as $\ell\neq k$ and also as $0 < \ell, \, k \leq n$ then $\ell - k \neq n$ for all $\ell, k$.
\end{enumerate}

\section{\textit{a}EIP(3): Local fidelity and probability}
\label{app sec 3errors}

\begin{figure}
    \centering
    \textbf{\textit{a}EIP(3) for Bell-diagonal rank-3 states}\par\medskip
    \subfloat[\centering]{\includegraphics[width=0.25\columnwidth]{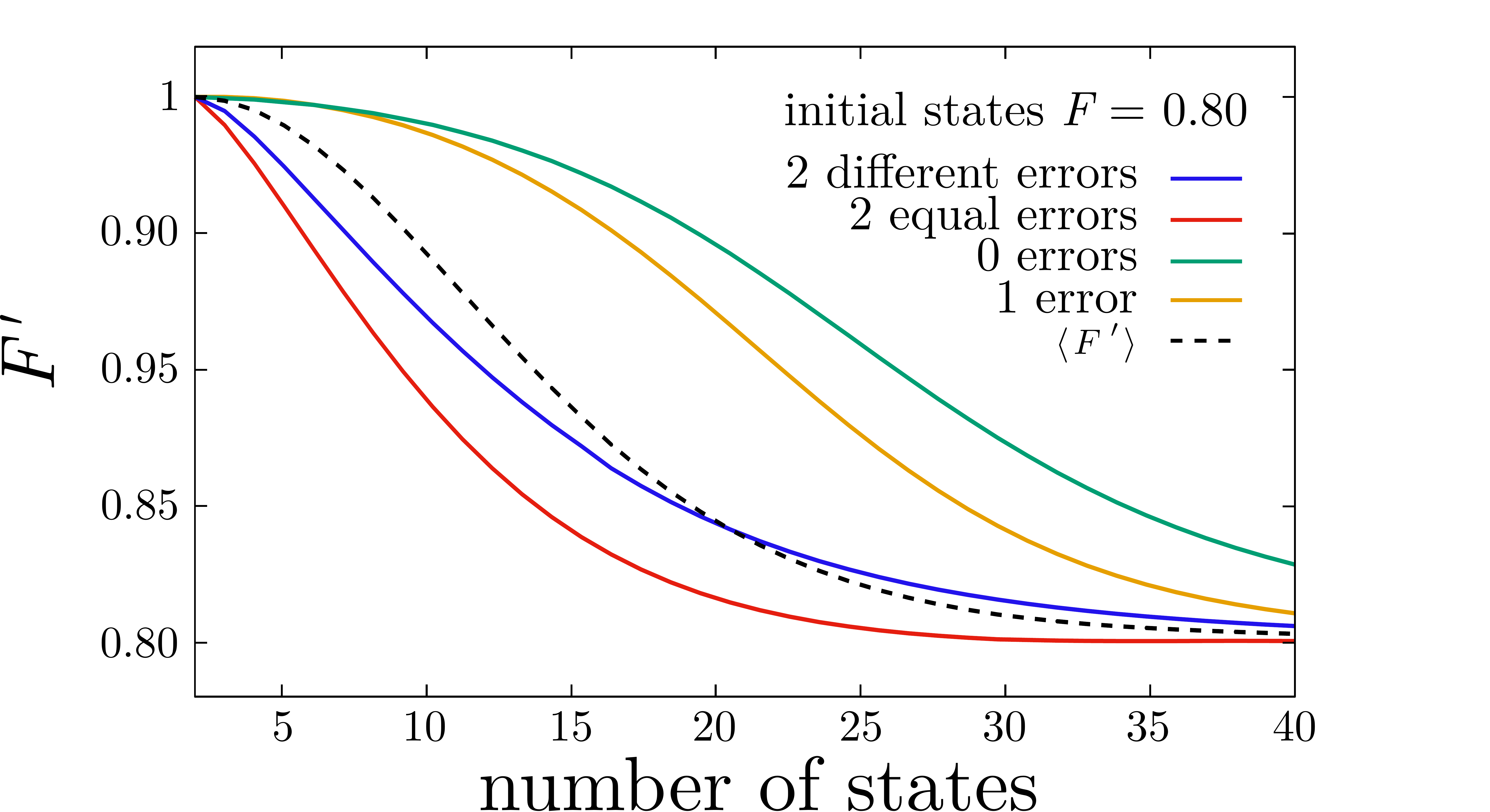} \label{fig r7 FiPa}}
    \subfloat[\centering]{\includegraphics[width=0.25\columnwidth]{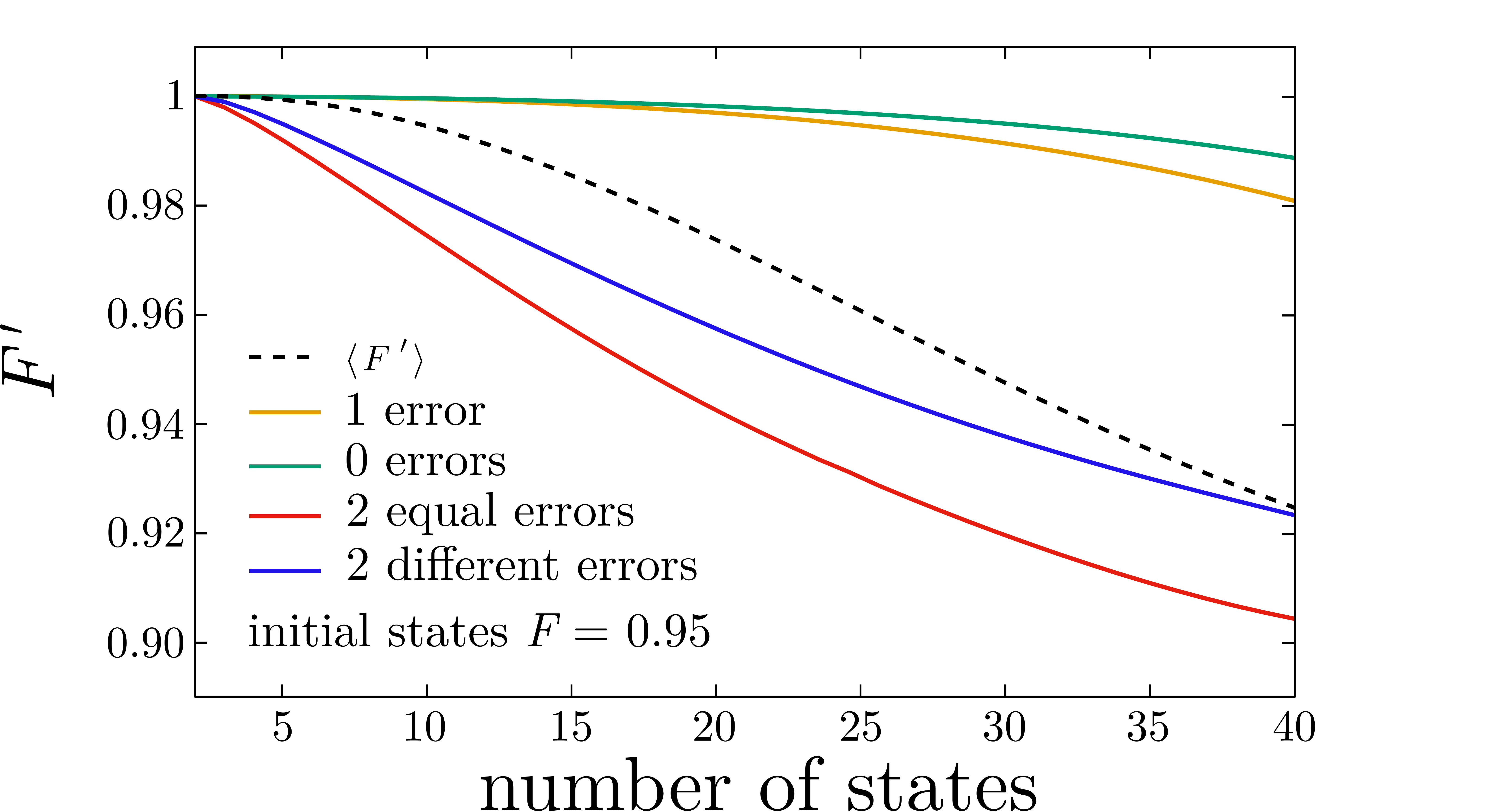} \label{fig r7 FiPb}}
    \subfloat[\centering]{\includegraphics[width=0.25\columnwidth]{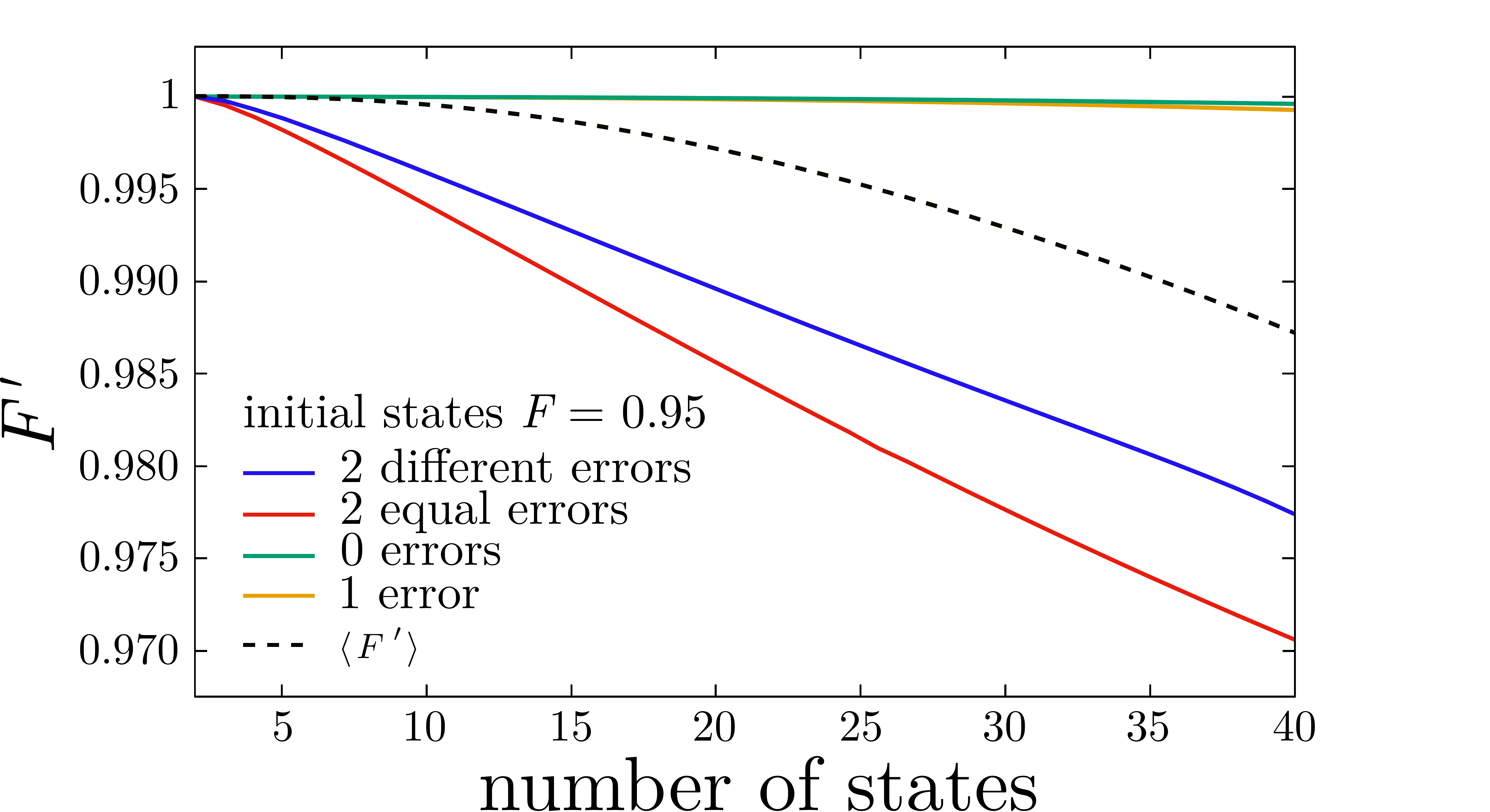} \label{fig r7 FiPc}}
    \subfloat[\centering]{\includegraphics[width=0.25\columnwidth]{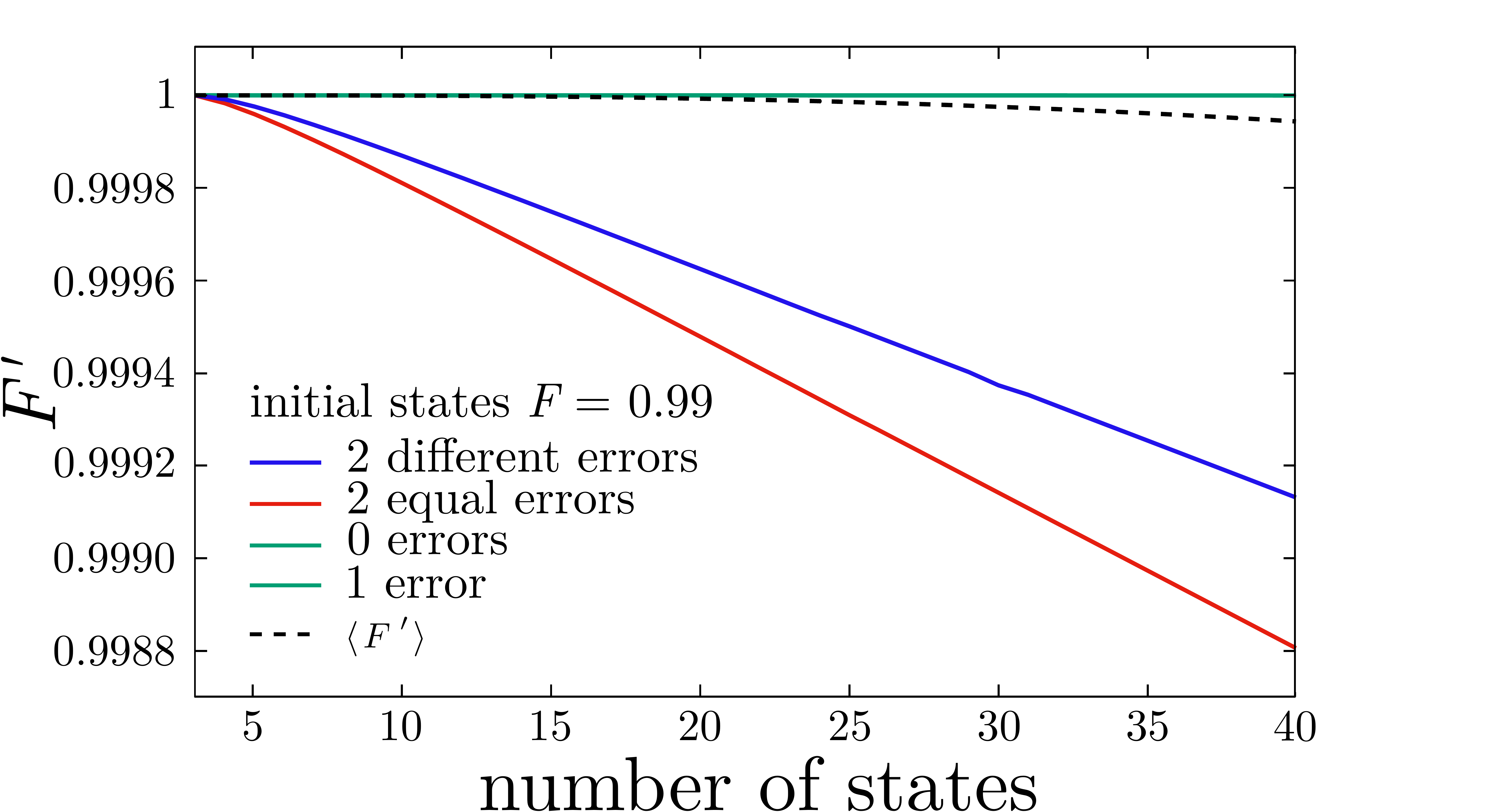} \label{fig r7 FiPd}}\hfill
    \subfloat[\centering]{\includegraphics[width=0.25\columnwidth]{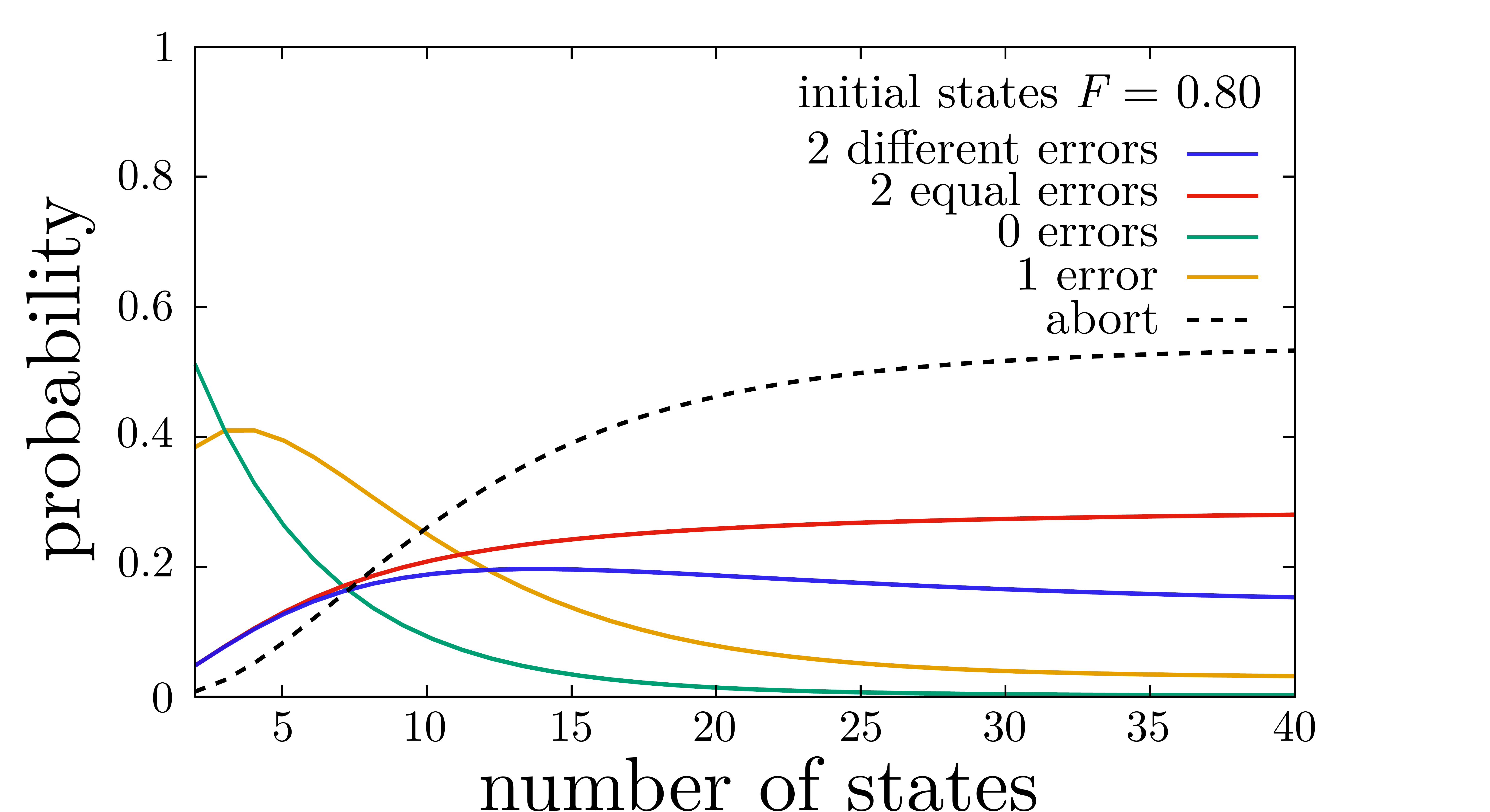} \label{fig r7 FiPe}}
    \subfloat[\centering]{\includegraphics[width=0.25\columnwidth]{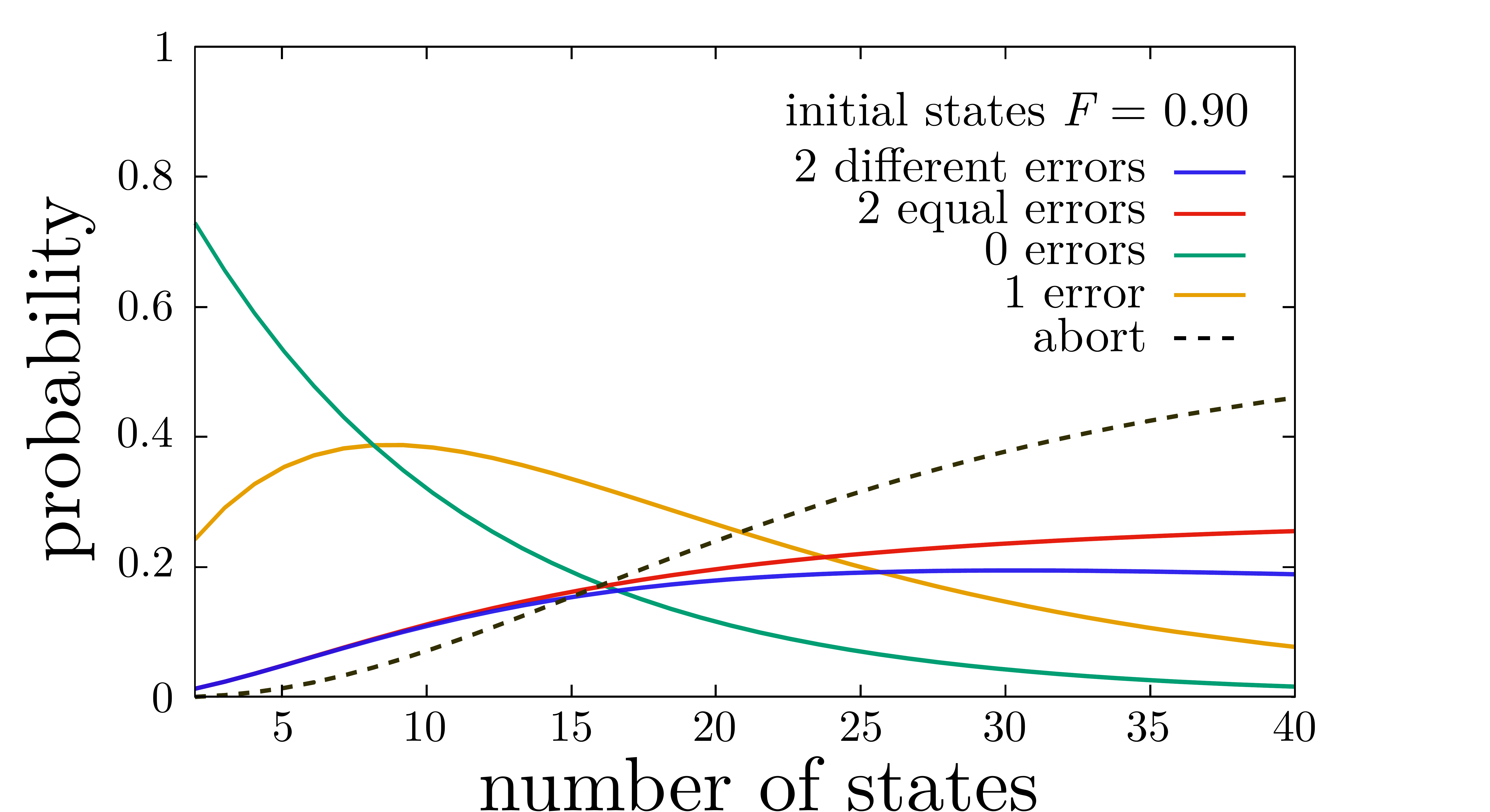} \label{fig r7 FiPf}}
    \subfloat[\centering]{\includegraphics[width=0.25\columnwidth]{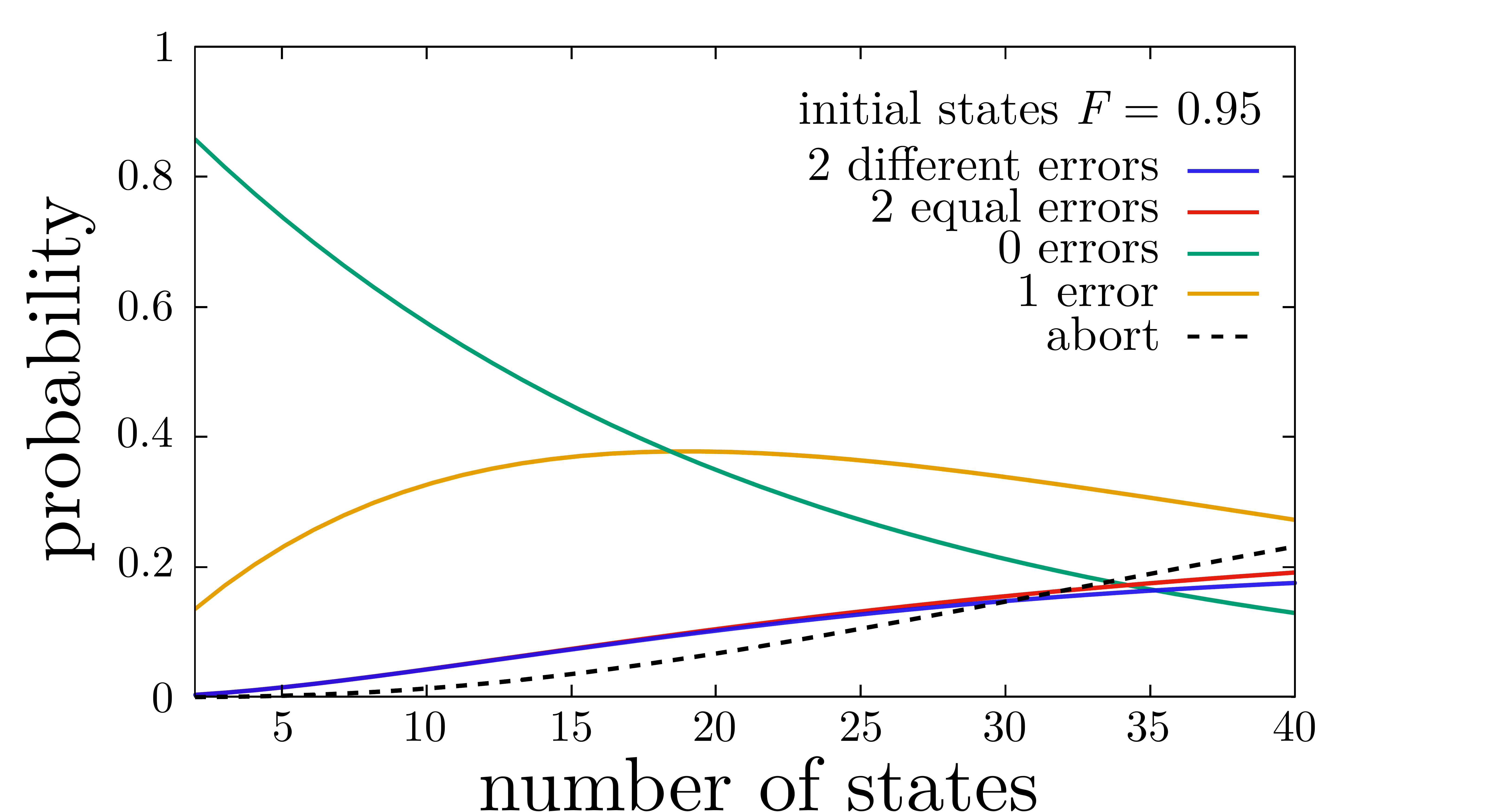} \label{fig r7 FiPg}}
    \subfloat[\centering]{\includegraphics[width=0.25\columnwidth]{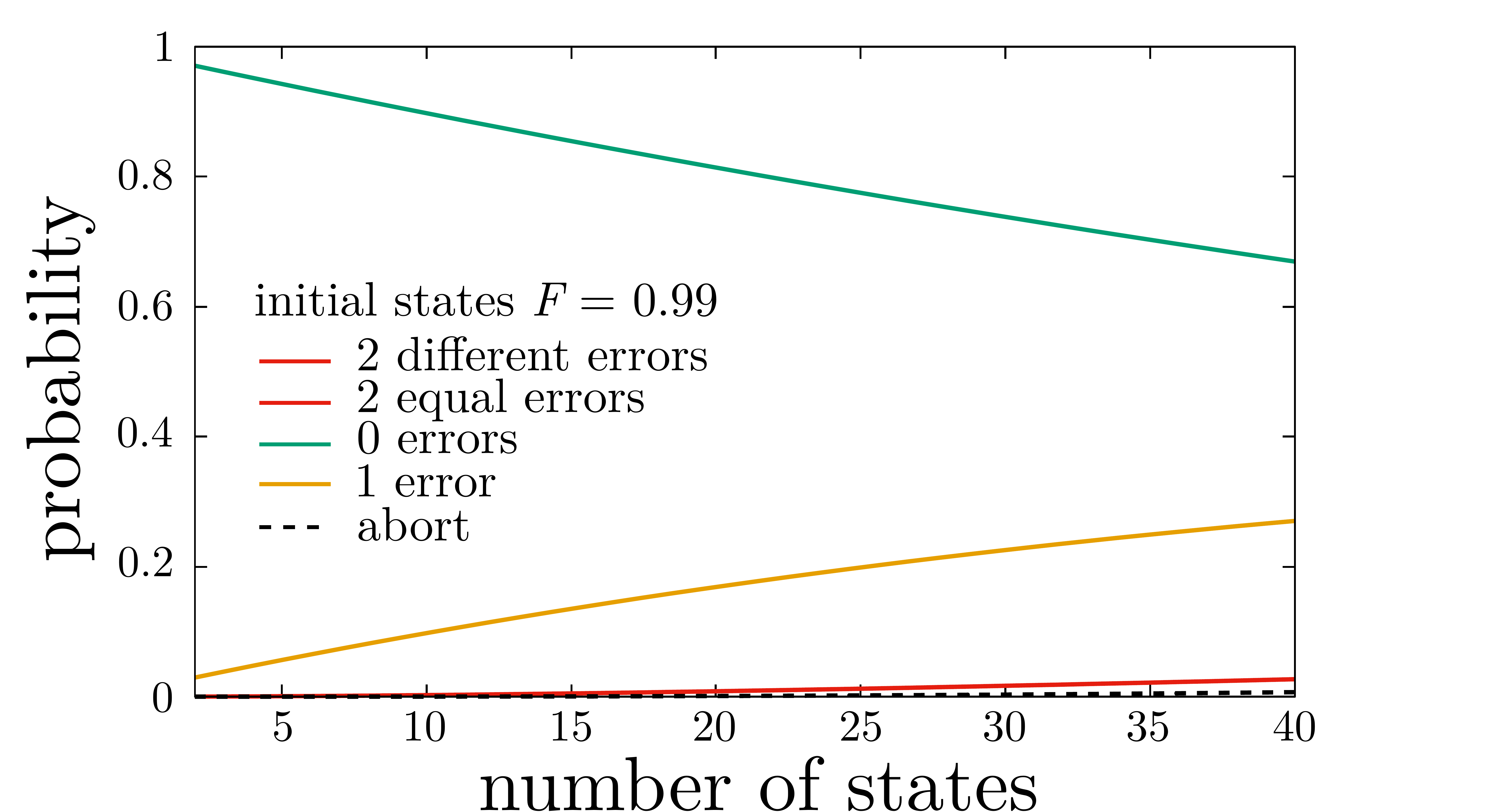} \label{fig r7 FiPh}}
    \caption{Purification of an ensemble of copies of a Bell-diagonal rank-3 state, Eq.~\eqref{eq rank3 state}, with the $a$EIP(3). If three errors are detected the protocol is over and all the ensemble is discarded. In a,b,c,d) we plot the local fidelity of the purified ensemble as a function of the number of states of the initial ensembles. The number of states of the purified ensemble depends on each possible scenario. Each different color corresponds to different scenarios given by the number and kind of errors identified (solid lines). Dashed curves correspond to the average of the different scenarios. In e,f,g,h) we plot the probability of detecting each of the five possible scenarios as a function of the number of states of the initial ensemble. In a,e) the initial local fidelity is $F=0.80$. In b,f) the initial local fidelity is $F=0.90$. In c,g) the initial local fidelity is $F=0.95$. In d,h) the initial local fidelity is $F=0.99$.}
    \label{fig r7 FiP}
\end{figure}

In Fig.~\ref{fig r7 FiP}, we plot the local fidelity of the purified ensemble as a function of the number of states. The purification is performed via the $a$EIP(3), which identifies the errors if two or less are detected. On the other hand, if three are detected the protocol is aborted and we discard the whole ensemble. The curves of the local fidelity are higher but are similar to the ones obtained with the EIP(2). In particular, the cases of one error and the two identical errors present a notably improving. This is because, in the EIP(2), when one error or two identical errors are detected configurations with three errors are still possible. Therefore, discarding the three error configurations leads to an increase in the fidelity especially in these two scenarios.

\section{Error identification protocol for Bell-diagonal states: Additional figure}

\begin{figure}[H]
    \centering
    \textbf{EIP(2) for Werner states}\par\medskip
    \includegraphics[width=0.8\columnwidth]{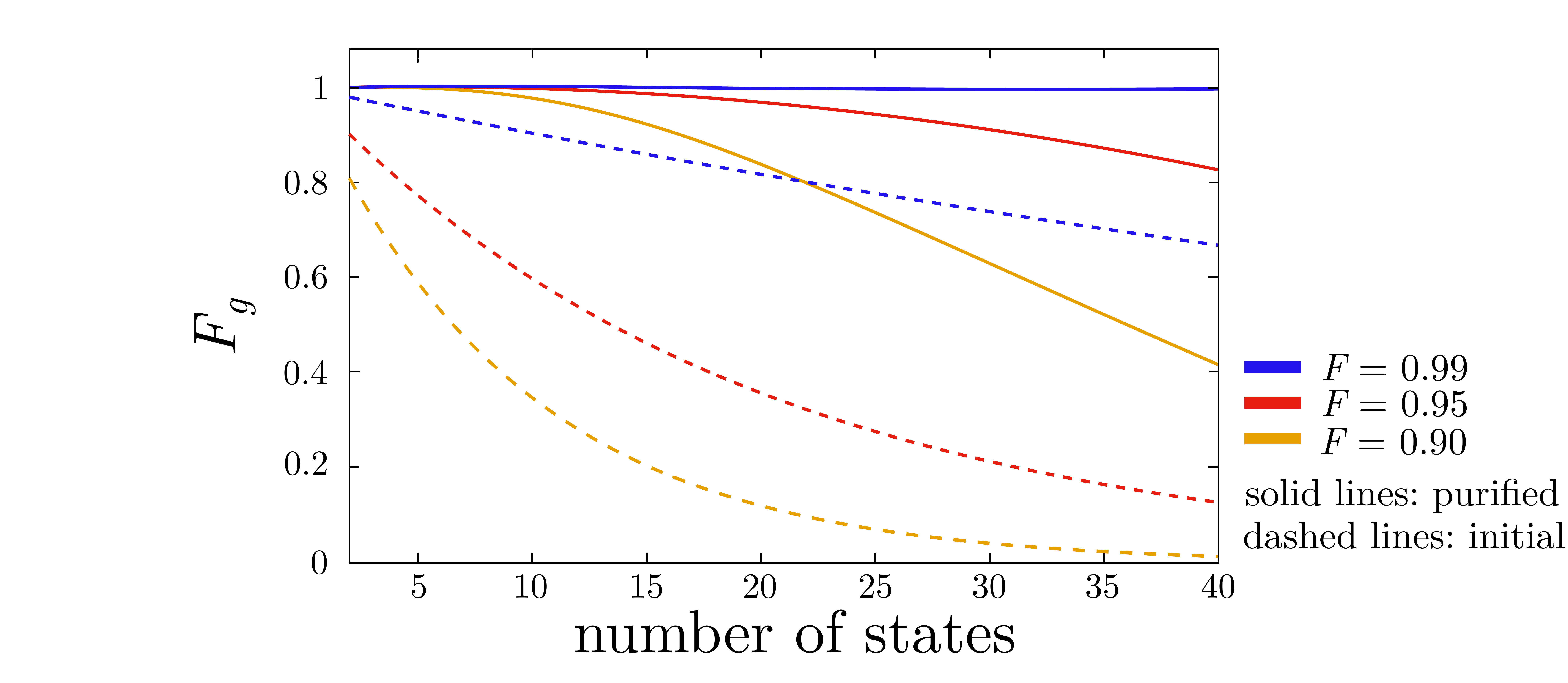}
    \caption{\label{fig Fg werner}Global fidelity of an ensemble of Werner states as a function of the purified ensemble. Solid lines correspond to the purified ensemble and dashed lines to the initial ensemble. The EIP(2) protocol is applied in two rounds.}
\end{figure}

In this section, we consider the purification of an ensemble of copies of a Werner state. Therefore, the initial state of the ensemble is given by
\begin{equation}
    \hat{\rho}_{AB}^{\otimes n} = \left[ F \, \big| \Psi_{00} \big\rangle \big\langle \Psi_{00} \big| + \frac{ 1 - F }{3} \Big( \big| 01 \big\rangle \big\langle 01 \big| + \big|10 \rangle\big\langle 10 \big| + \big| \Psi_{10} \big\rangle \big\langle \Psi_{10} \big| \Big) \right]^{\otimes n},
\end{equation}
where the initial local fidelity and global fidelity are given by $F$ and $F_g = F^n$. In Fig.~\ref{fig Fg werner}, we plot the initial and the final global fidelity. The purification procedure consists in first applying the EIP(2) protocol, followed by a depolarization and transformation of the errors. Then the EIP(2) protocol is applied again to identify the remaining errors.

\section{Error identification protocols vs hashing \& recurrence protocols: Additional details}
\label{app sec EIPvsDEJMPShash}

\begin{figure}
    \centering
    \textbf{EIP vs hashing \& DEJMPS protocols}\par\medskip
    \subfloat[\centering]{\includegraphics[width=0.5\columnwidth]{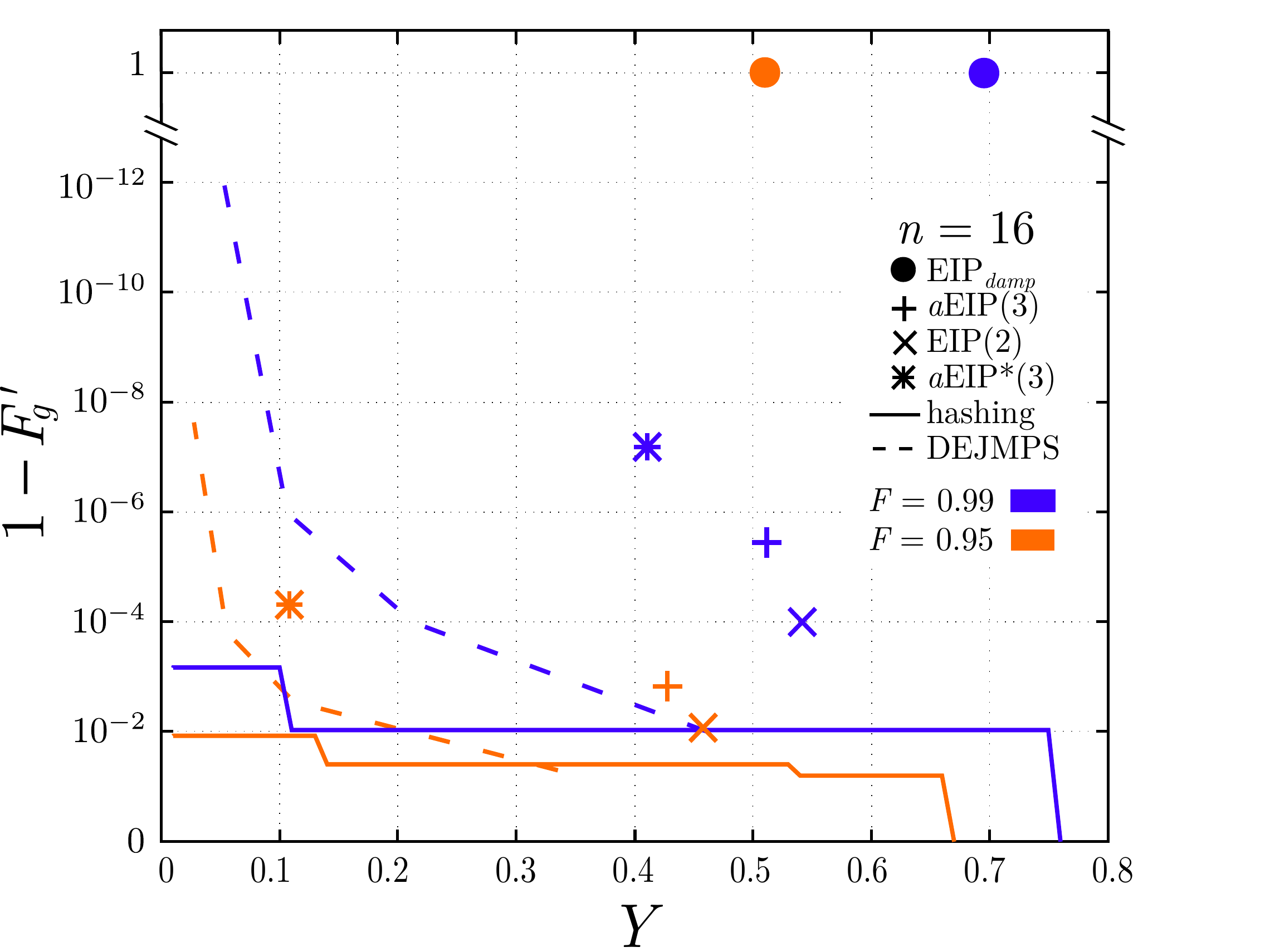}} \subfloat[\centering]{\includegraphics[width=0.5\columnwidth]{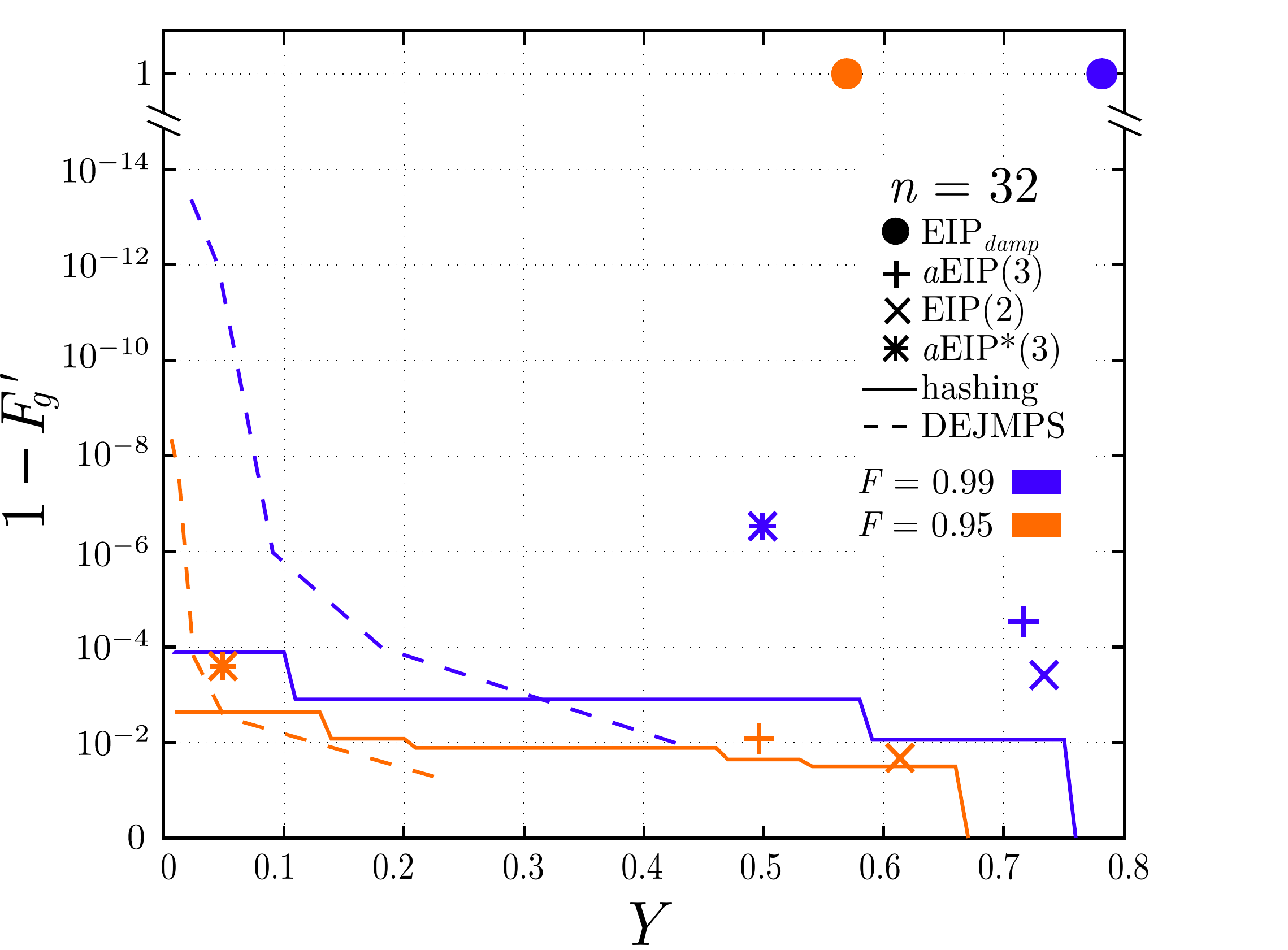}} 
    \caption{\label{app fig EIPvsHashDEJ 2}Local fidelity of the purified ensemble as a function of the yield of the purification. Each point corresponds to a different EIP, dashed lines to the DEJMPS recurrence protocol, and solid lines to the finite hashing protocol, where we use the upper bound for the output fidelity derived in Appendix~\ref{app sec upper bound hash}. Each color corresponds to a different initial fidelity. The initial ensemble consists of $n$ copies of the state of Eq.~\eqref{eq rank3 state}, except for EIP$_{damp}$ where we assume copies of the state of Eq.~\eqref{eq rank2}. In (a), the initial ensemble is of $n=16$ states, and in $b)$ of $n=32$.}
\end{figure}

In this section, we show an alternative way to compare the DEJMPS and the hashing protocols with the EIP. In the main text, we show a plot of the global fidelity of the purified ensemble $F_g'$ as a function of the yield. To understand these plots one has to take into account that the global fidelity quantifies the noise of the whole ensemble, and hence it depends on the number of states of the final ensemble. While the recurrence protocols can achieve a really high output global fidelity after a few iterations, only an exponentially small fraction of the ensemble is purified, as at least the half of the ensemble is lost in each iteration. Therefore, comparing protocols that provide ensembles with different numbers of states it is also of great interest to compare the output local fidelity.
\\ \\
The local fidelity is defined in Eq.~\eqref{eq local fidelty}. In Fig.~\ref{app fig EIPvsHashDEJ 2} we plot the output local fidelity as a function of the yield of the purification. For the hashing protocols, we do not have an expression to compute the local fidelity. Therefore, for this protocol, we plot $F \leq 1 - ( 1 - F_g ) / n$, i.e., the maximum local fidelity that can be reached given the global fidelity of the ensemble, see Appendix~\ref{app fidelities}. We recall that for the hashing protocol we use an unreachable upper bound for the output global fidelity, $F_g \leq 1 - p_1$ (see Appendix~\ref{app sec upper bound hash}) and the second upper bound for the local fidelity. These two assumptions make unreachable the results of Fig.~\ref{app fig EIPvsHashDEJ 2} corresponding to the hashing protocol.
\\ \\
In Table~\ref{tab EIPs} we summarize the main features of all different EIPs analysed in the paper.   
\renewcommand{\arraystretch}{1.5}
\begin{table}[h!]
    \begin{tabular}{|c|c|c|c|}
        \hline
        \textbf{ Protocol }   & \textbf{ Initial copies } & \textbf{ Possibility of abortion } & \textbf{ Output fidelity }  \\ \hline
        $\text{EIP}_{\text{damp}}$ & rank-2 states  & if $k>k^{\max}$ errors are found & $F'=1$ \\ \hline
        $\text{EIP}(2)$ & rank-3 states & no & $F'<1$ \\ \hline
        $a\text{EIP}(3)$ & rank-3 states & if 3 errors are found & $F'<1$ \\ \hline
        $a\text{EIP}^{*}(3)$ & rank-3 states & if 1, 2 or 3 errors are found & $F'<1$ \\ \hline
    \end{tabular}
    \caption{\label{tab EIPs}Comparison of the different EIPs ploted in Figs.~\ref{fig EIPvsHashDEJ} and \ref{app fig EIPvsHashDEJ 2}.}
\end{table}

\section{Embedding}
\label{app sec embedding}

In this section, we show how one can locally obtain amplitude noisy auxiliary states,
\begin{equation}
    \hat{\varrho}^{(d)}_{AB} = \mathcal{F} \, \big| \Psi^{(d)}_{ \, 00 } \big\rangle \big\langle \Psi^{(d)}_{ \, 00 }\big| + \frac{ 1 - \mathcal{F} }{ d - 1 }\sum_{ n = 1 }^{ d - 1 } \big| \Psi^{(d)}_{ \, 0n } \big\rangle \big\langle \Psi^{(d)}_{\,0n}\big|,
\end{equation}
embedding several copies of Bell-diagonal rank-2 states, i.e.,
\begin{equation}
    \label{app rank 2 states}
    \hat{\mu}_{AB} = F \, \big|\Psi_{00} \big\rangle \big\langle \Psi_{00} \big| + \big( 1 - F \big) \big|\Psi_{10}\big\rangle \big\langle \Psi_{10} \big|.
\end{equation}
First of all we have to see that the product state of two maximally entangled state with zero amplitude index of two arbitrary dimensions, i.e., $|\Psi^{(d_2)}_{k0}\rangle_{A_2B_2}\otimes|\Psi^{(d_1)}_{j0}\rangle_{A_1B_1}$, belong in the subspace $span\{\,|\Psi^{(d_1d_2)}_{\,i0}\rangle_{AB}\}_{i=0}^{d_1d_2-1}\in \mathcal{H}_{AB}=\mathcal{H}_{A_1}^{d_1}\otimes\mathcal{H}_{A_2}^{d_2}\otimes\mathcal{H}_{B_1}^{d_1}\otimes\mathcal{H}_{B_1}^{d_2}$. To this aim we use the definition of the generalized Bell states, Eq.~\eqref{eq generalized bell},
\begin{equation}
    \label{app embeding proces}
    \begin{aligned}
        \big| \Psi^{(d_2)}_{ k0 } \big\rangle_{A_2 B_2} \otimes \big| \Psi^{(d_1)}_{j0} \big\rangle_{A_1 B_1} & = \frac{1}{\sqrt{d_1 d_2}} \sum_{ n = 0 }^{d_2 - 1} e^{i \frac{2\pi}{d_1} nk} \big| n, n\big\rangle_{A_2 B_2} \otimes \sum_{m = 0}^{d_1 - 1} e^{i \frac{2 \pi}{d_1} mj } \big| m, m \big\rangle_{A_1 B_1} \\ & = \frac{1}{\sqrt{d_1 d_2}}\sum_{ n =0 }^{d_2 - 1} \sum_{m = 0}^{d_1 - 1} e^{ i \frac{ 2 \pi }{d_1} nk + i \frac{2 \pi}{d_2} mj} \, \big| n, m\big\rangle_{A_2 A_1} \otimes\big|n,m\big\rangle_{B_2 B_1} \\ & = \frac{1}{\sqrt{d_1 d_2}}\sum_{w = 0}^{ d_1 d_2 - 1} \! e^{i 2 \pi \varphi(w)} \big| \, w \, \big\rangle_{A} \otimes \big| \, w \, \big \rangle_{B},
    \end{aligned}
\end{equation}
where $w$ is a label to each pair $(n,m)$. We have chosen $w = n \cdot d_1 + m$, $\varphi(  w(n,m) ) = \frac{nk}{d_1} + \frac{mj}{d_2}$. We also define the basis $\{ |w\rangle \}_{ w = 0 }^{d_1 d_2 - 1}$ as the computational basis of the Hilbert space $\mathcal{H}_{AB}$. Therefore, we can use the expression obtained in Eq.~\eqref{app embeding proces} to see that
\begin{equation}
    \label{eq oli}
    \begin{aligned}
        \big\langle \Psi^{(d_1 d_2)}_{mn} \big| \Big( \big| \Psi^{(d_2)}_{k0} \big\rangle \otimes \big| \Psi^{ (d_1) }_{j0} \big\rangle \Big) & = \frac{1}{d_1 d_2} \left(\sum_{ v = 0 }^{d_1 d_2 - 1} e^{- i \frac{2 \pi}{d_1 d_2} mv} \big\langle v, v \ominus n \big| \right) \left( \sum_{w = 0}^{d_1 d_2 - 1} e^{ i2 \pi \varphi(w)} \big| w, w\big\rangle \right) \\ & = \frac{1}{d_1 d_2} \sum_{v, w = 0}^{d_1 d_2 - 1} e^{- i \frac{2\pi}{d_1 d_2} mv}e^{ i 2 \pi \varphi(w)} \underset{ \delta_{v,w} }{ \underbrace{\big\langle v \big| w \big\rangle }} \big\langle v \ominus n \big| w \big\rangle \\ & = \frac{1}{d_1 d_2}\sum_{ w = 0 }^{d_1 d_2 - 1} e^{- i \frac{2\pi}{d_1 d_2} mw } e^{ i2 \pi \varphi(w) } \delta_{ w \ominus n, w} = \alpha_{ mkj } \delta_{n, 0},
    \end{aligned}
\end{equation}
where $\alpha_{m,k,j}\in\mathbb{C}$. From Eq.~\eqref{eq oli}, we can see that the state of Eq.~\eqref{app embeding proces} is orthogonal to the state of the basis $\{|\Psi^{d_1d_2}_{mn}\rangle\}$ with nonzero amplitude index, i.e.,
\begin{equation}
    \label{resut 2 embedding}
    \big|\Psi^{(d_2)}_{k0} \big\rangle_{A_2 B_2} \otimes \big| \Psi^{(d_1)}_{j0} \big\rangle_{A_1 B_1} = \sum^{d_1d_2-1}_{m = 0} \alpha_{ mkj } \, \big|\Psi^{(d_1d_2)}_{ \, m0} \big\rangle_{AB}.
\end{equation}
\\ \\
Using Eq.~\eqref{resut 2 embedding}, if we take $k$ copies of the mixed state Eq.~\eqref{app rank 2 states} and we perform the same embedding as in Eq.~\eqref{app embeding proces}, the resulting state is
\begin{equation}
    \hat{\mu}^{\otimes k}_{AB} = F^k \, \big| \Psi^{(2^k)}_{ \, 00} \big\rangle \big\langle \Psi^{(2^k)}_{ \, 00} \big| + \big( 1 - F^k \big) \sum_{i, j = 1}^{2^k - 1} \eta_{ij} \big|\Psi^{(2^k)}_{ \, i0} \big\rangle\big \langle\Psi^{(2^k)}_{ \, j0} \big|,
\end{equation}
as $|\Psi^{(d_1)}_{00}\rangle|\Psi^{(d_2)}_{00}\rangle=|\Psi^{(d_1d_2)}_{00}\rangle$.
\\ \\
Then the state is depolarized to its Bell-diagonal form followed by a second depolarization step to equalize $\eta_{ii}=1/(d-1)$. Finally we apply $\text{QFT}\otimes\text{QFT}^\dagger$ to the state $\hat{\mu}_{AB}^{\otimes k}$ to transformed into to the amplitude noisy state, $\hat{\mu}^{\otimes k}_{AB}\to\hat{\varrho}^{(2^k)}_{AB}$
\begin{equation}
    \hat{\varrho}^{(2^k)}_{AB} = F^k \, \big| \Psi^{(2^k)}_{ \, 00} \big\rangle \big\langle\Psi^{(2^k)}_{ \, 00 } \big| + \frac{ 1 - F^k }{ d - 1 } \sum_{ n = 1 }^{ d - 1 } \big|\Psi^{(2^k)}_{ \, 0n } \big\rangle \big\langle\Psi^{(2^k)}_{ \, 0n } \big|.
\end{equation}
Note that only states with $d = 2^{k}$ for $k\in\mathbb{N}$ can be obtained and the fidelity of the final state is given by $\mathcal{F} = F^k$.

\section{Isotropic auxiliary state}
\label{app sec general noise}

The action of the counter gate between the maximally mixed state of two qudits and the $|\Psi_{00}\rangle$ is given by
\begin{equation}
    \begin{aligned}
       b\text{CX} \Big( \big| \Psi_{00} \big\rangle \big\langle \Psi_{00}\big| \otimes \frac{1}{d^2} \id_{d^2} \Big) b\text{CX}^\dag & = \big|00\rangle \big\langle00 \big| \otimes \frac{1}{d^2} \id_{d^2} + \sum_{m, n = 0}^{ d - 1 } e^{- i \frac{2\pi}{d} m} \big| 00 \big\rangle \big\langle 11 \big| \otimes \big| \Psi_{mn} \big\rangle \big\langle\Psi_{mn} \big| \\ & + \sum_{m, n = 0}^{ d - 1 } e^{i\frac{2\pi}{d} m} \big| 11 \big\rangle \big\langle 00 \big| \otimes \big| \Psi_{mn}\big\rangle \big\langle\Psi_{mn}\big| + \big| 11 \big\rangle \big\langle 11 \big| \otimes \frac{1}{d^2} \id_{d^2}, 
    \end{aligned}
\end{equation}
and the action with the computational basis by
\begin{equation}
    \begin{aligned}
        b \text{CX} \Big( \big| m \;\! n \big\rangle \big\langle m \;\! n \big| \otimes \frac{1}{d^2} \id_{d^2} \Big) b \text{CX}^\dag & = \big| m \;\! n \big\rangle \big\langle m \; \! n \big| \otimes\frac{1}{d^2} \id_{d^2}.
    \end{aligned}
\end{equation}
Therefore, when $b$CX is applied between the $\frac{1}{d^2} \id$ state and the $|\Psi_{00}\rangle \langle\Psi_{00}|$, they become correlated. If we trace out the qudits system the state of the qubits is given by
\begin{equation}
    \begin{aligned}
        \text{tr}_{\text{aux}} \left[ \, b\text{CX} \Big( \big|\Psi_{00} \big\rangle \big\langle \Psi_{00} \big| \otimes \frac{1}{d^2} \, \id_{d^2} \Big) b\text{CX}^\dagger \, \right] = \frac{1}{2} \Big( \, \big| 00 \big\rangle \big\langle 00 \big| + \big | 11 \big\rangle \big\langle 11 \big| \, \Big) &.
    \end{aligned}
\end{equation}
Therefore if we measure the qudits state, the state of the qubits becomes separable.

\end{document}